\documentclass[twocolumn]{aastex62}
\usepackage{amsmath}
\usepackage{epsf}
\usepackage{color}
\usepackage{color}
\usepackage{enumitem}

\shorttitle{Reionization with Low Galaxy Escape Fractions}
\shortauthors{Finkelstein et al.}

\newcommand{\sol}{$_{\odot}$}

\newcommand{\HI}{H\,{\sc i}}

\newcommand{\HeII}{He\,{\sc ii}}

\def\arcs{\hbox{$^{\prime\prime}$}}

\turnoffeditone

\begin{document}

\submitjournal{Astrophysical Journal}
\accepted{April 30, 2019}

\title{Conditions for Reionizing the Universe with A Low Galaxy Ionizing Photon Escape Fraction}
\author[0000-0001-8519-1130]{Steven L. Finkelstein}
\affiliation{Department of Astronomy, The University of Texas at Austin, Austin, TX 78712, USA}
\email{stevenf@astro.as.utexas.edu}
\author{Anson D'Aloisio}
\affiliation{University of California, Riverside, CA, USA}
\author{Jan-Pieter Paardekooper}
\affiliation{Universit\"{a}t Heidelberg, Zentrum f\"{u}r Astronomie, Institut f\"{u}r Theoretische Astrophysik, Albert-–Ueberle-–Str. 2, 69120 Heidelberg, Germany}
\author[0000-0003-0894-1588]{Russell Ryan Jr.}
\affiliation{Space Telescope Science Institute, Baltimore, MD, USA}
\author{Peter Behroozi}
\affiliation{University of Arizona, Tucson, AZ, USA}
\author[0000-0002-0496-1656]{Kristian Finlator}
\affiliation{New Mexico State University, Las Cruces, NM, USA}
\affiliation{DAWN Center for Reionization, Niels Bohr Institute, University of Copenhagen, Copenhagen, Denmark}
\author{Rachael Livermore}
\affiliation{University of Melbourne, Melbourne, Australia}
\affiliation{ARC Centre of Excellence for All Sky Astrophysics in 3 Dimensions (ASTRO 3D)}
\author{Phoebe R.  Upton Sanderbeck}
\affiliation{University of Washington, Seattle, WA, USA}
\author[0000-0002-2620-7056]{Claudio Dalla Vecchia}
\affiliation{Instituto de Astrof\'{i}sica de Canarias, La Laguna, Tenerife, Spain}
\affiliation{Departamento de Astrof\'{i}sica, Universidad de La Laguna, La Laguna Tenerife, Spain}
\author{Sadegh Khochfar}
\affiliation{Institute for Astronomy, University of Edinburgh, Royal Observatory, Edinburgh, UK}

\begin{abstract}
We explore scenarios for reionizing the intergalactic medium with
low galaxy ionizing photon escape fractions.  We combine simulation-based
halo-mass dependent escape fractions with an extrapolation of
the observed galaxy rest-ultraviolet luminosity functions to solve for
the reionization history from $z =$ 20 $\rightarrow$ 4.
We explore the posterior distributions for key unknown quantities, including the
limiting halo mass for star-formation, the ionizing
photon production efficiency, and a potential contribution
from active galactic nuclei (AGN).  We marginalize over the allowable parameter space using a Markov Chain
Monte Carlo method, finding a solution which satisfies the most
model-independent constraints on reionization.
Our fiducial model can match observational constraints with
an \emph{average} escape fraction of $<$5\% throughout
the bulk of the epoch of reionization if: \emph{i}) galaxies form
stars down to the atomic cooling limit before reionization and a
photosuppression mass of log ($M_\mathrm{h}$/M\sol) $\sim$ 9 during/after
reionization ($-$13 $<$ $M_\mathrm{UV,lim} < -$11); \emph{ii}) galaxies become more efficient producers of ionizing
photons at higher redshifts and fainter magnitudes, and \emph{iii})
there is a significant, but sub-dominant, contribution by AGN at $z
\lesssim$ 7.  In this model the faintest galaxies ($M_\mathrm{UV} > -$15)
dominate the ionizing emissivity, leading to an earlier start
to reionization and a smoother evolution of the ionized volume
filling fraction than models which assume a single
escape fraction at all redshifts and luminosities.  
The ionizing emissivity from this model is consistent with
observations at $z\!\!=$4--5 (and below, when extrapolated), in contrast
to some models which assume
a single escape fraction.
Our predicted ionized volume filling fraction at $z =$ 7 of $Q_{H_{II}} =$ 78\% ($\pm$ 8\%)
is in modest ($\sim$1--2$\sigma$) tension with observations of Ly$\alpha$ emitters
at $z \sim$ 7 and the damping wing analyses of the two known $z >$ 7 quasars, which prefer $Q_{H_{II},z=7} \sim$ 40--50\%.
\end{abstract}


\section{Introduction}\label{sec:intro}
The reionization of the intergalactic medium (IGM) was the last major phase
change in the universe, when high energy ultraviolet (UV) photons from the
first luminous sources in the universe ionized hydrogen (and singly
ionized helium) in the IGM.
Observational constraints on this epoch come from a variety of complementary
techniques, and are continuously improving in accuracy and growing in number.
Present-day observations constrain the bulk of reionization to be completed by
$z \sim$ 6 \citep[e.g.,][]{fan06,mcgreer15}, though some lines of
sight may remain somewhat neutral to $z
\lesssim$ 5.5 \citep[e.g.,][]{mcgreer15,kulkarni18,pentericci18}.  The beginning of
reionization is less well constrained, and depends sensitively on the
nature of the ionizing sources.  If rare objects such as quasars
provide the bulk of the ionizing photons, reionization likely didn't
get well underway until $z \sim$ 10 \citep[e.g,][]{madau15}.  On the
other hand, if young, massive stars dominated the ionizing
photon budget, reionization may have started much sooner, although the
constraints on the electron-scattering optical depth to the cosmic
microwave background (CMB) measured by \citet{planck16} imply
that the halfway point came at $z \lesssim$ 8.  
The apparent dichotomy between the sharp decline in the number density
of bright quasars at $z >$ 2  \citep[e.g.,][]{richards06,hopkins07} and the
relatively shallower decline in the UV luminosity
density from galaxies \citep[e.g.,][and references therein]{madau14}
  has led to the predominant theory that the bulk of the
  ionizing photon budget came from massive stars.  

Better understanding both
the temporal and spatial evolution of the process of reionization is key to
understanding a variety of unknown physical processes in the early
universe, including the time of the onset of the first stars and
galaxies, the effects of reionization heating on galaxy formation and
growth, and the escape of ionizing photons from galaxies.
Present-day efforts to reconstruct the progress of hydrogen reionization involves several
major uncertainties around the contribution of both massive stars in galaxies and
quasars.
Over the past decade advances in the capabilities of
near-infrared imaging on the {\it Hubble Space Telescope}, necessary
to measure rest-frame UV light in the epoch of reionization, have
led to robust constraints on the observable \emph{non-ionizing} UV
($\sim$1500 \AA) luminosity density from
galaxies in this epoch.

To understand how these galaxies contribute to reionization, one needs
convert this to an \emph{ionizing} emissivity ($\dot{N}_{ion}$; the number of ionizing photons produced per unit
time per unit volume which escape the galaxy) as
a function of redshift \citep[e.g.,][]{finkelstein12b, finkelstein15, robertson13,
  robertson15, bouwens15c, bouwens15}, which is dependent on three factors: the rest-UV non-ionizing specific luminosity density
($\rho_{UV}$), the ionizing photon production efficiency
($\xi_{ion}$), and the escape fraction of ionizing photons
($f_\mathrm{esc}$).  The product of the first two quantities produces the
intrinsic ionizing emissivity produced within galaxies
($\dot{N}_{ion,intrinsic}$), which when multiplied by $f_\mathrm{esc}$
produces the escaping ionizing emissivity $\dot{N}_{ion}$.
This quantity can be used to infer the
evolution of the IGM ionized volume filling fraction (denoted as
$Q_{H_{II}}$) by solving a set of ordinary differential
equations, which depend on this emissivity, the density of hydrogen,
and the recombination time (dependent itself on the clumping factor of
the gas and the temperature-dependent recombination coefficient; e.g., \citealt{madau99,robertson13}).  

The value of $\rho_{UV}$
is measured by integrating the observed
rest-UV luminosity function to some observationally unknown limiting magnitude.
This limiting magnitude is crucial as the steepening
faint-end slope with increasing redshift
means that the faintest galaxies dominate $\rho_{UV}$ \citep[e.g.,]{bunker04,yan04,bouwens15,finkelstein15}.  
Exactly how dominant these faint sources are depends on the shape of
the extreme faint-end, which should reverse its steep rise due to stellar feedback, the ability of halos to
atomically cool, and Jeans filtering due to the reionization-driven UV
background, as shown by a variety of simulations
\citep[e.g.][]{bullock00,gnedin00,iliev07,okamoto08,mesinger08,finlator11b,finlator12,alvarez12,onorbe17,jaacks18c}.
Informed by this theoretical work, observational studies have commonly
used $M_{UV}\!=-$13 as this integration limit
\citep[e.g.,][]{robertson15,finkelstein15}.  

This limit is
$\sim$100$\times$ fainter than that achievable in even the Hubble
Ultra Deep Field \citep{beckwith06} at these redshifts.  However, recent observations of much fainter galaxies rendered detectable via
gravitational lensing in the Hubble Frontier Fields \citep{lotz17}
have begun to provide empirical justification, with evidence that the observed luminosity functions maintain their steep slopes
down to $M_\mathrm{UV} > -$16 at $z =$ 6 \citep{atek15}, and possibly to
$M_\mathrm{UV} > -$15 \citep{livermore17,bouwens17,atek18}.  
We note the concept of a limiting magnitude is an approximation, as the luminosity function should gradually
roll over rather then exhibit a steep cut-off \citep[e.g.][]{jaacks13,weisz14,mbk15,jaacks18c},
and any cut-off or turnover point will evolve with redshift as
the halo masses evolve, the UV background ramps up, and feedback
effects manifest.  We refer the reader to the recent review by \citet{dayal18} for
further discussion on this topic.

The ionizing photon production efficiency $\xi_{ion}$ 
converts the (dust-corrected) rest-UV non-ionizing specific luminosity density
$\rho_{UV}$ [erg s$^{-1}$ Hz$^{-1}$ Mpc$^{-3}$]
to the intrinsic ionizing emissivity $\dot{N}_{ion,intrinsic}$ [s$^{-1}$
Mpc$^{-3}$].  This efficiency depends on the surface temperatures of
the massive stars, which in turn depends on the stellar metallicity,
age, and binarity, as well as the initial mass function \citep[e.g.,][]{eldridge09,stanway16,stanway18}.  If these quantities were known, one could then
measure $\xi_{ion}$ directly.  There are however large uncertainties,
thus until recently most studies assumed a value of $\xi_{ion} \sim$
25.2, expected from modestly metal-poor, but otherwise-normal, single-star
population models \citep[e.g.][]{finkelstein12b,robertson15}.  This is consistent with the observations that
stellar populations in faint $z \sim$ 7 galaxies are non-primordial
\citep[e.g.,][]{finkelstein10, finkelstein12a,wilkins11,
  bouwens12,bouwens14,dunlop13}, though it is
possible that fainter galaxies have much lower metallicities \citep{dunlop13,jaacks18b}.

\deleted{While full spectroscopic nebular modeling is needed to
directly measure $\xi_{ion}$,}
Recent work has
shown that the typically assumed conversion from observed non-ionizing
to ionizing UV is consistent with the inferred strength of H$\alpha$
emission deduced from IRAC photometric colors
for bright galaxies at $z \sim$ 4 \citep[e.g.,][]{bouwens16}.
However, this same work shows evidence that fainter/bluer galaxies, and
galaxies at $z \sim$ 5 have higher values of $\xi_{ion}$.  This
implies that $\xi_{ion}$ may vary both with galaxy luminosity (and/or
perhaps halo mass) as well as redshift, consistent with the high
values of $\xi_{ion}$ inferred from the few $z \sim$ 7 galaxies with
detectable C\,{\sc iii}] emission \citep{stark17}.  \deleted{This evidence
implies that a single value of $\xi_{ion}$ is not appropriate for all
galaxies at all high redshifts.}

Finally, 
\deleted{to convert the intrinsic ionizing emissivity
$\dot{N}_{ion,intrinsic}$ to that available to ionize the IGM
($\dot{N}_{ion}$)}
one needs to assume an escape fraction ($f_\mathrm{esc}$) for ionizing
photons, which is the dominant source of uncertainty.  A variety of
analyses have shown that when assuming a limiting magnitude of $M_\mathrm{UV}
= -$13 and $\xi_{ion} \sim$ 25.2, an escape fraction of 10--20\% produces
the requisite number of ionizing photons to complete reionization by
$z =$ 6 with no contribution from other sources
\citep[e.g.,][]{finkelstein12b,robertson13,finkelstein15,robertson15,bouwens15b}.
The assumption of a relatively high escape fraction at $z >$ 6 is
impossible to directly verify, as even a
predominantly ionized IGM produces an ionizing optical depth sufficient
to absorb all ionizing UV radiation at $z >$ 4 \citep[e.g.,][]{vanzella18}.
\deleted{(thus unsurprisingly
the most distant galaxy with observed ionizing photon
escape is at $z =$ 4.0.)}

We must observe galaxies at $z < 4$ to directly measure $f_\mathrm{esc}$, where
there is unambiguous observational evidence that \emph{most} studied galaxies have
low escape fractions \citep[e.g.,][though see \citealt{steidel18}]{siana10,sandberg15,rutkowski17,grazian17}.
\deleted{. In particular, studies which stack large samples of star-forming galaxies
in deep imaging datasets which probe $\lambda<$912 \AA\
ubiquitously find non-detections, with stringent upper limits of
$f_\mathrm{esc} <$5\%, and sometimes as low as $<$2\%}
Recent observational programs have improved at
identifying galaxies likely to exhibit higher escape fractions,
specifically those which exhibit intense ionizing environments as
traced by ratios of nebular emission lines, resulting a few dozen
direct detections of escaping ionizing photons
\citep[e.g.,][]{shapley16,debarros16,bian17,vanzella18,izotov18}.
However, the lack of significant ionizing photon escape from the bulk
of galaxies strongly implies that the escape
fraction from \emph{all} galaxies at \emph{all} redshifts cannot be as high
as 10--20\%.  

One way to reconcile this, suggested by a variety
of simulations, is if the escape fraction is dependent on
the halo mass, where lower-mass halos have higher escape fractions due
to lower gas covering fractions and an increased susceptibility of
starburst-driven escape routes \citep[e.g.,][]{paardekooper13,wise14,paardekooper15,anderson17,xu16},
while massive halos occasionally exhibit high escape fractions for
short periods due to extreme starburst-driven winds clearing channels
in the ISM \citep[e.g.][]{paardekooper15}.

\begin{figure*}[!t]
\epsscale{1.05}
\plotone{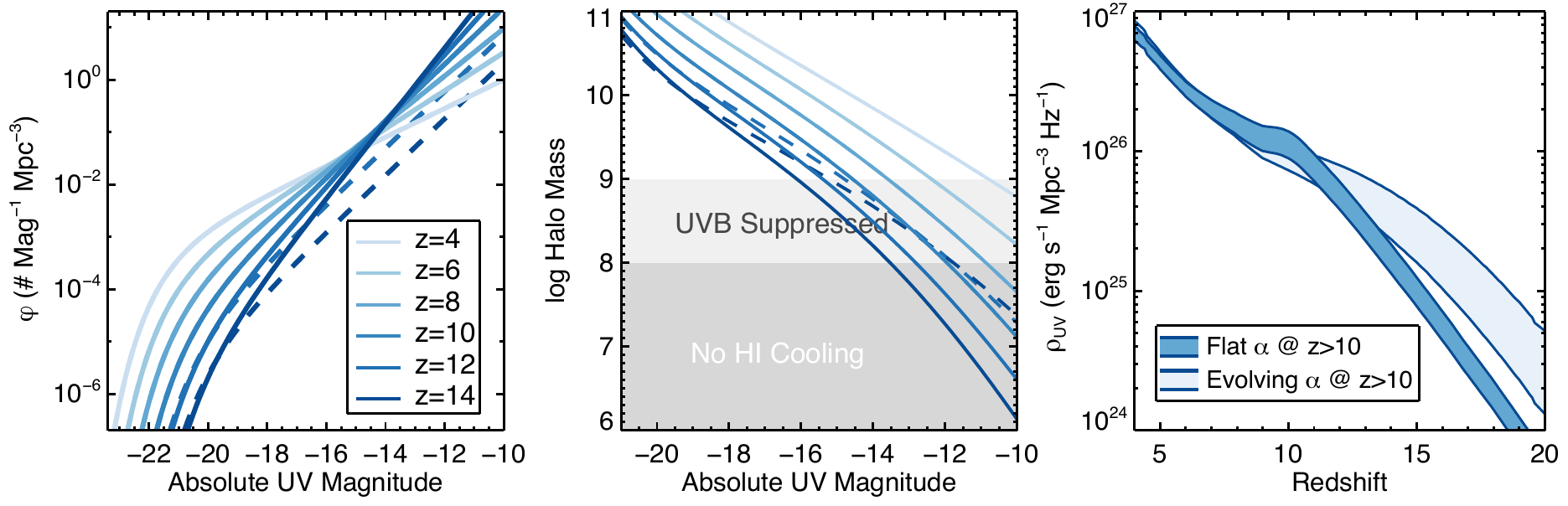}
\caption{Left) The reference luminosity functions from
  \citet{finkelstein16} used for this work, with
  the light-to-dark blue shading denoting $z =$ 4 to 14.  The dashed lines show the case when we fix the faint-end
slope at $z >$ 10 to be equal to the value at $z =$ 10.  Center) The
  relation between halo mass and UV magnitude obtained by abundance matching
  the luminosity functions from the left panel following \citet{behroozi13b}.  The dark gray region denotes the halo mass range
  where H\,{\sc i} cooling likely does not take place.  The lighter gray region
denotes the regime when the post-reionization UV background likely suppresses star
formation.  Right) The
non-ionizing UV luminosity density, highlighting the much shallower
evolution when the faint-end slope is allowed to evolve to extremely
steep values at $z >$ 10.}
\label{fig:fig1}
\end{figure*}   

In this work, we make use of halo-mass dependent escape fractions predicted from
simulations to explore scenarios for completing reionization with low
escape fractions for most observable galaxies. 
In \S 2, we focus on a critical examination of all
assumptions needed to solve for the reionization history, while in \S
3 we discuss our MCMC framework which we use
to probe the full parameter space for all such assumptions, using 
predominantly model-independant reionization observations to
constrain our analysis.  These results are given in \S 4, and
discussed in \S 5.  In \S 6 we explore the implications on the cosmic
star-formation rate density to faint luminosities and extremely high redshifts.  Throughout this paper, we assume AB magnitudes
\citep{oke83} and a Planck 2015
cosmology: H$_\mathrm{0}$=67.74 km s$^{-1}$ Mpc$^{-1}$, $\Omega_{m} =$
0.309, $\Omega_{\Lambda} =$ 0.691, $\Omega_b =$ 0.0486 and Y$_{He} =$
0.2453 \citep{planck15}.  We use the variable $M$ to
denote both halo mass and absolute magnitude, thus we distinguish these
by $M_\mathrm{h}$ and $M_\mathrm{UV}$, respectively.

\section{Defining the Total Ionizing Emissivity}\label{sec:totalnion}

To model reionization we must calculate
the evolution of the ionizing emissivity with redshift, which depends
on a number of variables discussed above.  Here we
attempt to gain new insight into reionization by pairing galaxy
observations with simulated, halo-mass dependent escape fractions.  As
shown in the Appendix, simply replacing a flat 10--20\% escape fraction
with results from simulations is destined to fail due to the
low escape fraction values for all but the smallest halos.
We must thus re-examine the assumptions for \emph{all}
critical variables. We allow this by flexibly exploring the dependence
of reionization on assumptions about the ionizing
photon production efficiency, limiting halo mass for star formation,
and a potential contribution from AGN to the ionizing emissivity.

As described in this section, our model includes seven free parameters
which, when combined with the observed UV luminosity function, define the
emissivity as a function of redshift.  In \S 3 we describe how we constrain the posterior
distribution of these parameters within a
MCMC framework constrained by several robust observations.  
We restrict our model to $z \geq$ 4, as this more than encompasses the
full epoch of hydrogen reionization, and at lower redshifts 
dusty star-forming galaxies (some of which could be absent from our UV luminosity
functions) may contribute to the ionizing
photon budget (e.g., \citealt{gruppioni13,cowie17,koprowski17}, though
see \citealt{casey18}).

\subsection{The Galaxy Ionizing Emissivity}\label{sec:galion}

\subsubsection{Luminosity Functions} \label{sec:mcmclf}
To understand the contribution from galaxies to the ionizing
emissivity, we adopt the
``reference'' luminosity functions of \citet[][hereafter
F16]{finkelstein16}, which were the result of a Markov Chain Monte
Carlo (MCMC) Schechter function fit to all recently published
data at $z =$ 4--10.  Rather than fitting luminosity
functions separately at each redshift, they fit all data
simultaneously, solving for the linear relations of the characteristic
magnitude M$^{\ast}$($z$), faint-end slope $\alpha$($z$) and
characteristic number density $\phi^{\ast}$($z$).  To incorporate the
uncertainties in these fits into our analysis, we make use of the
MCMC chains from F16, using a randomly chosen sample
of 10$^{4}$ chain steps (which were verified to be representative of the
full chain).  We note that while the F16
analysis did not include results from lensed galaxies in the Hubble
Frontier Fields, the faint-end slopes used here are consistent with
studies of those data, which reach to $M_\mathrm{UV} \gtrsim -15$ at
$z =$ 7, finding $\alpha \approx -$2 \citep[e.g.,][]{livermore17,bouwens17,atek18}.

To fully explore the epoch of reionization, it is necessary to
extrapolate these results to higher redshift.  In our analysis, we consider
redshifts from $z =$ 4 to 20.  As the data used to derive these luminosity functions
were limited to $z \leq$ 10, it is unknown if this extrapolation is
valid.  Specifically, the faint-end slope $\alpha$ is observed to
evolve somewhat steeply with redshift ($d\alpha/dz \propto -$0.11),
implying $\alpha(z\!\!=\!\!15)\!\!=\!\!-$2.90.  While this may be possible (indeed
the model results from \citealt{mason15b} find
$\alpha[z\!\!=\!\!16]\!\!=\!\!-$3.51), it is unknown if this is
actually the case.  To cover this
possibility, in our analysis we consider two scenarios: our fiducial
model is one in which
the faint-end slope ceases to evolve at $z >$ 10, and remains at the
$z =$ 10 value of $-$2.35, while in \S~\ref{sec:results_alpha} we also explore the case where $\alpha$
continues to steepen at $z >$ 10.  The
fiducial luminosity function parameters at the redshifts considered here are
given in Table 1, and they are shown in the left panel of
Figure~\ref{fig:fig1}.  

\begin{deluxetable}{cccc}
\vspace{2mm}
\tabletypesize{\small}
\tablecaption{UV Luminosity Function Parameters}
\tablewidth{\textwidth}
% \tablehead{
% \colhead{Redshift} & \colhead{M$^{\ast}$} & \colhead{$\alpha$} &
% \colhead{$\phi^{\ast}$}\\
% \colhead{$ $} & \colhead{(mag)} & \colhead{$ $}  &
% \colhead{(Mpc$^{-3}$)} 
\tablehead{
\multicolumn{1}{c}{Redshift} & \multicolumn{1}{c}{M$^{\ast}$} &
\multicolumn{1}{c}{$\alpha$} & \multicolumn{1}{c}{log $\phi^{\ast}$}\\
%\cmidrule(lr){2-4}\cmidrule(lr){5-7}\cmidrule(lr){8-10}\cmidrule(lr){11-13}
\multicolumn{1}{c}{$ $} & \multicolumn{1}{c}{(mag)} & \multicolumn{1}{c}{$ $} & \multicolumn{1}{c}{(Mpc$^{-3}$)}}
\startdata
4&$-$21.05$^{+0.05}_{-0.06}$&$-$1.69$^{+0.03}_{-0.04}$&$-$2.99$^{+0.04}_{-0.04}$\\
6&$-$20.79$^{+0.05}_{-0.04}$&$-$1.91$^{+0.04}_{-0.03}$&$-$3.37$^{+0.05}_{-0.04}$\\
8&$-$20.52$^{+0.06}_{-0.04}$&$-$2.13$^{+0.05}_{-0.03}$&$-$3.75$^{+0.06}_{-0.04}$\\
10&$-$20.25$^{+0.07}_{-0.06}$&$-$2.35$^{+0.06}_{-0.04}$&$-$4.13$^{+0.08}_{-0.06}$\\
12&$-$19.98$^{+0.09}_{-0.08}$&$-$2.57$^{+0.08}_{-0.06}$&$-$4.50$^{+0.10}_{-0.07}$\\
14&$-$19.71$^{+0.11}_{-0.10}$&$-$2.79$^{+0.10}_{-0.07}$&$-$4.88$^{+0.12}_{-0.09}$\\
16&$-$19.44$^{+0.14}_{-0.13}$&$-$3.01$^{+0.11}_{-0.09}$&$-$5.25$^{+0.14}_{-0.11}$\\
18&$-$19.17$^{+0.16}_{-0.15}$&$-$3.23$^{+0.13}_{-0.11}$&$-$5.63$^{+0.16}_{-0.12}$\\
20&$-$18.90$^{+0.19}_{-0.18}$&$-$3.45$^{+0.15}_{-0.12}$&$-$6.00$^{+0.18}_{-0.14}$
\enddata
\tablecomments{The assumed rest-frame UV luminosity functions used in
  this work, following the evolutionary trend derived via observations
  as discussed in \citet{finkelstein16}.  Our fiducial model keeps the
  faint-end slope $\alpha$ fixed at $z >$ 10 to the $z
=$ 10 value of $-$2.35.}
\vspace{-11mm}
\label{tab:tab1}
\end{deluxetable}

\subsubsection{Abundance Matching}\label{sec:mcmcam}
As discussed in the following two subsections, we must project the
observed UV luminosity function onto the underlying dark matter halo mass
function.  We use this mapping to obtain both the limiting
magnitude and the escape fraction
for a given UV luminosity.  We follow the abundance matching methods of
\citet{behroozi13b} to map the halo mass to each point on our
luminosity functions.
Following \citet{finkelstein15b}, we assume a log-normal UV magnitude
scatter at fixed halo mass of 0.2 dex, though we note that a scatter
as high as
0.4 dex does not affect the $M_\mathrm{halo}$--$M_{UV}$ relation at log
($M_\mathrm{h}$/M\sol) $<$ 11.  Our derived $M_\mathrm{halo}$--$M_{UV}$ relations
are shown in the middle panel of Figure~\ref{fig:fig1} for both luminosity function
evolution cases we consider.

\subsubsection{Minimum Halo Mass for Star Formation}\label{sec:mcmc1b}

A complete accounting of the available photon budget requires us to
include star formation in all galaxies, including those that are
too faint to be observed directly.  A recent analysis indicates that
current observations using lensing at $z =$ 6 probe galaxies hosted by log(M/M\sol)$=$9.5 halos
\citep{finlator16}, and theoretical models generally predict
that star formation in even lower-mass systems is expected \citep[e.g.,][]{paardekooper13,xu16}.
We thus must extrapolate beyond what is observed, yet as 
\deleted{This is especially crucial at $z \geq$ 6, where observations have shown that}
the $z \geq$ 6 luminosity function has a very steep faint-end slope \citep[e.g.,][]{bouwens15,finkelstein15,livermore17}, 
small changes in the minimum luminosity can have a large
impact on the total luminosity density.  

\deleted{Before reionization,} \edit1{Star formation should occur in any halo
which can both retain its gas, and cool it} to temperatures where it can
condense and form stars.  
Efficient cooling via collisional
excitation of H\,{\sc i} can occur in galaxies with halo virial
temperatures below $\sim$10$^4$ K.  This corresponds to log
($M_\mathrm{h}$/M\sol) $\approx$ 8 at $z =$ 6
\citep{okamoto08,finlator12}, and this critical mass shifts to lower values at
higher redshifts, as halos which collapse at earlier
times have steeper dark matter potential wells and thus
correspondingly higher virial velocities \citep[e.g.,][]{barkana01}.  
Lower mass halos can efficiently cool if they
have metals, as predicted by recent simulations which find
that significant star formation is happening down to log
($M_\mathrm{h}$/M\sol )$\approx$ 7 at $z \sim$ 10--15 due to the availability of
metal-line cooling in the immediate aftermath of the formation of
Pop III stars \citep[e.g.,][]{wise14,xu16}. 
Lacking metals, gas can cool inefficiently via molecular
hydrogen cooling, which is believed to be the dominant
cooling mechanism for the first generation of Population III stars
\deleted{which likely formed in minihalos with log
  ($M_\mathrm{h}$/M\sol) $=$ 7--8} forming at $z
=$ 15--30 \citep[e.g.][]{yoshida04,maio10,johnson13,wise14,jaacks18}.
Due to the inefficiency of this method, Population III stars are not predicted to contribute
significantly to the reionizing budget
\citep[e.g.,][]{ricotti04,greif06,ahn12,paardekooper13}, thus we do
not consider star formation in
molecular-cooling halos in this work \citep[though see][]{jaacks18c}.

Once the IGM begins to be photo-heated, even atomic cooling halos will
begin to have their star formation suppressed.  For the lowest mass
halos, ionization fronts in reionized regions will suppress
star-formation in mini-halos with log ($M_\mathrm{h}$/M\sol) $\lesssim$ 8 \citep[e.g.,][]{shapiro04}.  While more
massive halos may\deleted{be able to} self-shield against this
process, gas will not accrete onto dark matter halos with virial temperatures less than the IGM
temperature through Jeans filtering.  \deleted{In reionized
regions} Simulations predict that the halo mass where this process begins to dominate is around log
($M_\mathrm{h}$/M\sol) $=$ 9, \edit1{though the predictions are quite uncertain}
\citep[e.g.,][]{gnedin00,iliev07,mesinger08,okamoto08,alvarez12,ocvirk16,dawoodbhoy18,ocvirk18}.  Feedback
likely also plays a strong role \citep[see][and references therein]{somerville15}, as these small halos have relatively
shallow potential wells, allowing gas to easily be lost.  For example,
\citet{ceverino17} find that stellar feedback causes a flattening in the UV luminosity function at
$M_\mathrm{UV} > -$14, or log ($M_\mathrm{h}$/M\sol) $\approx$ 9.

The physics here are complicated, but in this analysis we wish only to capture the
broad trend of an evolving halo mass where star-formation is suppressed.
We allow star-formation to occur in halos above the redshift-dependent atomic
cooling limit, $M_\mathrm{h,atomic}$, which is given by Equation 26 in \citet{barkana01},
assuming a critical virial temperature for atomic cooling of 10,000 K.
After reionization begins, we implement photo-suppression below a
threshold halo mass due to the rising UV background.
However, there are a range of plausible limiting halo masses for this
photo-suppression to take effect \citep[e.g.,][]{iliev07,mesinger08,okamoto08,alvarez12}.  In addition, even once gas halts
accreting, these galaxies may still form stars for a period of time
until they use up all of their previously accreted gas \citep{sobacchi13}.  We
approximate these uncertainties by adding this photo-suppression mass as
a free parameter, $M_\mathrm{h,supp}$, with an adopted
flat prior of log ($M_\mathrm{h,supp}/M$\sol) $\in$ (8.5,10.5) encompassing the range
found in the literature.  The lower bound was set so that this mass
threshold was never lower than the atomic cooling limit at the
redshifts considered here; \edit1{this could have been avoided by
  allowing the photosuppression mass to be redshift-dependent, but we
  elected to choose a fixed value in the absence of evidence that this
  redshift dependance was needed, and to avoid adding another free
  parameter to our model.}
The non-ionizing specific UV luminosity density ($\rho_{UV}$) is calculated at
each redshift as the integral of the UV luminosity function down to
the magnitude corresponding to this limit, shown in the right-hand panel of Figure~\ref{fig:fig1}.\deleted{for both luminosity
function evolution cases we consider.}
In \S~\ref{sec:mcmc2a}
we describe how our model transitions from the atomic cooling limit to
the photosuppression mass as reionization progresses. 

We reiterate that while our model does not include star-formation
beyond the limits specified here, a number of recent simulations show
star-formation, especially in the pre-reionization universe, in very
low-mass halos of 7 $<$ log $M_\mathrm{h}/M$\sol $<$ 8.  However,
modern high-resolution simulations still predict a turnover in the UV luminosity function
at magnitudes corresponding to approximately the atomic cooling
limit \citep[e.g.,][]{wise14,jaacks18c}.  While the flat luminosity function beyond this turnover
implies star-formation activity is occurring in lower-mass halos, the
shallowing of the luminosity function slope results in these small
systems contributing little to the integrated UV luminosity density.
As \cite{jaacks18c} show in their Figure 17, although their UV
luminosity function continues to $M_\mathrm{UV} > -$8, the UV
luminosity density asymptotes to a constant value when integrating to 
$M_\mathrm{UV} > -$13. Future iterations of our model can better match these theoretical
results by including a turnover in our luminosity function, and the
results may not be inconsequential as these extreme low-mass halos
could have high ionizing photon escape fractions.

\subsubsection{Ionizing Photon Production Efficiency}\label{sec:mcmc1c}

To convert the total \emph{non-ionizing} ultraviolet luminosity density to
the ionizing emissivity, a value for the ionizing photon production efficiency
($\xi_{ion}$) needs to be assumed.  This parameter encompasses all of
the physics of the underlying stellar population, many which likely
evolve with redshift.  For example, the mean metallicity of young stars in
galaxies likely decreases from low-to-high redshift,\deleted{ something which
has been} observationally tracked by a decrease in the typical dust
attenuation \citep[e.g.,][]{bouwens12,finkelstein12a,bouwens14},
leading to\deleted{ Lower metallicity stars have} hotter stellar photospheres as the metal
opacity (mostly due to iron) is lower, and thus
a higher ionizing to non-ionizing UV photon ratio.
\edit1{Another factor to be considered is the effect of binary stars.}  \deleted{
Additionally, recent work on} Stellar population synthesis models which
include binary stars \citep{eldridge09} show that the ionizing flux
is boosted by $\sim$60\% (at low metallicities of $Z <$ 0.3$Z$\sol)
compared to models with isolated stars only \citep{stanway16} due to
both a harder ionizing spectrum from the primary star (which has its
envelope stripped), and an increase in mass for the secondary star,
allowing more massive stars to exist at later ages.

These effects certainly play a role in high-redshift galaxies, where
we cannot directly probe the ionizing flux.  However, the production
rate of ionizing
photons can be \emph{inferred} their via
the detection of nebular emission lines.\deleted{, which, when compared to the observed
non-ionizing UV continuum emission, places constraints on $\xi_{ion}$.}
\citet{bouwens16} inferred H$\alpha$
emission line fluxes from {\it Spitzer}/IRAC photometry at 3.8 $< z <$ 5.0, finding
log $\xi_{ion} =$ 25.34$^{+0.09}_{-0.08}$ erg$^{-1}$ Hz, consistent with typically
assumed values of log $\xi_{ion} \sim$ 25.2--25.3 erg$^{-1}$ Hz in
previous reionization studies
\citep[e.g.,][]{madau99,kuhlen12,finkelstein12b,robertson15}.
At 5.1 $< z <$ 5.4,
\citet{bouwens16} found log $\xi_{ion} =$ 25.48$^{+0.29}_{-0.23}$ Hz
erg$^{-1}$, hinting at evolution towards larger values at higher
redshift, though not at a significant level given the observational uncertainties.
\citet{bouwens16} also find evidence that the bluest galaxies exhibit
even higher values of $\xi_{ion}$, with log $\xi_{ion} =$
25.9$^{+0.4}_{-0.2}$ erg$^{-1}$ Hz for galaxies in their 5.1 $< z <$
5.4 sample with $\beta < -$2.3 (similar results are found for the
faintest galaxies in that sample, which, as shown by
\citet{bouwens14}, are likely also the bluest).

\edit1{\citet{stark15b} and \citet{stark17} measured $\xi_{ion}$ via ionized
carbon emission, finding $\xi_{ion} =$ 25.68$^{+0.27}_{-0.19}$
erg$^{-1}$ Hz in a lensed galaxy at $z =$ 7.045 with an intrinsic
$M_{UV} = -$19.3, and $\xi_{ion}\!\!=$25.6 for three
luminous ($M_{UV} = -$22) galaxies at $z =$ 7.15, 7.48 and 7.73.} 
\deleted{Further evidence for potential changes in $\xi_{ion}$ comes from
\citet{stark15b} and \citet{stark17}, who derive the ionizing
environment within galaxies by observing rest-frame UV emission from
species of ionized carbon.  \citet{stark15b} detected C\,{\sc iv} from
a lensed galaxy at $z =$ 7.045 with an intrinsic $M_{UV} = -$19.3,
inferring log $\xi_{ion} =$ 25.68$^{+0.27}_{-0.19}$ erg$^{-1}$ Hz.
\citet{stark17} published observations of Ly$\alpha$, C\,{\sc iii}],
and [O\,{\sc iii}] (with the latter inferred from {\it Spitzer}/IRAC photometry) from three
luminous ($M_{UV} = -$22) galaxies at $z =$ 7.15, 7.48 and 7.73.  They
used photoionization modeling to infer $\xi_{ion}\!\!=$25.6 for all three
galaxies.} 
Lastly, \citet{wilkins16} investigated the range of 
$\xi_{ion}$ expected from galaxies in the epoch of reionization based
on the BlueTides simulation, finding that simulated galaxies spanned
the range 25 $< \xi_{ion} <$ 26, with the highest values obtained when
assuming low-metallicity stellar population models which include binaries.

\edit1{Taken together, this evidence implies that $\xi_{ion}$ likely depends
on redshift and luminosity,} which we allow in our model via
\deleted{assuming a fixed value of
$\xi_{ion}$ for galaxies at all redshifts and luminosities is likely
not representative of the true ionizing production in distant
galaxies.}  
two free parameters, a redshift dependence $d$log$\xi_{ion}/dz$ and a magnitude dependence
$d$log$\xi_{ion}/dM_{UV}$.  We assume that at our lowest redshift
considered of $z\!\!=$4, galaxies brighter than $M_{UV} = -$20 have log $\xi_{ion} =$ 25.34
erg$^{-1}$ Hz, consistent with the results from \citet{bouwens16} for
this redshift and luminosity.  Galaxies at higher redshifts and/or at
fainter luminosities have values of $\xi_{ion}$ corresponding to
\begin{eqnarray}
log~\xi_{ion}(z,M_{UV}) =25.34 +(z-4)\frac{d\mathrm{log}\xi_{ion}}{dz}
  +\nonumber \\
(M_{UV}-M_{UV,ref})\frac{d\mathrm{log}\xi_{ion}}{dM_{UV}}
\end{eqnarray}
where $M_{UV,ref}$ is the reference magnitude
of $-$20.  We assume flat priors on both free parameters of
$d$log$\xi_{ion}/dz~\in$ (0.0,0.4) and $d$log$\xi_{ion}/dM_{UV}~\in$ (0.0,0.2).
$\xi_{ion}(z,M_{UV})$ was also constrained to have a maximum value of 26.0,
which corresponds to the highest value seen observationally or from
simulations \citep[e.g.,][]{bouwens16, wilkins16,izotov17}.

\subsubsection{Escape Fractions}\label{sec:mcmcfesc}

To derive the ionizing emissivity available to ionize the IGM, we must
combine our intrinsic ionizing emissivity ($\rho_{UV} \times \xi_{ion}$) with 
a model for $f_\mathrm{esc}$, which we anchor in the results from the
high-resolution First Billion Years (FiBY) simulations.
Previous observational studies typically consider only single values, ranging from
10-50\% \citep[e.g.,][]{finkelstein10,finkelstein12b,robertson13,robertson15,bouwens15,finkelstein15b}.
However, as discussed in \S 1, essentially all observations of
escaping ionizing radiation from galaxies (albeit at lower redshifts)
imply smaller escape fractions.  Therefore, rather than assume a
single arbitrary value, here we draw on information provided by
simulations.  While a number of simulations over the past several
years have derived this quantity
\citep[e.g.,][]{razoumov06,gnedin08,yajima11,kim13,kimm14,wise14}, here we
use the high-resolution radiative transfer simulations of ionizing photon
escape from \citet{paardekooper15}.  
These simulations were post-processed on outputs from the FiBY
simulation suite which follows the formation of the first stars and galaxies from
cosmological initial conditions and leads to a realistic galaxy
population at $z\!\!=$6 (Khochfar et al in prep). The escape fraction of
ionizing photons was determined in more than 75,000 haloes by
post-processing the highest resolution FiBY simulations with
high-resolution radiative transfer simulations. The radiative transfer
simulations were run at the same resolution as the hydrodynamics,
allowing the \edit1{average densities within giant molecular clouds in
  which stars are born} to be resolved. This is
essential for the determination of the escape fraction.

\begin{figure}[!t]
\epsscale{1.17}
\plotone{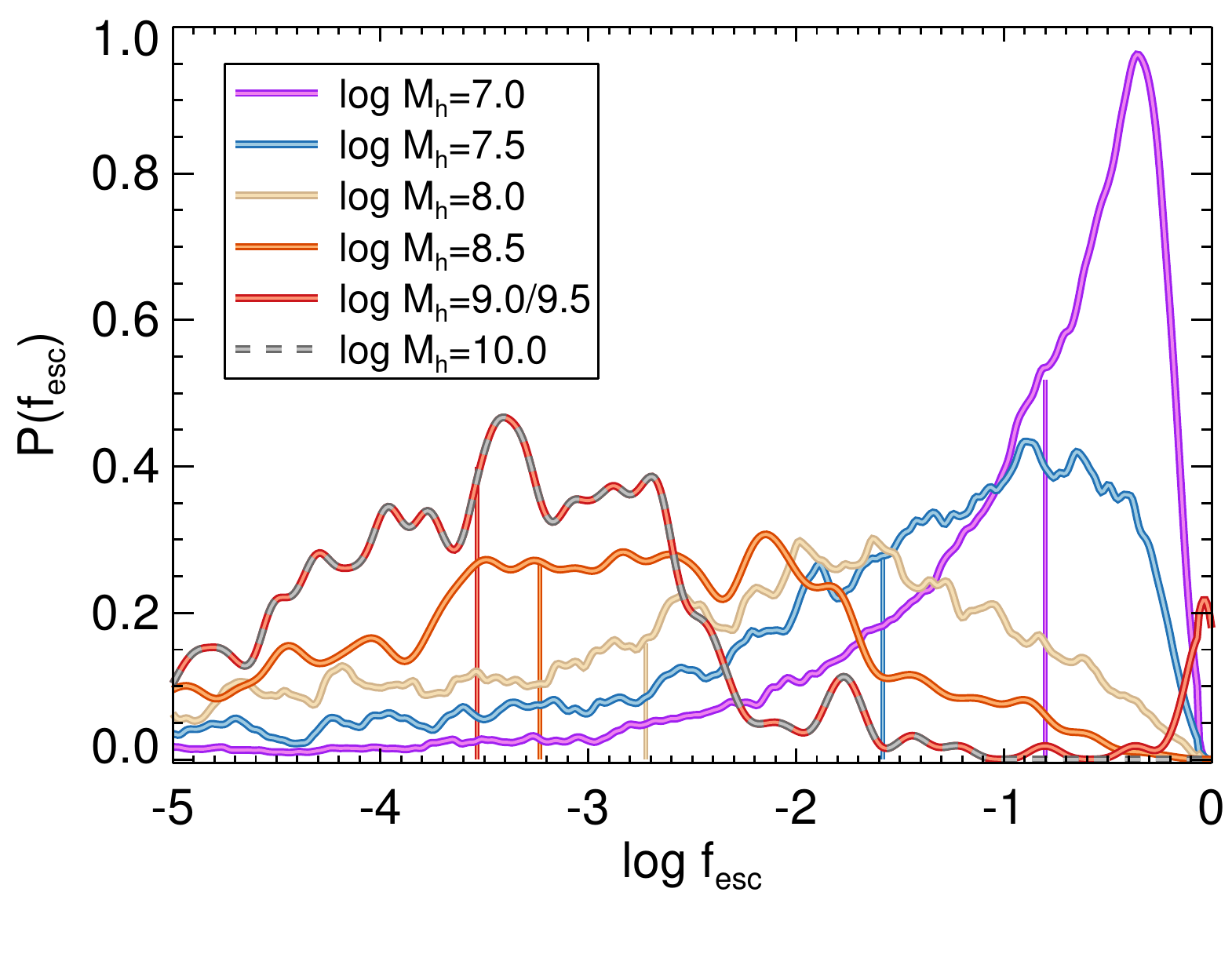}
\caption{The probability distribution functions of the ionizing photon
escape fraction for different halo masses.  These distributions come
from the simulations of \citet{paardekooper15}, based on
high-resolution radiative transfer modeling of 75,000 halos extracted
from the First Billion Years (FiBY) simulation (Khochfar et al., in prep).  While
the escape fraction does not appear to be heavily redshift dependent
in their explored epoch of 6 $< z <$ 15, as shown here it varies quite
strongly with halo mass, with only halo masses with log ($M_\mathrm{h}$/M\sol)
$\leq$ 8.0 having a majority of their probability distribution at
$f_\mathrm{esc} >$ 1\%.  Not shown in this figure is a small peak in
the distribution at log $f_\mathrm{esc} = -$10 for  log
($M_\mathrm{h}$/M\sol) $\leq$ 8.5, comprising $<$10\% of the
probability density.  The thin vertical lines denote the median
value of each distribution, ranging from 16\% for log ($M_\mathrm{h}$/M\sol)
$=$ 7, to $<$0.1\% for log ($M_\mathrm{h}$/M\sol) $\geq$ 8.5.}
\label{fig:fesc}
\end{figure}

Comparing to a number of galaxy properties, this study found that the ionizing
escape fraction is strongly anti-correlated with the halo mass.  
We use their results from all halos that are forming stars, which
results in an escape fraction versus halo mass relation which is
independant of redshift (we will address photoionization feedback
below).  At each halo mass (in steps of log[$M_\mathrm{h}$/M\sol] = 0.5) we compute the
distribution of the escape fraction via the Kernel Density Estimation
(KDE), using cross-validation to compute the optimal KDE
bandwidth.  \edit1{We used 20-fold cross-validation, optimizing how well the KDE fits the remaining data. In every step of the cross-validation, the KDE is constructed on 19/20th part of the data, and the log-likelihood of the remaining 1/20th part of the data fitting this KDE is computed. That is done 20 times (every time changing which part of the data is left out) and the result is averaged. This procedure is repeated for different values of the bandwidth and the bandwidth with the best score has been chosen.}
 While a larger bandwidth would result in a smoother
distribution, it would not fit the edges of the distribution as well
as our adopted bandwidth.  Our adopted escape fraction distributions
are shown in Figure~\ref{fig:fesc}.  
\edit1{These escape fractions are effectively time-averaged, as the
  distributions shown are the average of the instantaneous escape fractions of every halo in the simulation}.
This figure
highlights that the escape fraction distributions are quite broad, but
only halos with log($M_\mathrm{h}$/M\sol) $<$ 8.5 have more than half of their
distribution at $f_\mathrm{esc} >$ 1\%.

At log($M_\mathrm{h}$/M\sol) $=$ 9 there is a small probability that
$f_\mathrm{esc} \gg$ 10\%; these few halos are undergoing an extreme
starburst and the supernova feedback is able to evacuate almost all
of the gas.  As the simulation does not have a
representative sample of halos with log($M_\mathrm{h}$/M\sol) $>$ 9.5, we
assume that halos with log($M_\mathrm{h}$/M\sol) $>$ 11 have $f_\mathrm{esc}\!=$0, and
that halos with log($M_\mathrm{h}$/M\sol) $=$ 9-11 have a similar distribution
as those at halos with log($M_\mathrm{h}$/M\sol) $=$ 9 but without the small
peak at high $f_\mathrm{esc}$ (due to the increased potential making it more
difficult for supernovae to remove all the gas), as shown by the gray dashed line in Figure~\ref{fig:fesc}.
\edit1{We note that if we had treated halos with log($M_\mathrm{h}$/M\sol)
$>$ 11 the same as those with log($M_\mathrm{h}$/M\sol) $=$ 10, we
find almost no differences in the resulting ionization history
(completing at $z =$ 5.7, compared to $z =$ 5.5 for our fiducial model), although in
the post-reionization universe the galaxy emissivity drops off
slightly more shallowly, with a corresponding slight decrease in the
needed AGN emissivity (\S 2.2).}

The normalization of the escape fraction may be inaccurate in the \citet{paardekooper15}
simulations for several reasons. The resolution of the simulations is
insufficient to resolve the birth cloud of the star particles in great
detail, \edit1{potentially missing physics on the scale of individual stars that can affect the escape fraction}.  Simulations have shown that better resolving the ISM around
the stars results in a higher escape fraction because the porosity of
the gas is better accounted for \citep{paardekooper11}. In
addition, the stellar population model in their simulations does not
include the effects of binary interaction, such as mass transfer between
stars and mergers of binaries. These processes have been shown to
affect the average escape fraction in a halo because massive stars in
binaries live longer, and thus emit many ionizing photons when the birth cloud of the stars
has been dissolved by supernova explosions of the single massive stars
in the population \citep{ma16}.

We thus adopt an escape fraction ``scale factor'', where in a given
iteration of our model, the escape fractions at all halo masses are scaled
by the same factor, preserving the halo-mass-dependence of the
escape fraction.  We do not allow this scale factor to vary with
redshift, as the simulations find roughly constant escape fractions at
fixed halo mass through the epoch  6 $< z <$ 15, and the expected
physical reasons for this scale factor do not depend on redshift.  We denote this
parameter below as $f_{esc,scale}$, and adopt a flat prior on
$f_{esc,scale}$ over the range $f_{esc,scale}~\in$ (0,10).

The total ionizing emissivity from galaxies is thus calculated as 
$\dot{N}_{ion,gal} = \rho_{UV} \times \xi_{ion} \times f_{esc}$, where
the latter term includes this scale factor.

\subsection{Inclusion of an Active Galactic Nuclei Contribution}\label{sec:agnion}
\edit1{While quasars have been disfavored as the dominant source of the
reionization ionizing photon budget \citep[e.g.,][]{shapiro87}},
the low observed galaxy escape fractions leave room for
some contribution from active galactic nuclei.  This is not in violation of previous results, as most
observations at $z >$ 4 probe the bright end of
the AGN luminosity function (e.g., quasars only), thus, similar to
galaxies, it may be that the AGN luminosity function has a steepening
faint end slope, and that faint AGNs, and not the rare quasars, are
significant contributors.
\deleted{once quasars were a common inclusion when considering sources of
ionizing photons for reionization, the relative paucity of quasars at
$z >$ 4 found by SDSS, combined with the
observed steepening of the star-forming galaxy luminosity function
faint-end slope, has led many to conclude that quasars were not
significant contributors.  However,} 

\begin{figure}[!t]
\epsscale{1.15}
\plotone{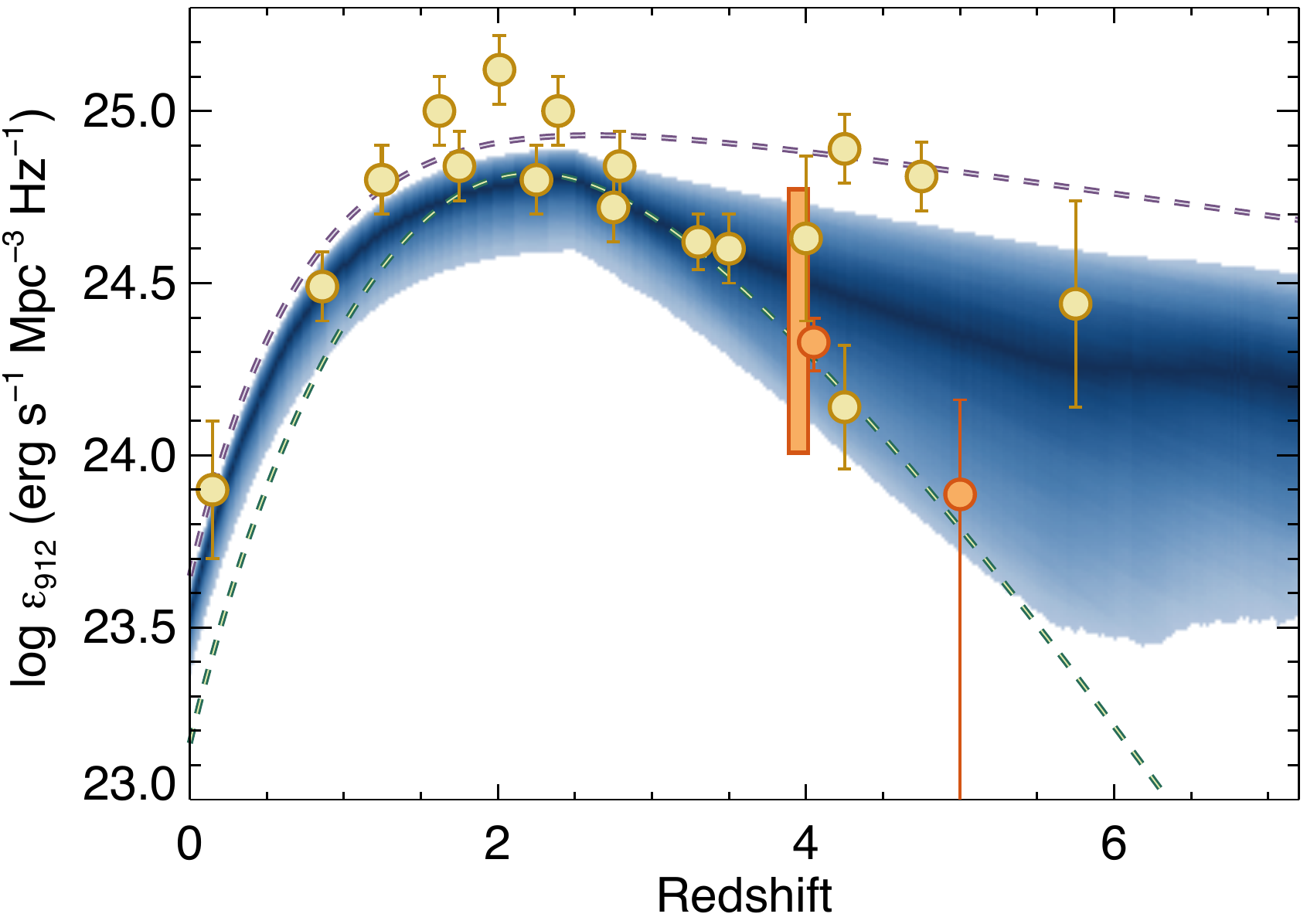}
\caption{The evolution of the AGN comoving monochromatic 912 \AA\ emissivity with
  redshift.  The dashed green line shows the results from the
  \citet{hopkins07} bolometric quasar luminosity function, while the
  dashed purple line shows the form proposed by \citet{madau15}, which
  allows quasars to complete reionization with no contribution from
  star-forming galaxies.  The circles denote results from the
  literature, using the compilation provided by \citet{madau15}, with
  orange symbols denoting the recent results from \citet{mcgreer18}
  and \citet{akiyama18}, and the orange bar denoting the range from \citet{stevans18}.  
The 68\% confidence range on our
  fiducial result is shown as the blue shaded region (with the shading
  density denoting the shape of the probability distribution function), which is
  consistent with the observed data, and roughly
between the two previous evolutionary trends,
at $z\!\!=$4.}
\label{fig:agn}
\end{figure}

There is observational evidence in favor of this possibility, as
\citet{giallongo15} discovered faint AGNs at $z\!\!=$4--6 by searching the
positions of known galaxies at those epochs in deep {\it Chandra}
X-ray data in the GOODS-S field.  At $z \sim$ 4, Giallongo et al.\
found ionizing emissivities nearly an order of magnitude
greater than those implied by the bolometric quasar luminosity function work
of \citet{hopkins07}, and a factor of a few higher than
\citet{glikman11}, with the evolution to $z\!\!=$5--6 shallower than
that of \citet{hopkins07}.
Taken at face value, ionizing photons
generated from AGN could account for the \emph{entire} reionization
photon budget, with no contribution from
galaxies at all \citep{madau15}.
\deleted{They found 22 AGN candidates, probing
several magnitudes fainter on the AGN luminosity function than
previous studies which rely on discovering sources in the X-ray
images.  This led to a followup analysis by
\citet{madau15}, who showed that if the Giallongo et al.\ results
were taken at face value,}
The Giallongo et al.\ results have been met with some
skepticism over the photometric redshifts of the sources
\citep[e.g.,][]{parsa18}, and also the apparent lower emissivity at
similar redshifts \citep[e.g.,][]{mcgreer18}.  Additionally,
Giallongo et al.\ note that they cannot rule out a significant stellar
contribution to the X-ray luminosity.
However, given the difficulties in isolating faint AGN at high redshift
\citep[e.g.,][]{stevans18}, and the fact that current observations
span a large range at $z \geq$ 4, we allow a
contribution from AGN to the ionizing budget in our
fiducial model.    Figure~\ref{fig:agn} shows the inferred evolution of the AGN comoving
ionizing emissivity both from the previous work by \citet{hopkins07},
and the ``quasars can do it all'' recent work by \citet{madau15},
along with a number of observational results from the literature.  

Our initial emissivity matches \citet{madau15} at $z
<$ 2.5, and at higher redshifts is a simple exponential with a slope
constrained to be between those of
\citet{hopkins07} at the low end, and \citet{madau15} at the high
end, spanning the full range of observational results.  Our emissivity is governed by three free parameters: a scale
factor $AGN_{scale} ~\in$ (0,1) applied to the emissivity allowing it to be lower than initially
assumed (due to a range of physical effects, including a non-unity AGN
ionizing photon escape fraction, which is likely the case for
less-luminous AGNs; \citealt{trebitsch18}); a redshift-evolution exponential slope $AGN_{slope} \in$
($-$1.05,$-$0.34), which approximately reproduces the \citealt{hopkins07} and \citealt{madau15}
respective AGN ionizing emissivity evolution in this formalism; and a
maximum redshift $z_{AGN,max}$, above which the AGN ionizing emissivity is assumed to
be zero.  The functional form for our monochromatic 912 \AA\ emissivity is given by
\begin{eqnarray}
\epsilon_{912,AGN}(z < z_{AGN,max}) = AGN_{scale}~e^{~z * AGN_{slope}} \nonumber \\ 
* \left(\frac{10^{25.15*e^{-0.0026*z_{eq}}-1.5*e^{-1.3*z_{eq}}}}{e^{z_{eq}*AGN_{slope}}}\right)
\end{eqnarray}
where the first exponential term is the initial emissivity, and the term in parentheses is a normalization
factor, normalizing our emissivity (prior to the application of a scale
factor) to be equal to that of \citet{madau15} at $z_{eq}\!\!=$2.5
(whose functional form is given in the numerator).  We note that while
this emissivity is included in our analysis, it is allowed within our
formalism to be negligibly low in the epoch of reionization, thus we
are not ``forcing'' AGNs to contribute significantly.  We discuss our
fiducial results in \S 4, but they are shown in
Figure~\ref{fig:agn}, falling roughly in the middle of the allowed
range.

The total emissivity from our model at a given redshift is the sum of
that from galaxies (\S~\ref{sec:galion}) and AGN: $\dot{N}_{ion}(z) = \dot{N}_{ion,gal}(z) + \dot{N}_{ion,AGN}(z)$. 

\begin{deluxetable*}{cccll}
\tabletypesize{\small}
\tablecaption{MCMC Model Parameters}
\tablewidth{0.9\textwidth}
\tablehead{
\colhead{Parameter Name} & \colhead{Initialization}  &
\colhead{Initialization} & \colhead{Flat prior}  & \colhead{Posterior} \\
\colhead{$ $} & \colhead{Central Value}  &
\colhead{$\sigma$} & \colhead{constraints} & \colhead{Median (68\% Confidence)}
}
\startdata
f$_{esc,scale}$$^{a}$&5.0&1.0&$\in$~0, 10&\phantom{$-$}5.2 (3.3 to 7.5)\\
log (M$_{h,supp}$/M\sol)$^{b}$&9.0&0.5&$\in$~8.5, 10.5&\phantom{$-$}8.9\phantom{0} ($<$9.5)\\%\phantom{$-$}8.9 (8.6 to 9.5)\\
$d$log$\xi_{ion}/dz^{c}$&0.10&0.05&$\in$~0, 0.4&\phantom{$-$}0.13 (0.05 to 0.25)\\
$d$log$\xi_{ion}/dM^{d}$&0.05&0.03&$\in$~0, 0.2&\phantom{$-$}0.07 (0.03 to 0.13)\\
AGN$_{scale}$$^{e}$&0.8&0.2&$\in$~0, 1&\phantom{$-$}0.77 ($>$0.47)\\%\phantom{$-$}0.77 (0.47 to 0.94)\\
$z_{AGN,max}$$^{g}$&10.0&2.0&$\in$~4, 12&\phantom{$-$}9.2\phantom{0} ($>$6.9)\\%\phantom{$-$}9.5 (6.8 to 11.3)\\
AGN$_{slope}$$^{f}$&$-$0.5&0.3&$\in$~$-$1.2, $-$0.1&$-$0.39 ($>-$0.93)\\%$-$0.32 ($-$0.84 to $-$0.14)
\enddata
\tablecomments{The free parameters for our fiducial model.  The
  initialization central value and initialization $\sigma$ define a
  normal distribution from which each walker draws an initial value.
  $^{a}$ The scale factor applied to the halo-mass-dependent escape fractions
 from the \citet{paardekooper15} simulations.  $^{b}$
 Post-reionization photosuppression  halo mass. $^{c}$  Evolution of
 ionizing photon production efficiency with redshift and
 $^{d}$absolute magnitude.  $^{e}$Scale factor applied to the AGN
 emissivity (mimicking an AGN ionizing photon escape fraction).
 $^{f}$ Exponential slope of the AGN emissivity with redshift,
 constrained to be zero at some $^{g}$maximum redshift.  The last column
 gives the median of the posterior distribution, and the central 68\%
 confidence range (or upper/lower 84\% confidence limits when the
 distribution is one-sided).}
\vspace{-5mm}
\label{tab:tab2}
\end{deluxetable*}

\subsection{Calculating $Q_{H_{II}}$} \label{sec:diffeq}
We calculate the IGM volume ionized fraction $Q_{H_{II}}$
by solving the differential equation
\begin{equation}
\dot{Q}_{H_{II}} = \frac{\dot{N}_{ion}}{\left<n_H\right>} - \frac{Q_{H_{II}}}{t_{rec,H}}
\end{equation}
where $\dot{N}_{ion}$ is the comoving ionizing emissivity derived above,
$\left<n_H\right>$ is the comoving hydrogen density, and $t_{rec,H}$ is
the IGM hydrogen recombination time.  The comoving hydrogen density is
calculated as the product of the hydrogen mass fraction $X_{p}$
(defined as $1-Y_{He}$, where $Y_{He}$ is the helium mass fraction), the
dimensionless cosmic baryon density $\Omega_b$, and the critical
density $\rho_c$ (defined as 3$H_0^2$/8$\pi$$G$).  The IGM
recombination time is given by
\begin{equation}
t_{rec,H} = \left[C_{H_{II}} \alpha_B(T) \left(1+ Y_{He}/4X_p\right) \left<n_H\right>(1+z)^3 \right]^{-1}
\end{equation}
where $\alpha_B(T)$ is the temperature-dependent case B recombination
coefficient for hydrogen using the functional form given by
\citet{hui97}.  Following \citet[][hereafter R15]{robertson15}, we
evaluate this at $T\!\!=$20,000 K (had we assumed 15,000 K, $\alpha_B(T)$
would be higher by a factor of 1.29).  We assume a redshift-dependent hydrogen clumping factor
$C_{H_{II}}$ from the simulations of \citet{pawlik15}, which
evolves from $C_{H_{II}}\!\!=$4.8 at $z =$ 6, to $C_{H_{II}}$=1.5 at $z =$
14.  We solve for $Q_{H_{II}}(z)$ by integrating the ordinary
differential equation in Equation 1 using the IDL routine
$\tt{ddeabm.pro}$ from $z\!\!=$20 to $z\!\!=$4.

\section{Exploring the Full Reionization Parameter Space with MCMC}\label{sec:mcmc}

Using the set of seven free parameters defined in \S 2 our
model can describe the escaping ionizing emissivity from both star-forming
galaxies and super-massive black hole accretion (AGN) activity.  In
this section we describe how we
use a MCMC framework to derive the posteriors on these free parameters
using a set of robust observational constraints.
We used an  IDL implementation of the affine-invariant sampler
\citep{goodman10} to sample the {\it posterior}, which is similar in
production to the Python \texttt{emcee} package \citep{foreman-mackey13}.  We
used the recommended stretch parameter of $a\!=\!2$ with 1000 walkers.
Each walker was initialized by choosing a starting position for each
of the free parameters, randomly drawn from a normal distribution with
a central value and width given in Table 2.  We assumed a flat
prior on each of our seven free parameters, with the prior bounds also
listed in Table 2.  If the log likelihood of a given set of parameters
was not finite (i.e., it violated the parameter flat priors), a new set of parameters was
drawn, until a set which gave a finite probability was drawn to initialize each
of the 1000 walkers.  The exact initialization values 
are not crucial as the burn-in process ensures that the starting
positions do not affect the results.

\subsection{Method}\label{sec:mcmc2a}

In this sub-section we describe in detail our MCMC analysis.  A flowchart
of this procedure is shown in Figure~\ref{fig:method}.
In each step of the chain, our routine used the chosen set of seven free
parameters to complete the following calculations:
\begin{enumerate}[leftmargin=*]

\item  For each redshift interval of $\Delta z$=0.1 from $z\!\!=$4 to
  20, we use a randomly drawn set of Schechter function parameters from
  the available F16 MCMC chains, where the drawn parameters are the redshift-evolution terms
$dM^{\ast}/dz$, $d\alpha/dz$, and $d\phi^{\ast}/dz$.
We use these parameterizations to calculate the non-ionizing specific
  UV luminosity density $\rho_{UV}(z,M_{UV})$ in absolute magnitude bin intervals of
  $\Delta M_{UV}$=0.1 from $-$6 to $-$24.  These were then corrected for
  dust attenuation ($\rho_{UV,corr}[z,M_{UV}]$) using the method described
  in Finkelstein (2016),
  which uses the relation between $M_\mathrm{UV}$ and $\beta$ from
  \citet{bouwens14}, the relation between $\beta$ and A$_{UV}$ from
  \citet{meurer99}, and the dust attenuation curve from
  \citet{calzetti00}.  A scatter in $\beta$ at fixed $M_\mathrm{UV}$ of 0.35
  was assumed, and zero dust attenuation was assumed at $z \geq$ 9
  \citep{bouwens14,wilkins15}.  We note that given the halo mass
  dependency of the escape fractions the bulk of ionizing photons come
  from faint galaxies with minimal dust, thus our final results are
  not sensitive to this correction.  \edit1{To validate this, we
    performed a model run with no dust correction, and found no
    significant change in the evolution of the ionization history}.

\begin{figure*}[!t]
\epsscale{1.15}
\plotone{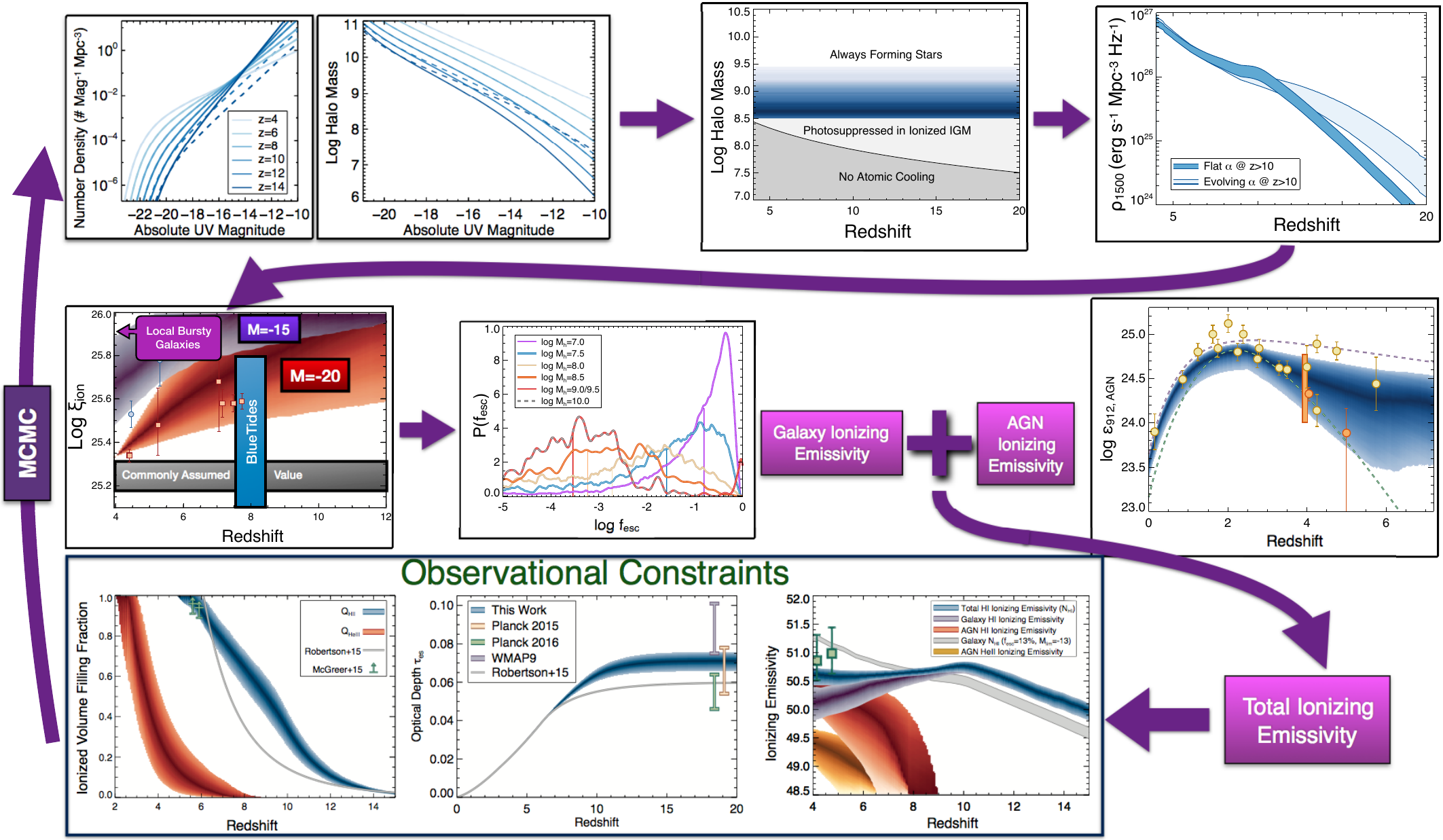}
\caption{A visual description of our Markov Chain Monte Carlo
  procedure for constraining the posteriors on our free parameters,
  described in full in \S~\ref{sec:mcmc2a}.  All figures appear full-size elsewhere in the
  paper.}
\label{fig:method}
\end{figure*}  

\item The intrinsic ionizing emissivity $\dot N_{gal,intrinsic}(z,M_{UV})$
  was calculated by multiplying $\rho_{UV,corr}(z,M_{UV})$ by the
  appropriate value of $\xi_{ion}(z,M_{UV})$ for the values of
  $d$log$\xi_{ion}/dz$ and $d$log$\xi_{ion}/dM_{UV}$ in a given step.  The escaping ionizing
  emissivity $\dot N_{gal}(z,M_{UV})$ was then calculated as $\dot
  N_{gal,intrinsic}(z,M_{UV})$ multiplied by the escape fraction, where the
  escape fraction is randomly drawn in each
  step of the chain for each absolute magnitude interval, from the
  $f_\mathrm{esc}$ probability distribution function (PDF) corresponding to the
  halo mass for the given absolute magnitude (from the $M_\mathrm{UV}$--$M_\mathrm{h}$
relations described above).  One feature of our process is that by
randomly sampling these PDFs over many MCMC chain steps, we
marginalize over the distribution of possible escape fractions,
such that this scatter is encompassed in our final results.

\item The IGM volume ionized fraction of hydrogen $Q_{H_{II}}(z)$ was calculated
  following \S~\ref{sec:diffeq}.  While solving the differential
  equation, we emulated the effects of
  photosuppression by calculating the emissivity down to the limiting
  UV magnitude corresponding to both the atomic cooling limit at a
  given redshift ($M_{UV,atomic}$; applicable for neutral regions),
  and also to $M_\mathrm{h,supp}$ for the given chain step ($M_\mathrm{UV,supp}$;
  for ionized regions).
The total value of
  $\dot N_{gal}$ for each redshift bin was then calculated as
\begin{equation}
\left.\begin{aligned}
\dot N_{gal}(z) = \dot N_{gal}(z,M_{UV}<M_{UV,supp})+(1-Q_{H_{II}})
\\ \times \dot N_{gal}(z,M_{UV,supp}<M_{UV}<M_{UV,atomic}),
\end{aligned}\right.
\end{equation}
where the first term accounts for ionizing photons from all
  galaxies above the photosuppression limit, while the second term
  accounts for those photons from halos between the photosuppression
  limit and the atomic cooling limit, but only in the fraction of the volume which is
  still neutral at a given redshift.  
We note that this is an approximation as we can only track the
globally-averaged ionized fraction, and thus it does not account for
the effects of spatial clustering of halos on the ionized fraction in
their proximity (e.g., the topology of reionization).
The total ionizing emissivity
  was then calculated as that from galaxies combined with that from
  AGN, $\dot N_{AGN}(z)$, where the latter was calculated from
  $\epsilon_{912}(z)$ as described in \S~\ref{sec:agnion}, assuming an AGN H\,{\sc I} ionizing spectral
  index of $\alpha_{AGN}\!\!=$1.7 \citep{lusso15}.  

\item While helium becomes singly ionized at a similar energy as hydrogen,
  high-energy photons from AGN can doubly-ionize helium.  We thus calculate
  the emissivity of He\,{\sc ii} ionizing photons (energies $>$ 4 Ryd) again assuming a spectral
  index of $\alpha_{AGN}\!\!=$1.7, and solve for the IGM volume ionized
  fraction of He\,{\sc iii} using
\begin{equation}
\dot{Q}_{He_{III}} = \frac{\dot{N}_{ion,HeII}}{\left<n_{He}\right>} - \frac{Q_{He_{III}}}{t_{rec,HeII}}
\end{equation}
and
\begin{equation}
\begin{split}
& t_{rec,HeII} = \\
& \left[C_{He_{III}} \alpha_{B,HeII}(T) (n_H + n_{He}) \left<n_{He}\right>(1+z)^3 \right]^{-1}
\end{split}
\end{equation}
The volume ionized fraction of He\,{\sc ii} was then assumed to be
$Q_{He_{II}} = Q_{H_{II}} - Q_{He_{III}}$.  We assume $C_{He_{III}}(z)
= C_{H_{II}}(z)$ (see \S~\ref{sec:discuss_thermal} for discussion).

\item Using the calculated volume ionized fractions for H\,{\sc ii}, He\,{\sc ii}, and
  He\,{\sc iii}, we calculated the electron scattering optical depth
  as measured from the CMB as
\begin{equation}\label{eqn:tau}
\tau_{es}(z) = \int^{z}_{0} c~\sigma_{T}~n_e(z^{\prime})~(1+z^{\prime})^2~H^{-1}(z^{\prime})~dz^{\prime}
\end{equation}
with
\begin{equation}
 n_e(z) =(n_{H_{II}}(z) + n_{He_{II}}(z) + 2n_{He_{III}}(z)),
\end{equation}
integrated from $z\!\!=$0 to 20.
\end{enumerate}

\subsection{Observational Constraints}\label{sec:mcmc2b}
The outcomes of these calculations are the volume ionized fractions
of H\,{\sc i}, He\,{\sc i} and He\,{\sc ii}, the galaxy and AGN ionizing emissivities, and the
electron scattering optical depth, all as a function of redshift.  To
calculate the likelihood for our model and constrain our free parameters, we used the following observations:

\begin{enumerate}[leftmargin=*]

\item The integrated hydrogen ionizing emissivity at $z\!\!=$4.0 and 4.75
  from \citet{becker13}, synthesized from a variety of measurements of
  the IGM based on spectroscopy of high-redshift quasars.  They find
  that this quantity rises from $z\!\!=$3.2 to 4.75, which is 
  consistent with the idea that a steepening galaxy UV luminosity
  function faint-end slope results in more ionizing photons at higher
  redshifts.  We use their measurements of 
  log $(\dot{N}/10^{51})\!\!=$$-$0.139$^{+0.451}_{-0.346}$ photons
  s$^{-1}$ Mpc$^{-3}$ at $z\!\!=$4.0 and   log $(\dot{N}/10^{51})\!\!=$$-$0.014$^{+0.454}_{-0.355}$ photons
  s$^{-1}$ Mpc$^{-3}$ at $z\!\!=$4.75.  These uncertainties include both
  the statistical errors, and the much larger systematic errors.  
We did not use measurements at $z <$ 4, as our luminosity functions were
  calculated only at $z \geq$ 4.  
We also did not include the upper limit on the
  integrated emissivity at $z\!\!=$6 from \citet{bolton07}, as the value
  of log ($\dot{N}/10^{51}) < -$0.585 was derived assuming that
  the ionizing background is uniform, while
  recent measurements imply that there are substantial spatial
  variations \citep[e.g.,][]{fan06,becker15,bosman18,becker18}.  Future results
  taking advantage of updated measurements of the spatial
  inhomogeneities in the ionizing background, the IGM temperature, and
  the mean free path of ionizing photons will both decrease these
  systematic uncertainties, and allow more robust results at higher
  redshifts, further constraining models such as the one we present here.
For each step in our chain, we calculated the goodness-of-fit
  $\chi^2$ statistic between both the $z\!\!=$4 and 4.75 observations
  and the summed galaxy and AGN ionizing emissivity from our model.

\begin{figure*}[!t]
\epsscale{0.9}
\plotone{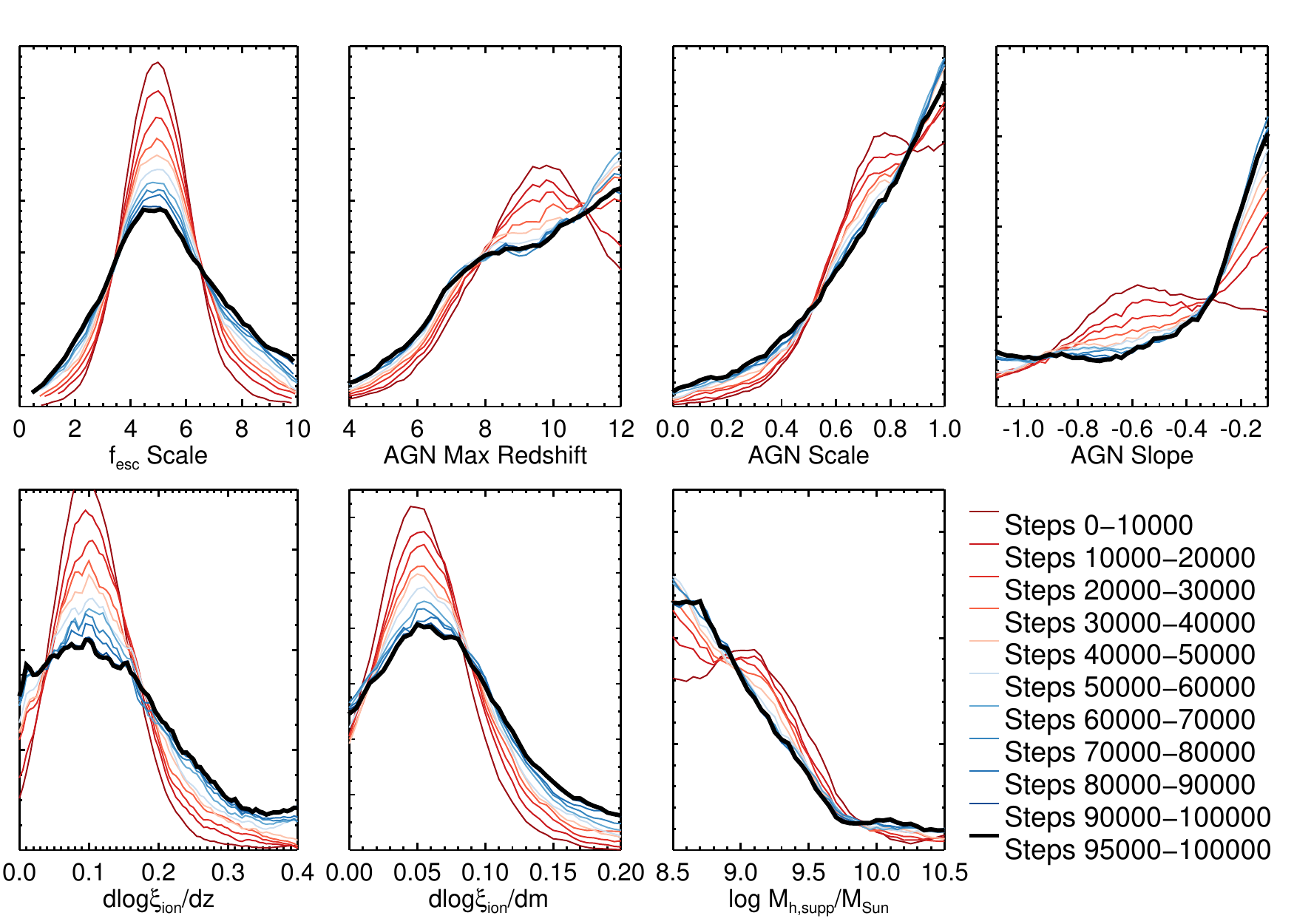}
\caption{Each panel shows the smoothed posterior distribution for each of our
  modeled free parameters.  The different colored lines denote which
  steps out of our 100,000-step chain were used (denoted by the legend
in the lower-right).  As these distributions do not appear to
significantly change after Step \#70,000, we adopt as our fiducial posteriors that from
the last 5,000 steps of the chain (corresponding to the last 5\%),
denoted by the black line.  We performed the Gelman-Rubin test, which
showed that this model is highly converged.  The majority of these parameters have a
relatively broad posterior distribution, which is propagated forward
into the uncertainties in our reionization model.  More robust
observational constraints are needed if we wish to further constrain
these parameters.}
\label{fig:converge}
\end{figure*}

\item The electron scattering optical depth to the CMB ($\tau_{es}$).  This quantity measures the integrated
  optical depth of Thomson scattering\deleted{ to the CMB.  This scattering is caused by the interaction}
of CMB photons with free electrons as they travel from the surface of last
  scattering to the present day, and thus it possesses constraining
  power on models of reionization.\deleted{ Models which reionize earlier will
  have correspondingly higher values of $\tau_{es}$, and vice versa.
  Assuming that reionization happens instantaneously, $\tau_{es}$ can be
  converted into a ``reionization redshift'' ($z_r$).  While most models
  predict reionization to be spatially and temporally inhomogeneous,
  this redshift is still a useful indicator of a rough midway point
  for the reionization process.}  
The first measurements of $\tau_{es}$ from the {\it Wilkinson
    Microwave Anisotropy Probe} ({\it WMAP}) Year 1 results showed $\tau_{es}
 \!\!=$0.17$\pm$0.06, which suggested an instantaneous ``reionization redshift'' of $z_r\!\!=$17$\pm$5
  \citep{spergel03}.  Additional data from {\it WMAP} revised these
  estimates immediately downward, from the WMAP Year 3 result of
  $\tau_{es}\!\!=$0.088$\pm$0.03 
\citep{spergel07} to the final WMAP Year 9 result of $\tau_{es} =$
  0.088$\pm$0.013 \citep{hinshaw13}, and $z_r\!\!=$10.5$\pm$1.1.
  The advent of the {\it Planck} satellite has revised these estimates
  again downward to $\tau_{es}\!\!=$0.066$\pm$0.012 and
  $z_r\!\!=$8.8$\pm$1.1 \citep{planck15}.
 Given the relatively large uncertainties of
  both the {\it WMAP}~\!9 and {\it Planck} measures, the discrepancy is
  only significant at the 1.3$\sigma$ level.  However, even more
  recent 2016 results have been published, highlighting improved
  removal of systematics from the {\it Planck} high-frequency data, showing
  $\tau_{es}\!\!=$0.055$\pm$0.009 \citep{planck16}, discrepant from the
  {\it WMAP}9 data at 2.1$\sigma$ significance.  

We elect use this newer 2016 {\it Planck} value of $\tau_{es}$ as
our fiducial constraint.  When comparing to R15, it is important to
remember that they used the 2015 value, though we note that
  the 2015 and 2016 {\it Planck} $\tau_{es}$ values differ only at the
  0.7$\sigma$ level.  For each step in our chain,
  we computed $\chi^2$ between the observational value of $\tau_{es}$ and
  that calculated from our model given in Equation~\ref{eqn:tau}, which
  includes the contribution to $\tau_{es}$ both from ionizing hydrogen,
  and singly and doubly ionized helium.\deleted{ In \S~\ref{sec:results_tau} we explore how our results change if we use the lower 2016 value.}

\item The model-independent lower limits on the IGM ionized fraction
  of Q$_{H_{II}} \geq 0.94 \pm 0.05$(1$\sigma$) at $z\!\!=$5.9, and
  Q$_{H_{I}} \geq 0.96 \pm 0.05$ (1$\sigma$) at $z\!\!=$5.6 from
  \citet{mcgreer15}.  This study measured the fraction of ``dark'' pixels in
  the Ly$\alpha$ and Ly$\beta$ forests of a sample of 22 bright
  quasars at $z\!\!=$5.7 -- 6.4.  Regions of the IGM containing any
  pre-reionization neutral hydrogen should result in completely saturated absorption
  in both of these transitions.  This method only provides a lower
  limit, as some absorption may also be due to collapsed systems, or residual H\,{\sc i}
  in ionized gas.  However, it is model independent as it does not
  depend on the intrinsic quasar spectral shape \citep[see discussion
  in][]{mcgreer15}.  Finally, as this method combines several objects
  (and, several locations along the line-of-sight to each object), it
  is far more robust against cosmic variance, and uncertainties due to
  inhomogeneous reionization than neutral fractions derived via
  effective optical depths to single quasars
  \citep[e.g.,][]{fan06,bolton11,greig17b,banados18}.

  For each step in our chain, we calculated the goodness-of-fit
  $\chi^2$ statistic between the ionized fraction in our model at each
  of these two redshifts, and these measurements.  As these are
  one-sided lower limits on the ionized fraction, if the model value was above these
  measurements ($Q_{H_{II}} >$ 0.94, 0.96 at $z\!\!=$5.9, 5.6), $\chi^2$ was set to
  zero; if the model was below these values, $\chi^2$ was calculated in
  the usual way using the published uncertainties in each
  redshift bin.  
\end{enumerate}
 
\deleted{Lastly, to rule out models where reionization ends
  too late, we added a conservative effective prior that H\,{\sc i} reionization must be
  complete by $z\!\!=$5.  While the \citet{mcgreer15} measurements
  suggest it is nearly completed by $z\!\!=$6, observations by
  \citet{becker15} have shown
  significant line-of-sight variation in Ly$\alpha$ forest
  transmission measurements, implying significant UV background fluctuations at $z =$
  5.6 -- 6 \citep{becker18}.  However, they find that by $z \sim$ 5, the data are
  consistent with reionization being fully completed.  Therefore,
  models which have $Q_{H_{II}} <$ 1.0 at $z\!\!=$5 have $\chi^2
  \rightarrow \infty$, and are rejected.}

\subsection{Deriving Posteriors}\label{sec:mcmc2c}
Rather than choose a pre-defined number of steps for the burn-in
period, we elected to run our chain for 10$^5$ steps, and then examine
the results to explore where the chain has converged, and select a
final set of samples to derive the posteriors.  In
Figure~\ref{fig:converge} we show the distributions of each of our
seven free parameters for different groupings of 10$^4$ steps.  One
can see that over the first few iterations of 10$^4$ steps, the
parameter distributions change, but that after 7 $\times$ 10$^4$ steps
the changes begin to stabilize, such that the distributions only
exhibit minor changes towards the end of the chain.  For this reason,
we defined the last 5000 steps of the chain as those used to sample the
posterior distribution of our free parameters.

\begin{figure*}[!t]
\epsscale{0.9}
\plotone{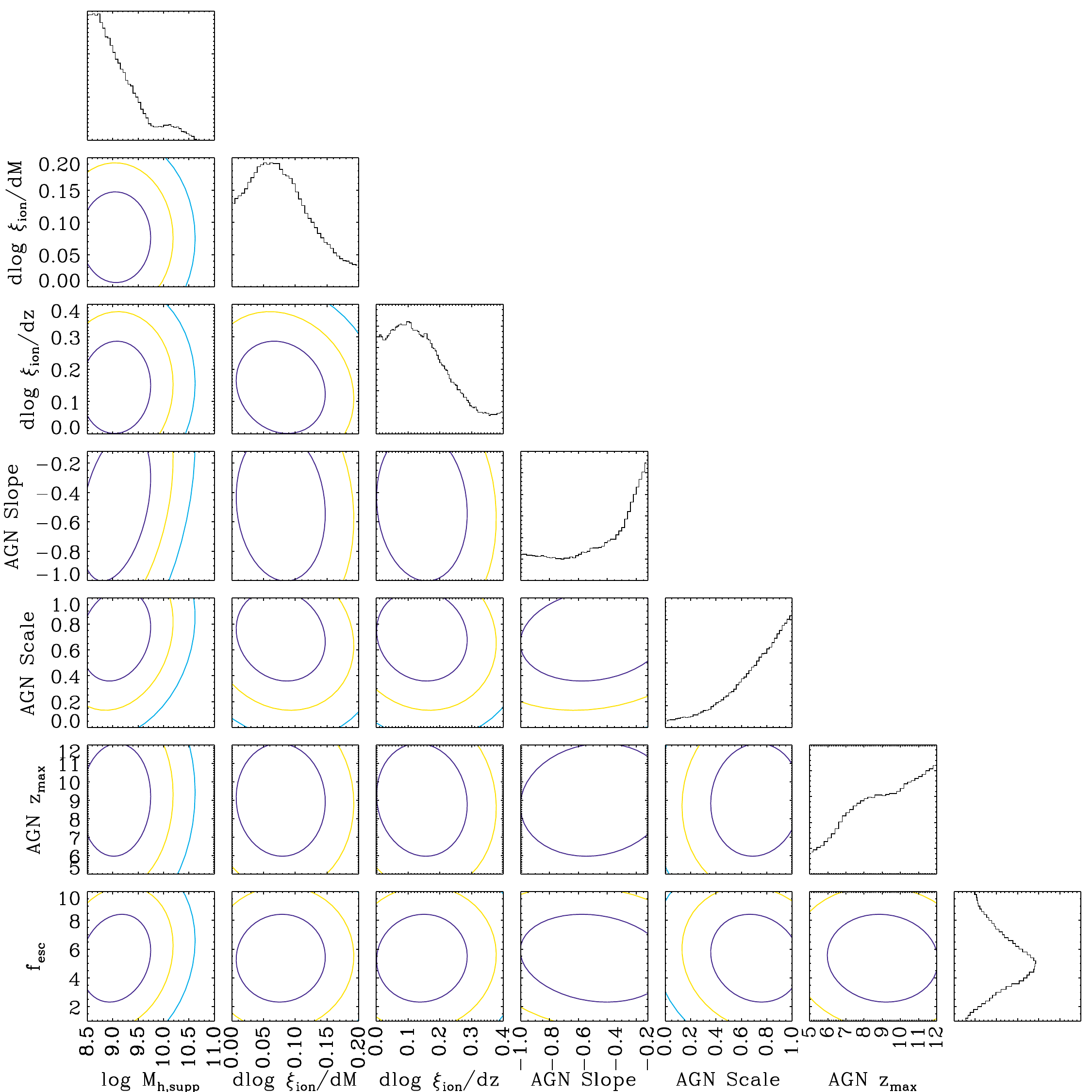}
\vspace{-2mm}
\caption{The covariances between our model parameters from the
  posterior distribution of our fiducial model.  The purple,
  yellow and blue contours denote the 68\%, 95\% and 99.5\% confidence
  range between the listed model parameters, while the histograms show
the smoothed distribution of a single parameter (similar to
Figure~\ref{fig:converge}).}
\label{fig:triangle}
\end{figure*}  

The acceptance fraction for this fiducial model was 14.5\%.  We also
ran this model ten independant times to check for convergence, finding
that the acceptances fractions spanned 14.1\% -- 14.6\%.  We tested for
convergence by comparing the posterior distributions of our free
parameters using the Gelman-Rubin test.  Specifically, we used the
$\tt{rhat.pro}$ IDL routine, which computes the Gelman-Rubin $\hat{R}$
statistic.  We found $\hat{R} =$ 1.00 for all seven of our free
parameters (from 1.00007 for $M_{h,supp}$, to 1.0060 for $AGN_{scale}$), showing that this model is highly converged.

When we compare versions of our model, we use the Deviance
  Information Criteria (DIC; rather than the full Bayes factors, which are
  prohibitive to compute for our high-dimensional parameter space).
  This is similar to the Bayesian Information criteria
  \citep{liddle04} in that it takes into account both the number of
  data points and the number of free parameters.  However, the DIC makes use
of the full chain, rather than just the median values.  The DIC is
defined as $DIC = -2 (L-P)$, where $L$ is the value of $ln(P)$ of our
model using the median of the posterior chains for each parameter, and
$P$ is defined as 
$P = 2 [L - \frac{1}{N} \sum_{s=1}^s ln P_s]$, where N is the number
of samples in the posterior ($N=$5000 here), and $s$ is the sample
index.  $P$ is thus twice the difference between $L$ and the average
value of $ln(P)$ for the full chain.  For a model to be preferred over
a competing model, it must have a lower DIC, with the
significance of the result determined qualitatively by the ``Jeffreys
Scale''.  Here we make use of the updated interpretation by
\citet{kass95}, where $\Delta$ DIC $>$ 2/6/10 is positive/strong/decisive evidence against
the model with the larger value of DIC.

\deleted{As the MCMC fitting routine did not save the results of the model,
such as $Q_{H_{II}}(z)$, $\tau_{es}(z)$ or $\dot{N}(z)$, it was necessary to
recompute those after completion of the chain for further analysis and
plotting.  This was done initially using
the final 10,000 steps, calculating everything in an identical manner
as was done in the chain.  We use this larger number as the random draw process
of our escape fraction parameterization causes the results of these
re-calculations to occasionally differ slightly than those from the MCMC process,
which can result in a particular parameter set violating the prior of reionization
completion by $z=$5 ($<$20\% of the trials for our fiducial set, but
$\sim$40\% for a run with minimal AGN contribution at $z>$6; see \S~\ref{sec:discuss_noagn}).  For this
reason, we use a randomly-selected 5000 of these final 10,000 steps which do not violate
the prior to calculate the distributions
of $Q_{H_{II}}(z)$, $\tau_{es}(z)$ or $\dot{N}(z)$.
We note that we have run these calculations several times, and differences in the
results from run-to-run are negligible due to the large number of
samples.}

\section{Results}\label{sec:results}

\subsection{Fiducial Model}\label{sec:results_fiducial}

\subsubsection{Posterior Distributions of Free Parameters}\label{sec:results_posterior}
The black lines in Figure~\ref{fig:converge} show the posterior
distributions of our free parameters for our fiducial model.  The median values and 68\% confidence ranges are
listed in Table 2, and we
show the covariances between our model parameters as a triangle plot
in Figure~\ref{fig:triangle}.  Of these seven distributions, three have a
clearly defined peak in the posterior distribution within our prior
range.  The parameter f$_{esc,scale}$ prefers a value near 5, with a
68\% confidence range of 3.3 $<$ f$_{esc,scale}$ $<$ 7.5.  This
implies that our model best matches observations when the escape
fractions from the simulations are scaled up by a factor of
$\sim$3--8.  As discussed in \S~\ref{sec:mcmcfesc}, this up-scaling is not
surprising, as the simulations were unable to resolve the birth cloud
of the star particles, and so the porosity of the gas may be
underestimated \edit1{(though see also \citealt{gnedin14}, who invoke
  a sub-unity scale factor to account for unresolved systems providing
  excess absorption)}.  In \S~\ref{sec:discuss_fesc} we will discuss the implications of this
up-scaling for the average galaxy escape fraction, and its evolution
with redshift.

The parameters related to $\xi_{ion}$ also show a clear peak, with
$d$log$\xi/dz$ $\sim$ 0.13, and $d$log$\xi/dM_{UV}$ $\sim$ 0.07, albeit with broad
tails to higher values.  These results indicate that, under our model, galaxies must
have higher ionizing photon production efficiencies both at higher
redshifts, and at lower luminosities.  These results are broadly
consistent with both the scarce observations at high redshift, but
also observations from local analogs for high-redshift galaxies, which
we discuss in \S~\ref{sec:discuss_xiion}.

The remaining four parameters had more one-sided distributions.  The
filtering mass has a clear preference for the lower end of our allowed
range, with a 1-sided 84\% upper limit of log ($M_\mathrm{h,supp}$/M\sol) $<$
9.5.  This indicates that the model requires star-formation in halos
as small as possible for as long as possible.  We note that our prior
of log ($M_\mathrm{h,supp}$/M\sol) $>$ 8.5 was set so that the filtering mass
did not drop below the atomic cooling mass throughout our redshift
range.  It is plausible that if we reduced this prior our model would
prefer even lower values, though, as discussed below in \S \ref{sec:discuss_limmag}, 
lower values would be in greater conflict with previous simulation
work on the \edit1{post-UV background} filtering mass.

The final three parameters govern the AGN contribution.  Combined,
they prefer an AGN contribution which evolves relatively shallowly
downward with increasing redshift (though not as shallowly as the
\citealt{madau15} model), with AGNs present to as high a
redshift as allowed, and with a high escape fraction for ionizing
photons produced by AGNs.  In \S \ref{sec:discuss_agn}, we compare the resultant emissivity from
these combined parameters to previous observations.  \edit1{The
  posterior results from this fiducial model corresponds to
  a DIC value of 5.3}.

\subsubsection{Comparison with $Q(z)$ Constraints}\label{sec:results_qcomp}
\begin{figure*}[!t]
\epsscale{0.9}
\plotone{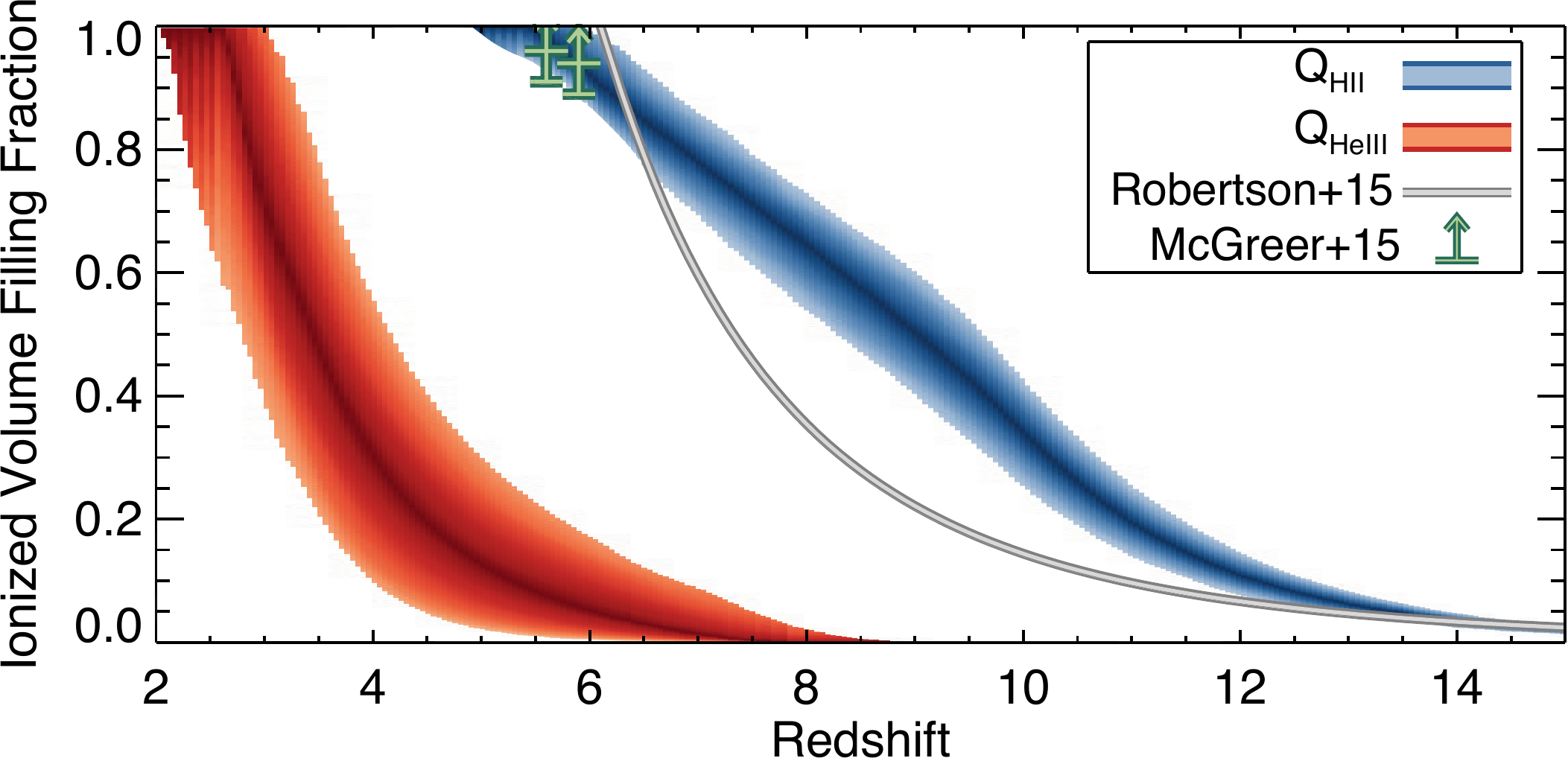}
\caption{The comparison of the ionized volume filling fraction of
  hydrogen (in blue) as a function of
  redshift to observations.  The blue shading denotes the shape of the probability
  distribution function within the 68\% central confidence range, with
  the darkest shading denoting the median (similar shading is used in
  many of the remaining figures).
The observations from \citet{mcgreer15} used to constrain
our model are shown in green, while in gray we show the results from
the model of \citet{robertson15}.  Our model completes reionization by
$z\!\!=$5.6 $\pm$ 0.5.  Although it was not used to
constrain our model, we also show the
ionized volume filling fraction of He\,{\sc ii} (in red), noting that
although AGNs are included in our model, He\,{\sc ii} does not
reionize too early (see \S~\ref{sec:results_he2}).}
\label{fig:q}
\end{figure*}

Figure~\ref{fig:q} shows the distribution of both $Q_{H_{II}}(z)$ and
$Q_{He_{III}}(z)$ for our fiducial model.  The blue-shaded region shows
the 68\% confidence range on our evolution of $Q_{H_{II}}(z)$.  This
result is consistent with the observations from
\citet{mcgreer15} used to constrain our model, as the lower 68\% of
our results fall near the lower 68\% confidence on their lower limits
at $z\!\!=$5.6 and 5.9.  Our model achieves $Q_{H_{II}}\!\!=$1 by $z\!\!=$5.6 $\pm$ 0.5.
Although previous studies have adopted a prior that reionization ended
by $z=6$, it is plausible that reionization ended later.
Indeed, significant neutral patches may be necessary to explain the
most opaque stretches of the Ly$\alpha$ forest at $z=5.5-6$ -- in
particular, the $\sim 110h^{-1}$ comoving Mpc Ly$\alpha$ trough
observed by \citealt{becker15} \citep{kulkarni18}.

\begin{figure}
\epsscale{1.15}
\plotone{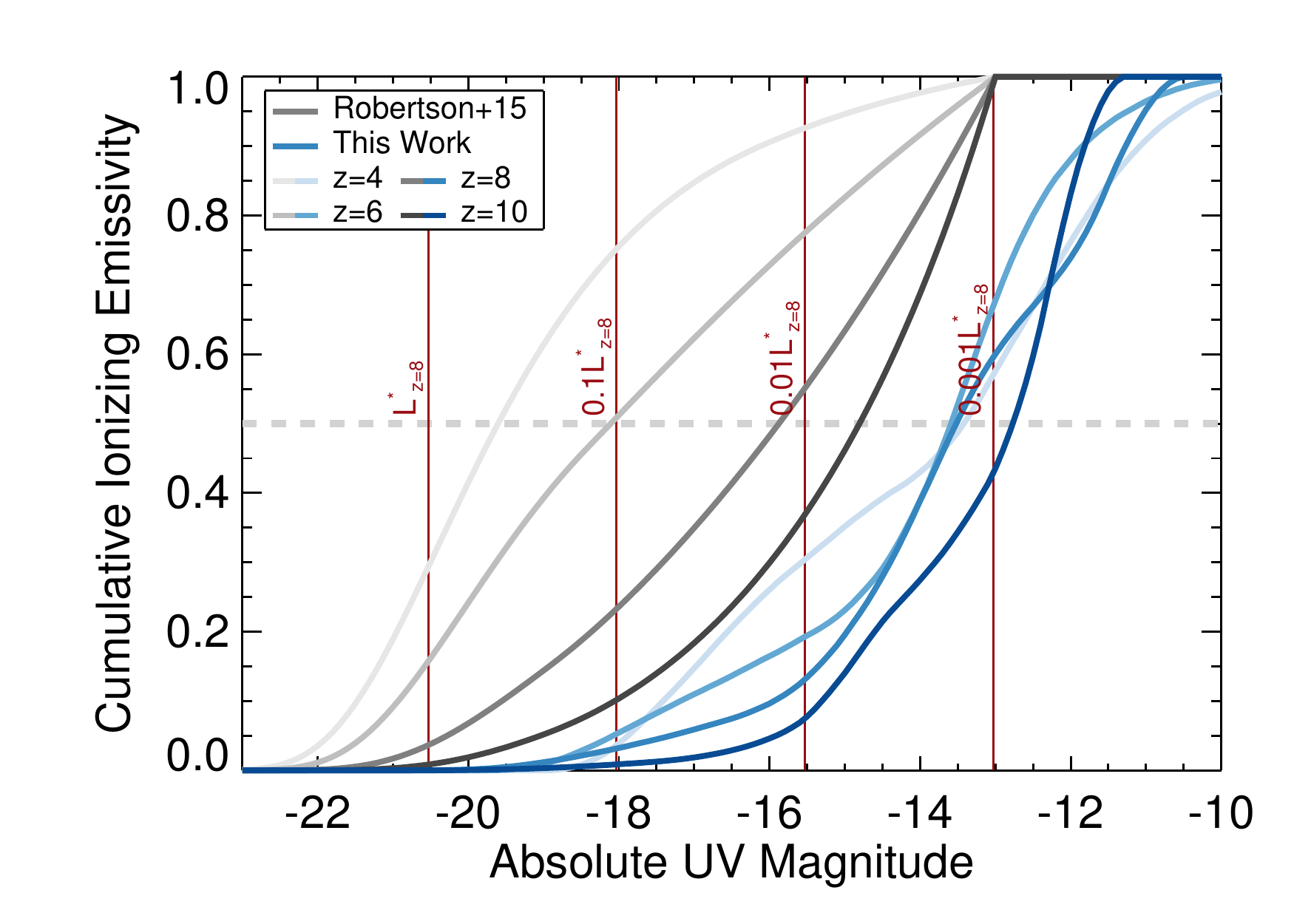}
\vspace{-3mm}
\caption{A comparison of the cumulative ionizing emissivity from this
  work to that using the assumptions of a fixed escape fraction and
  ionizing photon production efficiency from Robertson et al.\ (2015).  We indicate with red
  lines the values of [1, 0.1, 0.01. 0.001] L$^{\ast}$ at $z\!\!=$8 (values at
other redshifts are comparable; this work assumes
dM$^{\ast}$/dz=0.13).  In our model, where the lowest-mass halos have
the highest escape fractions, the extreme faint-end of the luminosity
function dominates the ionizing emissivity.}
\label{fig:ncum}
\end{figure}

The shape of the evolution of $Q_{H_{II}}(z)$ from our fiducial model shows a roughly linear evolution with redshift from
$z \sim$ 6 to $z \sim$ 10, with a slight acceleration in the evolution
at $z >$ 10 as the faint-end slope of the galaxy luminosity function
in our fiducial model stops evolving.  This evolution is in sharp
contrast to the results of Robertson et al.\ (2015), which imply that
reionization starts very slowly at $z >$ 10, and then undergoes a
rapid acceleration at $z <$ 10.  This stark difference can be
understood by the differences in these two models.  The R15 model fits a variable star-formation rate (SFR) density to the
observations of the galaxy UV luminosity function from a variety of
sources, and uses as an additional constraint the 2015 value of
the \textit{Planck} optical depth.  They
convert their derived SFR density to an ionizing emissivity by
assuming a single value for the escape fraction (20\%) and ionizing photon
production efficiency (log[$\xi_{ion}$] $=$ 53.14 Lyc photons s$^{-1}$ M\sol$^{-1}$
yr$^{-1}$; 25.24 in the units used here of Hz
erg$^{-1}$).  These assumed values are comparable to those used for
the purple shaded region in Figure~\ref{fig:appendix} in the Appendix ($f_\mathrm{esc}$=13\%; log[$\xi_{ion}$] $=$ 25.34), which give
near-identical results to those from Robertson et al.\ when using our
assumed luminosity functions, highlighting that the minor differences
in the luminosity functions assumed (or resultant SFR density) do not
play a large role in the differences in $Q_{H_{II}}(z)$. 

The differences in $\xi_{ion}$ and $f_\mathrm{esc}$ must therefore be
responsible.  While both certainly play a role, the differences in
$Q_{H_{II}}(z)$ are easy to understand when simply exploring $f_\mathrm{esc}$.
In the Robertson et al.\ model, galaxies at all redshifts and
luminosities have the same escape fraction of 20\%.  This means that
the ionizing emissivity from a given luminosity range is directly
proportional to the non-ionizing UV luminosity density in that same
range.  The result of this assumption is that the faintest galaxies
($M_\mathrm{UV} < -$14; $<10^{-3} L^{\ast}$ at $z\!\!=$6) do
not play a major role.  This is illustrated in Figure~\ref{fig:ncum},
where the gray shaded lines show the cumulative ionizing emissivity
using the assumptions from Robertson et al (albeit with our luminosity
functions used here; as discussed in the preceding paragraph, this
results in minimal differences).  At $z\!\!=$6, at the end of
reionization, $\sim$half of the ionizing emissivity comes from rather
bright galaxies, with L $>$ 0.1L$^{\ast}$ in the Robertson et al.\ model.  While the evolving faint
end slope changes this with redshift, even by $z\!\!=$10, a near majority
of the ionizing emissivity comes from galaxies with L $>$
0.01L$^{\ast}$.  As massive/bright galaxies are building up with time,
the relative insignificance of L $\ll$
0.01L$^{\ast}$ galaxies to the ionizing photon budget means that
reionization gets a late start.  As the brighter/more massive galaxies
build up, the ionizing emissivity increases rapidly, resulting in a
reionization history that starts late, but finishes quickly.

The blue shaded lines in
Figure~\ref{fig:ncum} show the results of our fiducial model.  While the same
luminosity functions are assumed as the gray lines, in our model the
escape fraction is halo-mass dependent, so even modestly-bright
galaxies contribute very little, while the faintest galaxies
dominate.  This, combined with the steepening faint-end slope at
higher redshift, allows reionization to begin earlier.  
This difference is enhanced by our evolution in $\xi_{ion}$ to higher
values at higher redshifts and lower luminosities, allowing those
galaxies which have significant escape fractions to have larger
ionizing photon budgets. However, as
the faint-end slope shallows with decreasing redshift, the emissivity
from galaxies from our model decreases, resulting in a very
constrained ionizing photon budget towards the end of reionization,
discussed in \S~\ref{sec:results_ndot}.

While both models can successfully complete reionization, the
different assumptions on the escape
fraction  result in modest differences in the reionization history.
These differences are greatest at $z \approx$ 9 where the Robertson
et al.\ model, driven by modestly bright galaxies, predicts $Q_{H_{II}}
\sim$ 0.2, and our model, driven by the faintest galaxies, predicts $Q_{H_{II}}
\sim$ 0.5.  Future observations may be able to distinguish between
these scenarios, and can thus potentially constrain the luminosity
range of the galaxies driving reionization.  In \S~\ref{sec:discuss_qcomp} and Figure~\ref{fig:qhii_compare}
we further compare our results for $Q_{H_{II}}(z)$ to several
observational and theoretical results in the literature.

\begin{figure*}[!t]
\epsscale{0.49}
\plotone{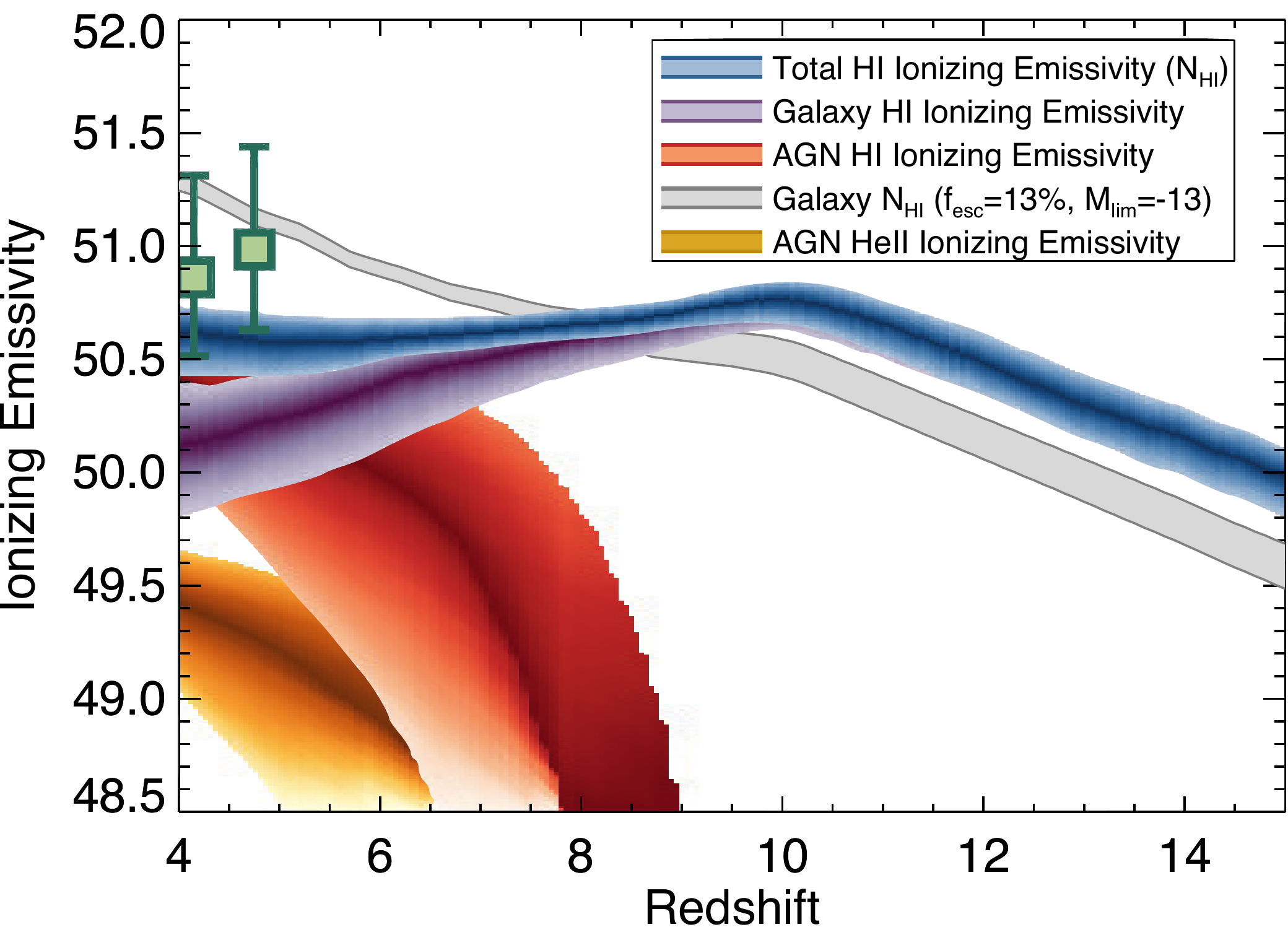}
\hspace{10mm}
\epsscale{0.52}
\plotone{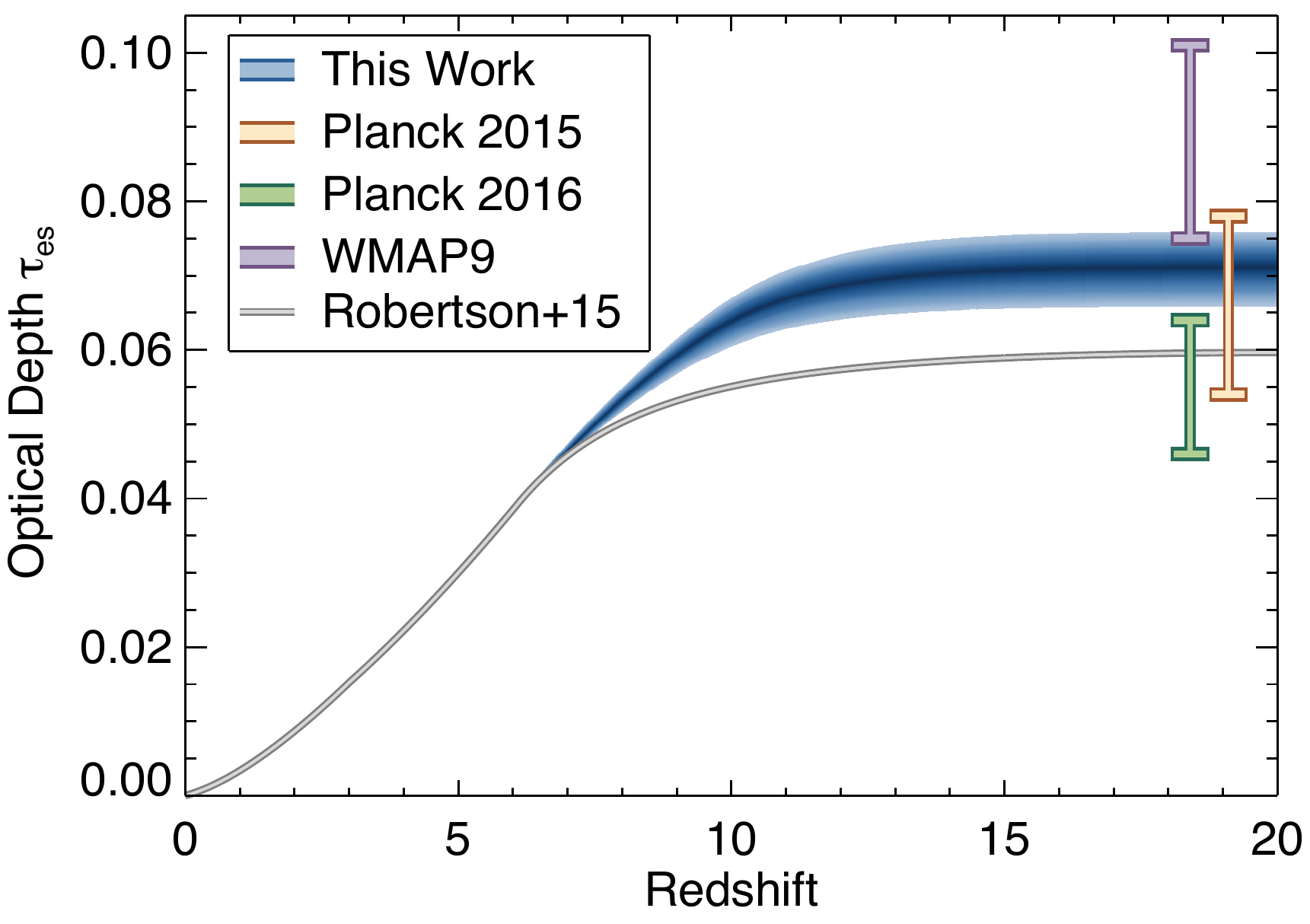}
\caption{\emph{Left:} The ionizing
emissivity ($\dot{N}_{ion}$) as a function of redshift.  The H\,{\sc i}
ionizing emissivity from galaxies and AGNs are shown in purple and
red, respectively, with their sum shown in blue.  The He\,{\sc ii}
ionizing emissivity from AGNs is shown in yellow.  The observations
from \citet{becker13} used to constrain our model are shown in green.
For comparison, the gray region denotes the hypothetical galaxy ionizing emissivity if
a fixed limiting magnitude of $-$13 and ionizing photon escape
fraction of 0.13 are assumed (similar to assumptions used in
\citealt{finkelstein12b}, \citealt{robertson13} and
\citealt{robertson15}, and the purple curve in Figure~\ref{fig:appendix}).  \emph{Right:}  The electron scattering
optical depth to the CMB ($\tau_{es}$).  The observational {\it Planck}
2016 observations used to constrain the model are shown in green,
while the {\it WMAP}9 and {\it Planck} 2015 results are shown in
purple and tan, respectively.  The results from \citet{robertson15}
are shown by the gray line.  For all plots, the shading denotes
the 68\% confidence range, with the darkest color indicating the
median of a given quantity.}
\label{fig:ndottau}
\end{figure*}  

\subsubsection{He~II Reionization}\label{sec:results_he2}
While we did not include any constraints on the ionization of
He\,{\sc ii} in our model, here we comment on the resultant
distribution of $Q_{He_{III}}(z)$, shown as the red-shaded region in
Figure~\ref{fig:q}.  Our fiducial model results in a He\,{\sc ii}
reionization history which gets started at a low level at $z \sim$ 6,
as the AGN help to complete hydrogen reionization.  The  volume
ionized fraction of He\,{\sc iii} hits 50\% at $z\!\!=$3.4 $\pm$ 0.6,
and He\,{\sc ii} reionization completes at $z\!\!=$2.7 $\pm$ 0.4.  This is
consistent with observations of the \HI\ and \HeII\ Ly$\alpha$
forests.  Current observations of the latter show a strong evolution
in its mean opacity and dispersion at $z\gtrsim 2.8$
\citep[e.g.][]{2011ApJ...733L..24W, 2016ApJ...825..144W}.
Additionally, the analogue of Gunn-Peterson troughs observed in the
$z\gtrsim 2.7$ \HeII\ Ly$\alpha$ forest have been used to place limits
of $Q_{\mathrm{HeII}} \gtrsim 10\%$ at the cosmic mean density, which
implies that \HeII\ reionization was likely still in progress at $z\sim 2.7$
\citep{2009ApJ...704L..89M, 2010ApJ...722.1312S, 2014ApJ...784...42S}.
Finally, recent \HI\ Ly$\alpha$ forest measurements have found
evidence for a bump in the thermal history of the IGM at $z\sim 2.8$,
which has been interpreted to coincide with the end of the \HeII\
reionization process \citep[e.g.][]{2000MNRAS.318..817S,
  2011MNRAS.410.1096B, 2017arXiv171000700H, 2015MNRAS.450.4081P,
  2016MNRAS.460.1885U}.  

Together, these measurements provide evidence
that \HeII\ reionization may be ending at $z\sim 2.5$, consistent with our findings
\citep[see however][for alternative interpretations of the \HeII\
Ly$\alpha$ forest data]{2014MNRAS.440.2406M, 2014MNRAS.437.1141D,
  2017MNRAS.465.2886D}.  We note that our results also suggest a more
extended \HeII\ reionization process than is found in existing
simulations -- \edit1{with a 1$\sigma$ upper bound of} as much as $18\%$ complete by $z\approx 6$
\citep{2009ApJ...694..842M, 2013MNRAS.435.3169C, 2017ApJ...841...87L}.
This is qualitatively consistent with the conclusion of
\citet{2016ApJ...825..144W}, who find evidence for extended
reionization in the statistics of the \HeII\ Ly$\alpha$ forest
opacity\footnote{See however \citet{2017MNRAS.468.4691D} for a
  discussion of potential caveats in simulating the impact of \HeII\
  reionization on the transmission statistics of the \HeII\ Ly$\alpha$
  forest.}.

\subsubsection{Comparison with $\dot{N}_{ion}(z)$ Constraints}\label{sec:results_ndot}

In the left panel of Figure~\ref{fig:ndottau} we show the evolution of
the ionizing emissivity.  The blue-shaded region denotes the total
H\,{\sc i} ionizing emissivity, which is consistent with the
observational constraints used for our model.  Notably, the emissivity in our model exhibits a
slight rise from $z\!\!=$4 to 10, consistent with the observations from
\citet{becker13} of an increasing emissivity from $z \sim$ 2 to 5.  The
purple and red shaded regions denote the components of this emissivity
contributed by galaxies and AGN, respectively.  Although the non-ionizing emissivity from
galaxies increases with decreasing redshift from $z\!\!=$10 to 4 (Figure~\ref{fig:fig1}), the ionizing emissivity
does the opposite.  This is due to a combination of the faint-end slope
of the UV luminosity function becoming shallower, resulting in less
luminosity coming from faint galaxies, while it is those very faint
galaxies which have the highest escape fractions.  This can be better
understood by examining Figure~\ref{fig:ncum}.  While the cumulative
ionizing emissivity at $z\!\!=$4 is not highly dissimilar to that at
$z\!\!=$10, Figure~\ref{fig:fig1} highlights that the $z\!\!=$10 universe
harbors a greater abundance of extremely faint ($M_{UV} > -$15) galaxies
than the $z\!\!=$4 universe.  These factors combine to create a galaxy
ionizing emissivity which counterintuitively increases with increasing
redshift at $z <$ 10.  This emissivity peaks at $z \sim$ 10, and then procedes to
decline to higher redshift.  This transition point is set by the
assumption in our fiducial model that the faint-end slope does not get
any steeper at $z >$ 10, while $M^{\ast}$ and $\phi^{\ast}$ continue
to decline to higher redshifts, lowering the overall emissivity.  

The gray curve shows
the emissivity if one assumed a fixed limiting magnitude of $M_{UV} =
-$13 and escape fraction of 13\%.  The difference in the ionization
history from our model and that from previous results
\citep[e.g.][]{robertson15,finkelstein15} can be understood by
comparing this to the blue curve.  Our model has a greater emissivity
at $z >$ 9, allowing an earlier start to reionization.  The emissivity
at $z \leq$ 7 flattens out -- just enough to complete reionization by
$z \sim$ 5.6, but not enough to exceed the emissivity observations at lower redshifts.
We note that this gray curve, when extrapolated to $z <$ 4, will
exceed the emissivity measurements from \citet{becker13}.  This
indicates that some previous models with fixed large escape fractions
may have not matched all observational data when considering $z <$ 6
\citep[see also][]{stanway16}.

The AGN emissivity, shown by the red shaded region, increases with
decreasing redshift.  This is by construction as our method assumes an
AGN emissivity which rises with decreasing redshift, although the
exact slope of this increase and the normalization are set by the
posterior distributions of these two free parameters.  The AGN
emissivity, and specifically how it compares to that from galaxies, is
discussed further in \S~\ref{sec:discuss_agn} below.

\subsubsection{Comparison with $\tau_{es}$ Constraints}\label{sec:results_tau}
The right panel of Figure~\ref{fig:ndottau} shows the posterior
distribution on $\tau_{es}$ from our fiducial model.  The median value
from our model is $\tau_{es}\!\!=$0.071 $\pm$ 0.005 (integrated to $z
=$ 20).

This is higher than the recent 2016 value published by
the {\it Planck} collaboration of $\tau_{es}\!\!=$0.055 $\pm$ 0.009,
which we used as our constraint.  However, the tension is not high,
with the difference being significant at only the 1.55$\sigma$ level
(we note that our result is 0.4$\sigma$ higher than the 2015 {\it Planck}
value of $\tau_{es,Planck15}\!\!=$0.066 $\pm$ 0.012, and 1.2$\sigma$
\emph{lower} than the final {\it WMAP}9 value of $\tau_{es,WMAP9}\!\!=$0.088 $\pm$ 0.013).
Comparing to R15 (who used the 2015 {\it Planck} value), while our model prefers an earlier start to
reionization, the large uncertainties
on $\tau_{es}$ result in both our model and the results from
R15 maintaining consistency with the observational constraints.

\begin{figure*}[!t]
\epsscale{1.1}
\plotone{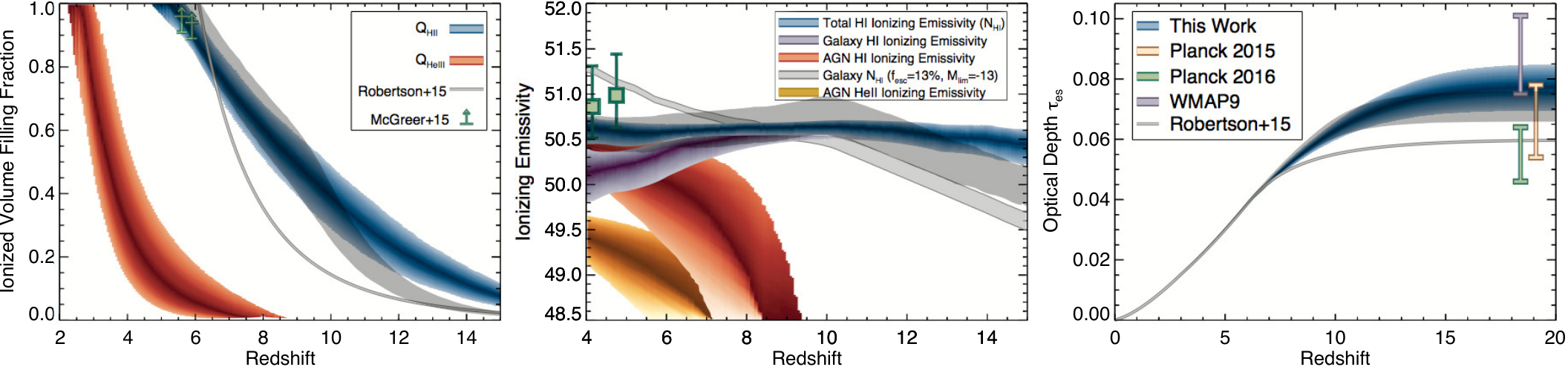}
\caption{Results from our model when we allow the faint-end slope of
  the luminosity function to continue evolving to steeper slopes at $z
>$ 10.  The transparent gray shading denotes our fiducial model, where
the faint-end slope stays fixed at $z >$ 10 to $\alpha = -$2.35.  The
evolving faint-end slope results in a higher emissivity
at $z >$ 10, which begins reionization slightly sooner. This earlier ionization results in a 
higher electron scattering optical depth of $\tau_{es}$ = 0.077 $\pm$ 0.008  (0.7$\sigma$
higher than that from our fiducial model, which was already higher
than observations, though not statistically discrepant).  This model
is a significantly worse fit to the observational constraints than our
fiducial model, with the DIC comparison showing strong evidence that the fiducial
model is preferred.}
\label{fig:evolvealpha}
\end{figure*}   

Lowering $\tau_{es}$ would require removing electrons somewhere along the line-of-sight.
One could do this by, for example, slightly lowering the galaxy
emissivities, resulting in a slightly later end to reionization.  However, the emissivities from the model are already at
the lower-end of the observations, and the ionization history is also
pushing the limits of the observations.  Thus, while this would
decrease the tension between the model $\tau_{es}$ and the
observations, it would increase the tension on our other two
constraints.  
Clearly reduced observational
uncertainties on $\tau_{es}$ will help, though the outlook for
significant future improvement is uncertain.
We present tabulated values of the results from this fiducial model in
Table 3.  As listed in the table, the ratio of the galaxy-to-AGN ionizing
emissivity becomes infinite at redshifts when there is no AGN contribution.

\subsection{Evolving Versus Flat Faint-End Slope}\label{sec:results_alpha}

Our model uses a redshift-dependent parameterization of the
UV luminosity function, which stipulates that the characteristic
luminosity and number density decrease to increasing redshift, while
the faint-end slope becomes steeper.  While these parameterizations were
developed by fitting to the observational data at 4$<z<$10, in our
model we extrapolate to higher redshift.  As discussed in \S~\ref{sec:mcmclf}, our
fiducial model assumes the faint-end slope ceases to become steeper at
$z >$ 10, assuming $\alpha = -$2.35 (the $z\!\!=$10 value) at higher
redshifts.  This choice was made as the redshift evolution of $\alpha$
would result in extremely steep slopes of $\sim -$3.5 by $z\!\!=$20.  As
shown in Figure~\ref{fig:fig1}, this results in a \emph{higher} abundance of $z >$
10 galaxies at $M_{UV} > -$15 compared to lower redshifts.  Another
reason to potentially disfavor such steep values is that the halo mass
function is expected to asymptote to a slope of $\sim-$2 at very low
masses.  However, at $z\!\!=$15 the
halo mass function exponential decline begins at log
($M_\mathrm{h}$/M\sol) $\gtrsim$ 8, near to the atomic cooling limit, thus any potential luminous emission relevant
here likely originates from a mass regime where the slope is steeper
(though of course the luminosity function is further affected by
galaxy physics).

Therefore as there is nothing preventing these steep slopes, we consider the results of our model if
we allowed $\alpha$ to continue to evolve at $z >$ 10.  In this model
run, all free parameters were fit with the same priors as our fiducial
run, and are free to develop different posteriors to satisfy (if
possible) the observational constraints in light of the additional
photons available at $z >$ 10.  The right panel of Figure~\ref{fig:fig1} shows how the
evolution of the non-ionizing 1500 \AA\ luminosity density evolves for
these two scenarios, highlighting that when the faint-end slope
continues to steepen, the non-ionizing luminosity density evolves much
more shallowly to higher redshift.

We show the results of this analysis in Figure~\ref{fig:evolvealpha}.
This model is still consistent with the observational constraints on
the ionized volume filling fraction and ionizing emissivity, though
the increased number of galaxies at higher redshifts results an
overall emissivity which is essentially flat at all redshifts,
resulting in a slightly earlier start to reionization, though with a similar midpoint of
$z_{reion}$ = 8.9 $\pm$ 0.9. This earlier ionization results in a higher electron scattering optical depth of $\tau_{es}$ = 0.077 $\pm$ 0.008 (0.7$\sigma$
higher than that from our fiducial model). The right-hand panel of
Figure~\ref{fig:evolvealpha} highlights that this value is in tension with the Planck
2016 constraints at the 1.9$\sigma$ level. \edit1{Unsurprisingly, this model with an evolving faint-end
slope has a worse DIC value of 13.7 than our fiducial model, which has
5.3.  This difference is large enough to statistically
differentiate between these models, with our fiducial model with the
fixed faint-end slope at $z >$ 10 showing ``strong'' evidence of
being preferred ($\Delta$DIC $>$ 2).
We conclude that our fiducial model is a more plausible scenario, which we discuss in
more detail below.}

\begin{deluxetable*}{ccccccccccccc}
\tabletypesize{\footnotesize}
\vspace{-2mm}
\tablecaption{Results of Fiducial Model}
\vspace{-2mm}
\tablehead{
\multicolumn{1}{c}{Redshift} & \multicolumn{3}{c}{Q$_{HII}$} & \multicolumn{3}{c}{$\tau_{e}$} & \multicolumn{3}{c}{log $\dot{N}_{total}$ (s$^{-1}$ Mpc$^{-3}$)} & \multicolumn{3}{c}{log $\dot{N}_{gal}$/$\dot{N}_{AGN}$} \\
\multicolumn{1}{c}{$ $} & \multicolumn{1}{c}{16\%} & \multicolumn{1}{c}{Median} & \multicolumn{1}{c}{84\%} & \multicolumn{1}{c}{16\%} & \multicolumn{1}{c}{Median} & \multicolumn{1}{c}{84\%} & \multicolumn{1}{c}{16\%} & \multicolumn{1}{c}{Median} & \multicolumn{1}{c}{84\%} & \multicolumn{1}{c}{16\%} & \multicolumn{1}{c}{Median} & \multicolumn{1}{c}{84\%}
}
\startdata
4.0&1.000&1.000&1.000&0.0222&0.0225&0.0226&50.42&50.64&50.76&-0.81&-0.24&0.32\\
4.2&1.000&1.000&1.000&0.0237&0.0240&0.0242&50.40&50.62&50.76&-0.78&-0.21&0.44\\
4.4&1.000&1.000&1.000&0.0253&0.0255&0.0258&50.38&50.62&50.76&-0.72&-0.12&0.53\\
4.6&1.000&1.000&1.000&0.0268&0.0271&0.0274&50.36&50.59&50.74&-0.78&-0.07&0.63\\
4.8&1.000&1.000&1.000&0.0284&0.0287&0.0290&50.35&50.60&50.74&-0.72&0.01&0.78\\
5.0&1.000&1.000&1.000&0.0300&0.0304&0.0307&50.38&50.60&50.74&-0.66&0.06&0.92\\
5.2&0.993&1.000&1.000&0.0317&0.0320&0.0323&50.34&50.59&50.74&-0.58&0.13&1.01\\
5.4&0.963&1.000&1.000&0.0333&0.0337&0.0340&50.37&50.60&50.75&-0.60&0.18&1.16\\
5.6&0.938&1.000&1.000&0.0350&0.0354&0.0358&50.35&50.58&50.76&-0.57&0.23&1.29\\
5.8&0.905&0.967&1.000&0.0366&0.0370&0.0375&50.39&50.60&50.75&-0.53&0.33&1.41\\
6.0&0.870&0.930&1.000&0.0382&0.0387&0.0392&50.41&50.62&50.79&-0.44&0.42&1.51\\
6.2&0.832&0.899&0.996&0.0398&0.0404&0.0409&50.42&50.62&50.78&-0.33&0.44&1.69\\
6.4&0.795&0.866&0.959&0.0413&0.0420&0.0425&50.44&50.64&50.80&-0.29&0.56&1.93\\
6.6&0.756&0.836&0.925&0.0428&0.0435&0.0442&50.38&50.61&50.78&-0.27&0.58&2.16\\
6.8&0.722&0.809&0.892&0.0442&0.0450&0.0458&50.39&50.63&50.79&-0.20&0.66&2.44\\
7.0&0.694&0.781&0.862&0.0456&0.0465&0.0475&50.43&50.64&50.82&-0.14&0.81&$\infty$\\
7.2&0.659&0.753&0.836&0.0469&0.0479&0.0490&50.42&50.65&50.82&-0.04&0.93&$\infty$\\
7.4&0.628&0.724&0.808&0.0482&0.0493&0.0506&50.41&50.65&50.82&0.03&1.11&$\infty$\\
7.6&0.598&0.699&0.780&0.0494&0.0507&0.0521&50.43&50.66&50.83&0.07&1.21&$\infty$\\
7.8&0.570&0.672&0.753&0.0506&0.0520&0.0535&50.44&50.67&50.84&0.16&1.42&$\infty$\\
8.0&0.540&0.647&0.730&0.0517&0.0534&0.0550&50.47&50.69&50.86&0.21&1.74&$\infty$\\
8.2&0.513&0.617&0.703&0.0529&0.0546&0.0564&50.45&50.68&50.87&0.24&1.92&$\infty$\\
8.4&0.485&0.590&0.678&0.0539&0.0559&0.0577&50.49&50.70&50.86&0.30&2.25&$\infty$\\
8.6&0.458&0.563&0.654&0.0549&0.0571&0.0590&50.46&50.68&50.85&0.38&2.46&$\infty$\\
8.8&0.432&0.533&0.629&0.0558&0.0582&0.0603&50.49&50.72&50.90&0.45&2.83&$\infty$\\
9.0&0.403&0.502&0.603&0.0568&0.0593&0.0615&50.51&50.72&50.90&0.51&3.24&$\infty$\\
9.2&0.373&0.476&0.574&0.0577&0.0603&0.0628&50.51&50.74&50.92&0.58&4.42&$\infty$\\
9.4&0.344&0.443&0.543&0.0584&0.0613&0.0639&50.52&50.74&50.92&0.63&$\infty$&$\infty$\\
9.6&0.314&0.413&0.507&0.0591&0.0623&0.0650&50.55&50.77&50.95&0.80&$\infty$&$\infty$\\
9.8&0.285&0.377&0.468&0.0598&0.0632&0.0661&50.57&50.79&50.98&0.90&$\infty$&$\infty$\\
10.0&0.254&0.342&0.428&0.0605&0.0639&0.0670&50.55&50.79&50.98&1.04&$\infty$&$\infty$\\
10.2&0.227&0.309&0.389&0.0610&0.0646&0.0679&50.53&50.79&50.96&1.20&$\infty$&$\infty$\\
10.4&0.202&0.278&0.352&0.0615&0.0653&0.0687&50.51&50.75&50.94&1.32&$\infty$&$\infty$\\
10.6&0.180&0.247&0.318&0.0620&0.0659&0.0694&50.48&50.73&50.92&1.58&$\infty$&$\infty$\\
10.8&0.159&0.220&0.287&0.0624&0.0664&0.0701&50.44&50.68&50.90&2.15&$\infty$&$\infty$\\
11.0&0.140&0.196&0.256&0.0628&0.0669&0.0707&50.42&50.68&50.87&2.78&$\infty$&$\infty$\\
11.2&0.125&0.174&0.230&0.0632&0.0674&0.0712&50.38&50.62&50.83&4.34&$\infty$&$\infty$\\
11.4&0.112&0.156&0.207&0.0634&0.0677&0.0717&50.36&50.60&50.81&$\infty$&$\infty$&$\infty$\\
11.6&0.099&0.139&0.185&0.0637&0.0681&0.0721&50.31&50.57&50.77&$\infty$&$\infty$&$\infty$\\
11.8&0.087&0.124&0.166&0.0639&0.0684&0.0725&50.29&50.55&50.75&$\infty$&$\infty$&$\infty$\\
12.0&0.077&0.110&0.147&0.0641&0.0687&0.0729&50.24&50.50&50.71&$\infty$&$\infty$&$\infty$\\
12.2&0.068&0.098&0.131&0.0643&0.0689&0.0732&50.17&50.44&50.66&$\infty$&$\infty$&$\infty$\\
12.4&0.060&0.088&0.117&0.0645&0.0691&0.0735&50.18&50.44&50.63&$\infty$&$\infty$&$\infty$\\
12.6&0.054&0.078&0.104&0.0646&0.0693&0.0738&50.11&50.39&50.60&$\infty$&$\infty$&$\infty$\\
12.8&0.048&0.070&0.094&0.0647&0.0695&0.0740&50.08&50.33&50.54&$\infty$&$\infty$&$\infty$\\
13.0&0.043&0.062&0.084&0.0648&0.0697&0.0742&50.05&50.30&50.52&$\infty$&$\infty$&$\infty$\\
13.2&0.038&0.056&0.075&0.0649&0.0698&0.0744&49.99&50.25&50.47&$\infty$&$\infty$&$\infty$\\
13.4&0.034&0.051&0.067&0.0650&0.0699&0.0745&49.97&50.22&50.45&$\infty$&$\infty$&$\infty$\\
13.6&0.031&0.045&0.060&0.0651&0.0701&0.0747&49.98&50.23&50.44&$\infty$&$\infty$&$\infty$\\
13.8&0.027&0.040&0.053&0.0652&0.0702&0.0748&49.94&50.19&50.40&$\infty$&$\infty$&$\infty$\\
14.0&0.024&0.036&0.048&0.0653&0.0703&0.0749&49.90&50.14&50.36&$\infty$&$\infty$&$\infty$\\
14.2&0.021&0.032&0.043&0.0654&0.0704&0.0751&49.88&50.14&50.34&$\infty$&$\infty$&$\infty$\\
14.4&0.019&0.028&0.038&0.0655&0.0704&0.0752&49.87&50.10&50.29&$\infty$&$\infty$&$\infty$\\
14.6&0.017&0.025&0.033&0.0655&0.0705&0.0752&49.81&50.05&50.23&$\infty$&$\infty$&$\infty$\\
14.8&0.015&0.022&0.030&0.0655&0.0706&0.0753&49.76&50.00&50.20&$\infty$&$\infty$&$\infty$\\
15.0&0.013&0.020&0.026&0.0656&0.0706&0.0754&49.75&50.01&50.19&$\infty$&$\infty$&$\infty$\\
---&---&---&---&---&---&---&---&---&---&---&---&---\\
20.0&0.000&0.000&0.000&0.0658&0.0710&0.0759&48.70&48.93&49.10&$\infty$&$\infty$&$\infty$\\
\enddata
%\tablecomments{}
\label{tab:tab2}
\end{deluxetable*}

\section{Discussion}

\subsection{Comparison of $Q_{H_{II}}(z)$ to Model-Dependent Analyses}\label{sec:discuss_qcomp}

Observationally constraining the ionization fraction of the IGM
towards the end of reionization is a very active area of
astrophysics.  Here we compare our results for $Q_{H_{II}}(z)$ to those from several
recent studies, focusing on complementary observational methods
involving Ly$\alpha$ emission and quasars at $z
>$ 6, as well as theoretical studies, which were not used as
constraints on our analysis, summarized in
Figure~\ref{fig:qhii_compare}.

\subsubsection{Contraints from Ly$\alpha$ Emission -- Clustering and
  Luminosity Functions}\label{sec:discuss_lyalf}
First we consider constraints inferred via observations of
Ly$\alpha$ emission.  Ly$\alpha$ can be an
excellent tracer of a neutral IGM, as neutral H\,{\sc i} gas resonantly
scatters Ly$\alpha$ photons, attenuating their observable surface
brightness \citep[e.g.][]{miralda-escude98, santos04, malhotra04,
  verhamme06, mcquinn07, dijkstra14}.
\deleted{This potential decrease can be observable both through the evolution
of global measures such as the Ly$\alpha$ line luminosity function
typically measured from narrowband imaging surveys
%\citep[e.g.,][]{malhotra04,ouchi08,ouchi10}, 
as well as statistics on the Ly$\alpha$ emission
from spectroscopic followup of individual sources.}
%\citep[e.g.,][]{stark10,stark11}.}
Constraints on the evolution of the Ly$\alpha$ luminosity function
from $z\!\!=$5.7 to $z\!\!=$6.5 have been debated for over a decade,
as \citet{malhotra04} found no evidence for significant evolution, while
\citet{ouchi10} found evidence for a significant, albeit mild
($\sim$30\% in line luminosity) decrease to $z\!\!=$6.6.  

The advent of
wider-area narrowband searches has recently led to improved
statistics, notably the HyperSuprimeCam (HSC) Subaru Strategic
Program (SSP). The larger areas covered now allow constraints on
reionization from not only the luminosity function, but also the
angular clustering.  The first HSC results at $z >$ 6 were published by \citet{ouchi17}, who studied
the evolution of $\sim$2000 Ly$\alpha$ emitting galaxies (LAEs) at $z
=$ 5.7 and 6.6 over 14--21 deg$^2$.  They specifically compared the
evolution in the angular clustering across this redshift range,
finding that LAEs at $z\!\!=$6.6 appeared slightly more biased than at
$z\!\!=$5.7, inferring constraints on the ionized IGM gas fraction of
$Q_{H_{II}}\!\!=$0.85 $\pm$ 0.15 at $z\!\!=$6.6, consistent with model
inferences from previous clustering results \citep{ouchi10,sobacchi15}.
This same sample was used
by \citet{konno17} to study the evolution of the Ly$\alpha$ luminosity
function.  They found a slight evolution in the luminosity function
(primarily a factor of $\sim$2 decrease in $\phi^{\ast}$) and inferred
$Q_{H_{II}}\!\!=$0.7 $\pm$ 0.2 at $z\!\!=$6.6.

Studies of the Ly$\alpha$ luminosity function at higher redshift are
being pursued with a variety of telescopes, though they are still
nascent. \deleted{so results are ongoing.}  \citet{krug12} published results from
a narrowband search for LAEs at $z\!\!=$7.7, finding four candidate
galaxies.  They concluded that if at least one candidate was a true $z
=$ 7.7 LAE, there was no evidence for significant evolution of the LAE
luminosity function.  Similar results were found at this redshift
using data with similar depths by \citet{tilvi10} and \citet{hibon10}.
However, \citet{konno14} used a
Subaru SuprimeCam survey for LAEs at $z\!\!=$7.3 a factor of
$\sim$2 deeper and wider that the $z\!\!=$7.7 surveys to find 
seven LAEs over 0.5 deg$^2$, while 65 would have been expected in the case of no
evolution from $z\!\!=$6.6.  They concluded this was evidence for
significant evolution, inferring $Q_{H_{II}}\!\!=$0.55 $\pm$ 0.25 at $z
=$ 7.3.  Similar results are found when modeling the evolution of the
observed Ly$\alpha$ luminosity functions and correlation functions
across 5.7 $< z <$ 7.3 from \citet{inoue18}.

While tenuous, these surveys combine to suggest that the IGM is not
completely ionized by $z \sim$ 7, consistent with our model predictions.  However, while the Ly$\alpha$
luminosity function appears to decline, this may not be uniform at all
luminosities, as there has been evidence for a bright-end ``bump''
seen in multiple surveys, where the abundance of LAEs at log $L$
$\gtrsim$ 43 erg s$^{-1}$, \edit1{most spectroscopically confirmed}, is higher than expected from a Schechter
function fit to lower luminosities
\citep[e.g.,][]{matthee15,bagley17,zheng17,hu17}.  \deleted{Due to these bright
line luminosities, the majority of these objects have been
spectroscopically confirmed, thus this excess appears real.}  One
proposed physical explanation is that the neutral fraction inferred by
the decline in the overall luminosity function is real, while these
bright LAEs live in ionized bubbles, and thus the photons suffer reduced
attenuation.  Another proposed scenario is that these extremely bright
emitters are powered by AGN.  At a much lower redshift of
$z\!\!=$0.3 and 2.2, \citet{wold17} and \citet{konno16}, respectively, found that with a wide-enough dynamic range,
one could simultaneously measure the Ly$\alpha$ luminosity function
from star-forming galaxies as a Schechter function, and those from AGN
as a power-law.  Whether this extends to such distant redshifts is
unknown, though interestingly three highly luminous (log $L$ $>$ 43 erg
s$^{-1}$) $z \gtrsim$ 7 LAEs have tenuous detections of N\,{\sc v}
\citep{tilvi16,hu17,laporte17,mainali18}.

\subsubsection{Contraints from Ly$\alpha$ Emission --
  Spectroscopy}\label{sec:discuss_lyaspec}

Constraints on the IGM ionization state have also been made using
spectroscopic followup of Ly$\alpha$ emission from 
continuum-selected distant star-forming galaxies.  For brevity we
focus our discussion on more recent results which have ever-increasing
statistical confidence, but
acknowledge that a significant amount of work has been done in this
area \citep[e.g.,][]{stark10,fontana10,stark11,pentericci11,schenker12,ono12,rhoads12,rhoads13,caruana14}.
\citet{pentericci14} observed 12 $z \sim$ 7 galaxy candidates in the
CANDELS EGS field with the FORS2 optical spectrograph on the VLT.
They included data on similarly selected sources from previous observations
\citep{castellano10,fontana10,pentericci11,vanzella11,ono12,schenker12,bradac12}
to amass a total sample of 68 candidate $z \sim$ 7 galaxies
spectroscopically observed with 8--10m telescopes.
Over this combined sample, they find that Ly$\alpha$ is significantly
detected with rest-frame EW $>$ 25 \AA\ in 7 of 39 galaxies at
$-$21.25 $<$ $M_\mathrm{UV} < -$20.25, and 5 of 25 galaxies at $-$20.25 $<$
$M_\mathrm{UV} < -$18.75.  By comparing to reionization models, they find that this detection
fraction is consistent with $Q_{H_{II}} <$ 0.49 (1$\sigma$) at $z\!\!=$7, under the
assumption that the IGM is fully ionized at $z\!\!=$6.  

More recently, \citet{pentericci18} used spectroscopic observations of 167
z$\sim$6--7 candidates with a large ESO FORS2 program to revisit the
Ly$\alpha$ detectability at $z \sim$ 7, but also to improve
constraints at $z \sim$ 6.  Interestingly, \deleted{with this updated sample of
$z \sim$ 6 galaxies,} this study finds a similar detection fraction in
bright galaxies ($M_\mathrm{UV} < -$20.25) at $z \sim$ 6 and 7 of
10\%, while the fraction drops from 35\% at $z \sim$ 6 to 20\% at $z
\sim$ 7 in fainter galaxies.  Taken together with
previous results at $z \sim$ 5 from \citet{stark10}, this implies a
flat evolution in the Ly$\alpha$ detectability from $z
=$ 5 to 6, followed by a milder decline to $z\!\!=$7 (see also \citealt{debarros17}).  This is in
contrast to previous results which found a sharper decrease at $z >$ 6
due to the previous rise in the detectability from $z\!\!=$5 to 6 \citep[e.g.][]{stark10}.  This is
consistent with their observation of only a slight decrease in the
residual flux on blue side of the line at $z \sim$ 6 compared to $z
\sim$ 7, as well as only a modest evolution in the EW distribution
from $z\!\!=$6 to 7.  Taken together with previous measurements of a
rise in the Ly$\alpha$ detectability from $z\!\!=$4 to 5, this result
indicates that the IGM may not be fully reionized by $z \sim$ 6,
necessitating a smaller change in the neutral fraction to $z\!\!=$7 than when assuming
the IGM is fully ionized at $z\!\!=$6.

\citet{mason18}, building on a Bayesian framework introduced in
\citet{treu12} and \citet{treu13}, use this new sample of $z \sim$ 7
galaxies from \citet{pentericci18} to infer the neutral fraction of
the IGM.  In an advance over previous studies, this paper measures the magnitude-dependent
evolution of the Ly$\alpha$ EW distribution rather than just the
Ly$\alpha$ detection fraction (see also \citealt{jung18}).  They also include the
effects of Ly$\alpha$ velocity offsets, which may evolve to lower
values at higher redshifts or for lower-mass galaxies
\citep[e.g.][]{song14,stark15,pentericci16,bradac17}, allowing a
smaller amount of neutral IGM gas to obscure a given Ly$\alpha$ line.
They use the sample of $z \sim$ 6 galaxies from \citet{debarros17} and
\citet{pentericci18} to set a baseline $z \sim$ 6 EW distribution
measurement, and then explore the evolution to $z \sim$ 7 using the
sample from \citet{pentericci14}, finding Q$_{HII}\!\!=$0.41$^{+0.15}_{-0.11}$ at $z\!\!=$7.
This result is model dependent as it draws sightlines from cosmological simulations to
realistically model the impact on Ly$\alpha$ on all scales.
Additionally, \citet{mason18} assumed that the IGM was fully ionized at $z \sim$ 6, thus if the
implication from \citet{pentericci18} is correct, the 
neutral fraction at $z \sim$ 7 needed to explain the observations may be lower.  Additionally, this paper
relied on the relatively few observations of the Ly$\alpha$ velocity
offset at high redshift to calibrate their model; larger samples of non-resonant line
redshifts are needed to increase the robustness of these
calibrations.  Finally, improved statistical power on the IGM at $z \sim$ 7 can be obtained by
fitting the evolution of EWs down to lower redshift, while also making
use of larger samples of $z \sim$ 7 galaxies \citep[e.g.][]{pentericci18}.

\begin{figure*}[!t]
\epsscale{0.9}
\vspace{2mm}
\plotone{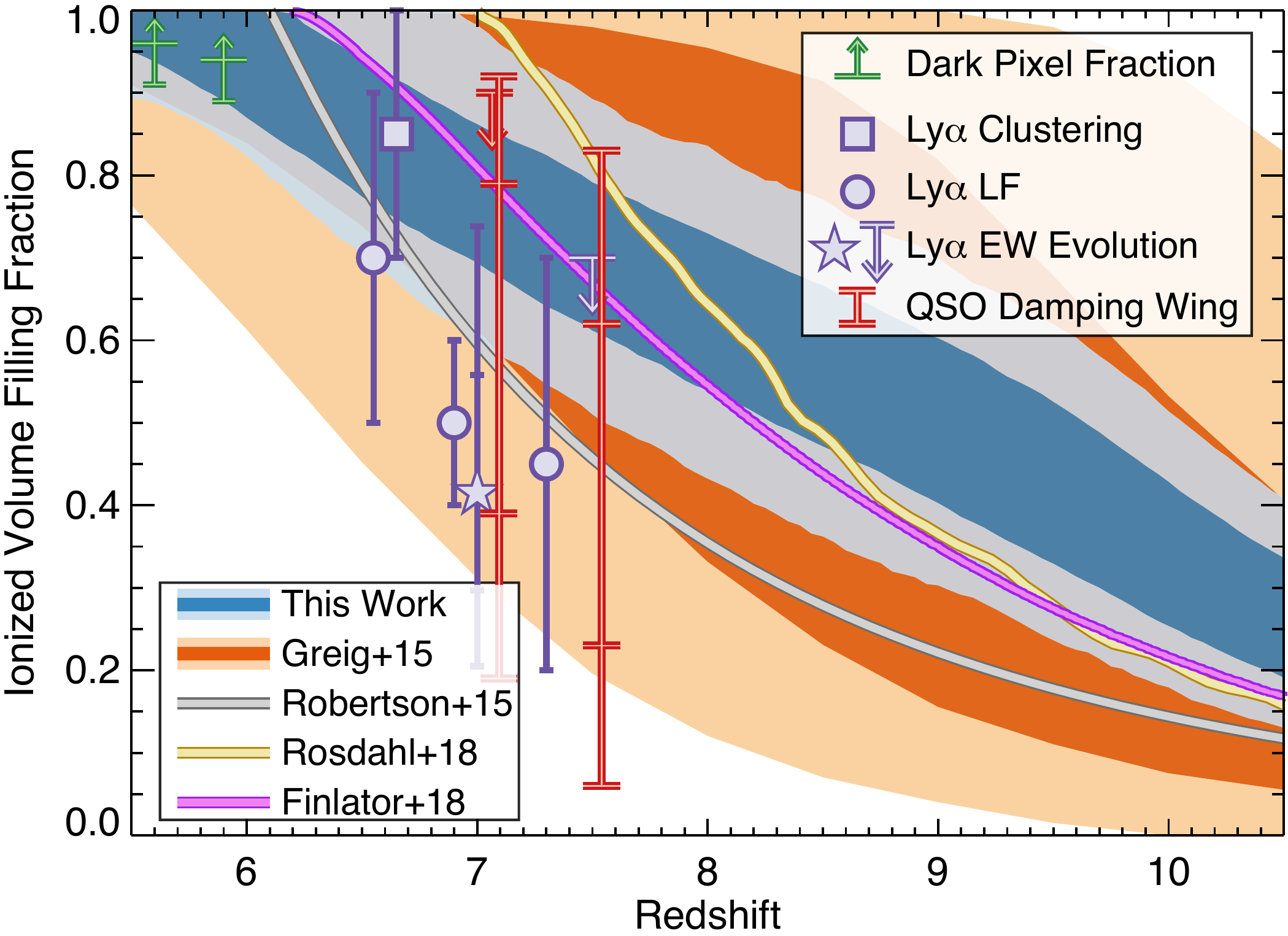}
\caption{A comparison of our fiducial model to recent observational
  and theoretical results.  The
  dark and light shading for our model denote the 1$\sigma$ and
  2$\sigma$ confidence ranges, respectively, and a similar shading is
  used in orange for the results of the \citet{greig17} model when
  using constraints similar to our model.  Purple
symbols denote constraints from Ly$\alpha$ emission from clustering
\citep{ouchi17}, luminosity function evolution
\citep{konno14,konno17,zheng17}, and Ly$\alpha$ spectroscopic followup
\citep{tilvi14,mason18}.  The green arrows denote the limits from
\citet{mcgreer15} used to constrain our model, while the red lines
denote confidence ranges from $z >$ 7 quasar damping wing measurements
\citep{bolton11,greig17b,banados18}.  While some measurements lie below the
posterior of our model, the significance of the difference is not
large ($<$2$\sigma$ in most cases), and there are a variety of
reasons to be cautious when interpreting these results, as discussed
in the text.  Specifically, the result which differs most
significantly from our model is that from \citet[][purple star]{mason18} based on the evolution of the Ly$\alpha$ EW distribution.  This
model assumes a fully ionized IGM at $z\!\!=$6.  However, recent results
imply a small neutral fraction at $z\!\!=$6 \citep[e.g.,][]{kulkarni18,pentericci18}, which when folded through
the \citet{mason18} model, will increase the inferred ionized
fraction at $z\!\!=$7.}
\label{fig:qhii_compare}
\end{figure*}   

As the number of Ly$\alpha$ detections at $z >$ 7.5 is small, less
work has been done on the implied neutral fraction at such epochs.
Using deep Keck/MOSFIRE integrations of 48 candidate $z \sim$ 7--8
galaxies in the CANDELS GOODS-N field, \citet{finkelstein13} explored how the single detection at
$z\!\!=$7.51 compared to the expectation from the predicted Ly$\alpha$ EW
distribution from \citet{stark11}, finding that this EW distribution
could be ruled out at 2.6$\sigma$ significance.  
\citet{song16b} used Keck/MOSFIRE observations of 12 $z \sim$ 7--8
candidates in the CANDELS GOODS-S field to perform a similar analysis,
finding that their single detection at $z\!\!=$7.66 also ruled out no
evolution in the EW distribution from $z\!\!=$6 (at 1.3$\sigma$
significance).  A more detailed Bayesian analysis with the data from
\citet{finkelstein13} was performed by \citet{tilvi14}, finding
conservatively that Q$_{HII} <$ 0.7 at $z \sim$ 7.5.  

Together, these observational results on Ly$\alpha$ emission (luminosity functions,
clustering, and spectroscopic followup) tell a coherent story -- the
IGM must be significantly neutral by $z \sim$ 7.  Taken at face value, 
they imply tension with our fiducial model, which gives Q$_{HII} =$
0.78 $\pm$ 0.08 at $z\!\!=$7, though we note our result at $z\!\!=$6 of
Q$_{HII} >$0.87 is consistent with the slightly later end to
reionization discussed by \citet{pentericci18}.  This comparison is
highlighted in Figure~\ref{fig:qhii_compare}, which shows our fiducial model
compared to a number of the Ly$\alpha$-based constraints discussed
here.  

However, there are a few reasons to be cautious about these
interpretations, especially those which infer very large neutral
fractions (Q$_{HII} <$ 0.5 at $z\!\!=$7) from an apparent deficit of
observable Ly$\alpha$ emission.
First, not all galaxies have been difficult to detect.  Several
publications have noted that bright ($M_\mathrm{UV} < -$21.5) galaxies with a
non-zero IRAC color indicative of strong [O\,{\sc iii}] emission
have a very high Ly$\alpha$ success rate
\citep[e.g.][]{finkelstein13,oesch15,zitrin15}.  Notably,
\citet{roberts-borsani15} identified four such galaxies, which,
together with spectroscopy from \citet{zitrin15} and \citet{stark17}, all
have detected Ly$\alpha$ emission, with \citet{stark17} arguing that
these galaxies have boosted Ly$\alpha$ transmission due to inhabiting
ionized bubbles, consistent with similar inferences from the
high-luminosity ``bump'' seen in $z >$ 6 Ly$\alpha$ luminosity
functions.  Similar results are also seen for bright galaxies at $z =$
6--6.5 \citep{curtis-lake12}.  Such a physical scenario may also explain the
Ly$\alpha$ line discovered by \citet{larson18} at $z\!\!=$7.45, which
has an EW of 140 \AA, more than double that of any other detected
Ly$\alpha$ line at $z >$ 7.  

\deleted{While we are awaiting inferences on the
neutral fraction accounting for these observations,} \citet{mason18b}
recently considered this from a modeling perspective,
finding that the data are consistent with bright galaxies having a
greater Ly$\alpha$ transmission fraction through the IGM (by
$\sim$2$\times$).  They found that if the samples are large enough,
these bright galaxies can accurately constrain the neutral fraction,
which presents a compelling argument for targeted followup of bright
galaxies at $z >$ 7.  However, while the modeling
results by \citet{weinberger18} agree that the IGM transmission
fraction is higher for brighter galaxies, they also found that self-shielded neutral
gas in the CGM can attenuate Ly$\alpha$, and that this occurs
preferentially in more massive
halos.  They thus argue that \emph{fainter} LAEs are a more reliable probe of
the IGM state, and that current observations are consistent with
$Q_{H_{II}} =$ 45--75\% at $z\!\!=$7 (for their preferred ``very late''
and ``late'' reionization models, respectively).

Second, there is also evidence from recent observations that the previous
evolution of Ly$\alpha$ detectability from \citet{stark10}, which have
been used to invoke large neutral fraction evolution, may need to
be revised.  In addition to the result from \citet{pentericci18},
\citet{caruana18} revisited the Ly$\alpha$ fraction evolution from $z
=$ 3--5 with deep VLT/MUSE IFU spectroscopy, finding no significant
evolution from $z \sim$ 3 to 5 (X$_{Ly\alpha} \sim$ 30\% across this
range).  Several spectroscopic surveys are underway to improve these
measurements, thus the near future should yield more
well-characterized Ly$\alpha$ fractions and EW distributions at $z =$
3--6.  

Finally, it is worth discussing the several factors working
\emph{against} detecting Ly$\alpha$ in this epoch, even if the
galaxies are copiously producing these photons and the IGM is
ionized.  First, the galaxies targeted spectroscopically can have
broad photometric-redshift probability distributions, which at $z\!\!=$7
results in a significant probability that Ly$\alpha$ would be observed
both in the optical and in the near-infrared.  This would necessitate
two observations with different instruments to fully sample the
redshift PDF, which is not commonly done.  Even within one instrument,
an individual grating/filter pair may not fully sample the PDF.  While
this incompleteness effect can be accounted for, it is reliant on the
shape and width of the photometric redshift PDFs, which have not been calibrated at the
redshifts of interest.  \citet{jung18} explore this effect, and find
that if the redshift uncertainties are increased by 50\%, the number
of expected Ly$\alpha$ detections for a fixed EW distribution is
reduced by more than a factor of two.  This doesn't include the
possibility that some galaxies are contaminants from lower redshift,
though this is explored by \citet{pentericci11} and
\citet{pentericci14}, who find that significant contamination is
unlikely (though this depends strongly on the quality and amount of
photometric data available).  Secondly, at $z >$ 5, Ly$\alpha$ is observed at wavelengths
significantly contaminated by night-sky emission, rendering much
($>$50\% at R$\sim$2000) of the wavelength space unavailable (except
for extremely bright lines). \citet{finkelstein13} found that the
presence of night-sky emission lines alone reduce the expected number
of detections by a factor of $\sim$ 3 (see their Figure S5).
Last, and easiest to model, is the effect of the depth of the
observations, though this could be impacted by inaccurate flux
calibration or slit-loss correction.

Nonetheless, as the current observations do indicate some tension with
our model, in \S~\ref{sec:discuss_lyaqso} below we explore how our results change if we
include recent Ly$\alpha$-based constraints during our fitting process.

\subsubsection{Contraints from Quasars}
In addition to the statistical measures of the Ly$\alpha$ and
Ly$\beta$ dark pixel fraction over multiple combined quasars used by
\citet{mcgreer15} to constrain the neutral fraction, individual
quasars can also be constraining.  Spectroscopy of the region near the
Ly$\alpha$ transition can allow one to measure excess absorption redward of
the Ly$\alpha$ transition indicative of the presence of neutral gas,
and thus a strong Ly$\alpha$ absorption damping wing.  Here we
consider the two presently available damping wing measurements at $z
>$ 7.

The first quasar discovered at $z >$ 7 was published by
\citet{mortlock11}, at $z\!\!=$7.085.  The spectrum of this object, and
specifically the Ly$\alpha$ damping wing, was
further analyzed by \citet{bolton11}, who found a lower limit on the
neutral fraction of $>$10\%, or Q$_{HII} <$ 90\%.  These data were
further analyzed by \citet{greig17b}, who found 0.39 $<$ Q$_{HII}
<$0.79 at 1$\sigma$ (and 0.19 $<$
Q$_{HII} <$0.92 at 2$\sigma$).  These constraints are consistent with
the 1$\sigma$ confidence range from our fiducial model at this
redshift of 0.68 $<$ Q$_{HII} <$0.85, although the allowable range of
ionized fractions from the quasar analysis extends to lower values, as
shown in Figure~\ref{fig:qhii_compare}.
More recently \citet{banados18} published the
discovery of a quasar at $z\!\!=$7.54, finding a spectral damping wing
consistent with 0.23 $<$ Q$_{HII} <$0.62 at 1$\sigma$ (and 0.06 $<$
Q$_{HII} <$0.83 at 2$\sigma$).  While the $z\!\!=$7.085 quasar implies a
neutral fraction consistent with our fiducial model, this
higher redshift quasar has a damping wing which implies a higher neutral
fraction than our model, although the
discrepancy is only at 1.3$\sigma$.

However, we note that these
measurements are for single lines-of-sight to these cosmologically-biased
sources, and with only two lines-of-sight, it is difficult to make
robust conclusions on the global neutral fraction.  
Additionally, these measurements rely on the ability to model the
intrinsic quasars spectrum near the Ly$\alpha$ line, which cannot be
directly observed.  Finally, the
observed signature of neutral gas could also be created if there was a
nearby damped Ly$\alpha$ absorber along the line-of-sight.  As
discussed by \citet{bolton11}, if the $z\!\!=$7.085 quasar had a nearby
absorber, the spectrum would be consistent with Q$_{HII} >$ 0.999
\citep[though see][]{finlator13}.
Combined with the low significance of the difference between our
fiducial model and the quasar results, we conclude the quasar
observations do not rule out our fiducial model, though future
observations may necessitate a revision.

Further confidence will be gained as the sample of $z >$ 7 quasars
increases, in particular if they all show a significant damping wing.
Similar studies can be pursued with fast followup of gamma ray bursts \citep[e.g.,][]{totani14},
potentially to even higher redshifts.
A variety of wide-field surveys are underway, and the
future for quasar discovery is bright with the launch in the next
decade of the wide-field near-infrared survey telescopes
\textit{Euclid} and \textit{WFIRST}.

\subsubsection{Predictions from Theoretical Models}\label{sec:discuss_theory}
The past few decades have seen tremendous advances in our ability to
theoretically model the process of reionization using a variety of
methods \citep[e.g.][]{shapiro94,trac07,kuhlen12,haardt12,gnedin17}.  In
Figure~\ref{fig:qhii_compare} we compare our fiducial results to a few
recent simulation results using different techniques.  The gray line
shows the result from R15, who showed that one could
match both the \citet{planck15} optical depth values and the observed
SFR density from high-redshift galaxies with a model where galaxies
all have a uniformly high escape fraction of 20\%, and modest
ionizing photon production efficiency of log $\xi_{ion}\!\!=$25.2.  As
discussed in \S~\ref{sec:results_qcomp}, this results in L $>$ 0.1L$^{\ast}$ galaxies
dominating the photon budget, a late beginning to reionization,
followed by a rapid ramp-up in the ionized volume filling fraction at
$z <$ 8.  While this model has the benefit of exhibiting larger
consistency with the Ly$\alpha$ and quasar constraints, it relies on a
uniformly high escape fraction which is unlikely (see \S~\ref{sec:discuss_fesc}).  Also shown are the results from the model of \citet{greig17},
using only Planck $\tau_{es}$ and the \citet{mcgreer15} dark pixel
fraction as constraints.  While the \citet{greig17} model is less
empirical than our own, using a set of
analytical formulae with three free parameters, they find similar
reionization histories as our model when they use similar constraints.

\citet{bouwens15b} use a 
two-parameter model to model the emissivity from galaxies,
constraining the redshift evolution of this quantity using a variety
of observational constraints.  They found that the emissivity must
decrease by 0.15 dex per unit increasing redshift over $6 < z < 10$, and that this
emissivity is consistent with being produced fully by galaxies if log$(f_\mathrm{esc}\xi_{ion})\!\!=$24.53.  For the commonly used fiducial value of
log$\xi_{ion}\!\!=$25.2, this corresponds to an average escape fraction
of $\sim$20\%, similar to that used by R15.  For a
higher log$\xi_{ion}\!\!=$25.5, this can be accomplished with $f_\mathrm{esc}
=$ 10\%.  One key difference between the results of \citet{bouwens15b}
and our own is that our inferred emissivity \emph{increases} with increasing
redshift over $6 < z < 10$, which is a consequence of our lower
average escape fraction over this epoch (see \S~\ref{sec:discuss_fesc}).

Recently \citet{rosdahl18} explored reionization with the
SPHINX simulations, which simultaneously resolves the small scale physics
regulating the ionizing emissivity and begins to approach the larger scales needed to
solve for global reionization.  This paper focuses specifically on the
impact of binary stars, finding that by including binary stars their
full volume ionizes by $z \approx$ 7; this is due to both the increase
in ionizing photon production by massive binaries, but also the
increase in escape fraction as well (see \S~\ref{sec:discuss_fesc} for further discussion).
Similar to our fiducial model, their simulation begins reionization
relatively early.  However, their reionization history completes by $z\!\!=$7.  Examining their results, this is likely because their typical
escape fractions are $\sim$10\%, whereas the globally averaged escape fractions in our model are
rapidly evolving from $>$10\% at $z \sim$ 15, to $<$5\% at $z <$ 10
(\S~\ref{sec:discuss_fesc}).  Therefore, while the early phases of reionization are
similar between this model and our own, the end of reionization is
more extended in our fiducial model due to the declining galaxy emissivity.
Finally, the Technicolor Dawn simulations \citep{finlator18} combine
an updated model for galaxy formation and feedback with a
multifrequency moment-based radiation transport solver that models
reionization and photoionization heating in detail.  The
redshift-dependent ionizing escape fraction is calibrated to match
observations of the optical depth to Thomson scattering as well as
ionizing emissivity at $z=5$.  As shown in
Figure~\ref{fig:qhii_compare}, this calibration leads to a
reionization history which predicts a similar ionized fraction as our
model at $z \sim$ 7, though predicts a somewhat lower value of
$Q_{H_{II}} \sim$ 35\% at $z =$ 9 (compared to $\sim$50\% for our model).

\subsection{Escape Fraction}\label{sec:discuss_fesc}

\subsubsection{Comparison to Observations}\label{sec:discuss_fesca}
\begin{figure}[!t]
\epsscale{1.15}
\vspace{4mm}
\plotone{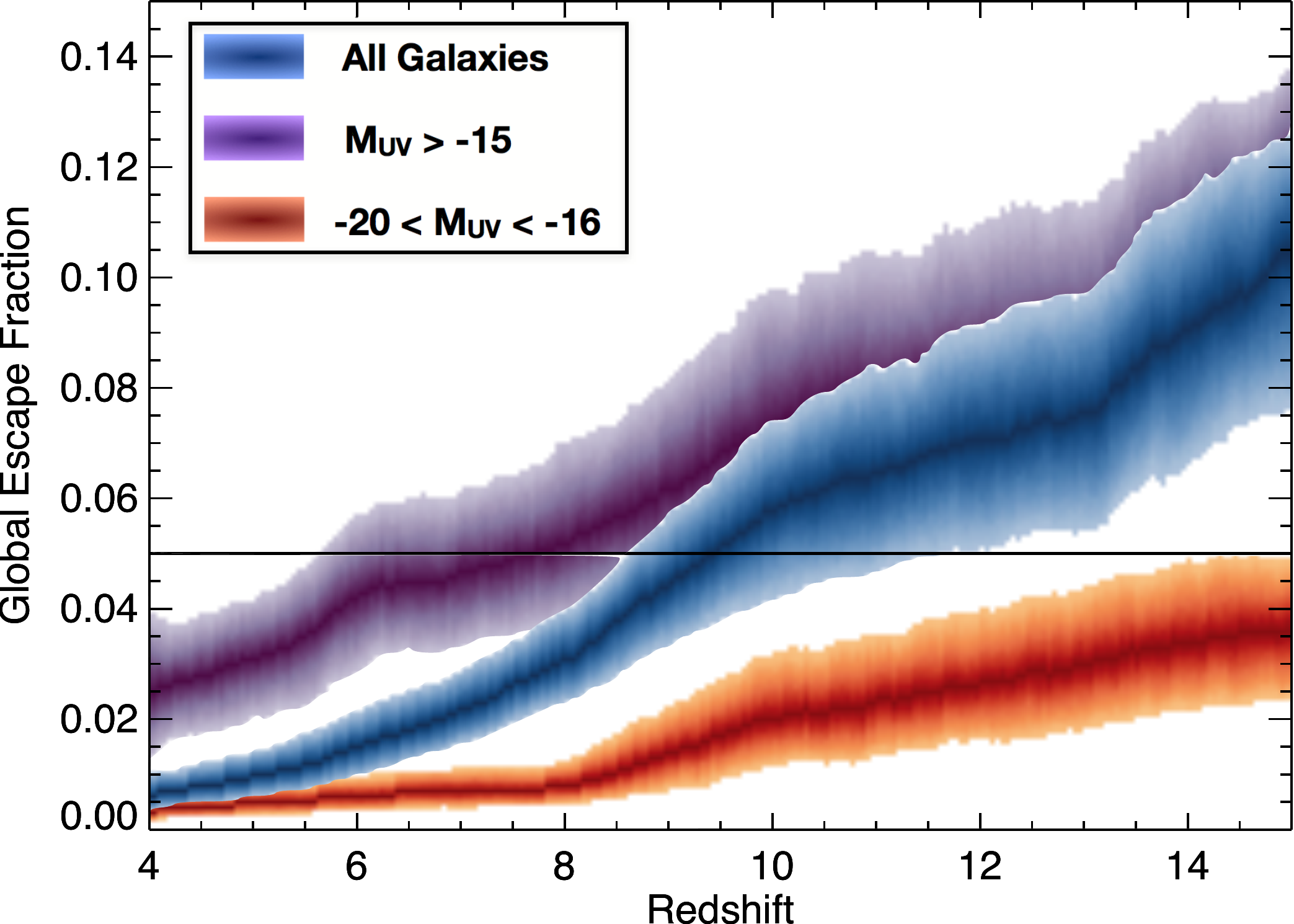}
\caption{The evolution of the global ionizing photon escape fraction,
  defined as the total number of escaping ionizing photons divided by
the total number created at each redshift.  The shading is defined in
the same way as Figure~\ref{fig:agn}.  The horizontal line denotes an
escape fraction of 5\%.  At the highest redshifts, the steep
luminosity function faint-end slope results in a larger abundance of
low mass/extremely faint galaxies, which have the highest escape
fractions, leading to a typical escape fraction of $\sim$10\% at
$z\!\!=$15.  As the halo mass function evolves and the faint-end slope
of the luminosity function shallows towards lower redshift, the global escape
fraction drops, to $<$5\% at $z\!\!=$9, and only $\sim$1\% at $z\!\!=$4,
consistent with the majority of unbiased observations which constrain
average escape fractions to be $<$5\% at $z \sim$ 2--4.  This is seen
as the global value for all galaxies transitions from being similar to
that for faint galaxies only at $z >$ 12, to that for brighter
galaxies at $z <$ 6.}
\label{fig:fesc_global}
\end{figure}

One of the primary differences of our model from previous analyses is the escape fraction
distribution, which is skewed highly towards low halo masses.  As
shown in Figure~\ref{fig:ncum}, this results in the bulk of ionizing
photons coming from extremely faint galaxies.  Here we examine what
effect this has on the redshift evolution of the global escape
fraction, which is shown in Figure~\ref{fig:fesc_global}.  The blue
shading in this figure
shows the posterior distribution at each redshift of the total number
of escaping ionizing photons divided by the total number created over
all galaxies (i.e., over the full luminosity function), and highlights
that this globally-averaged escape fraction evolves significantly, from
$<$1\% at $z\!\!=$4, to $>$10\% at $z\!\!=$15.  In this figure, we
also show separately the global escape fraction from faint galaxies
with $M_\mathrm{UV} > -$15, and brighter galaxies with $-$20 $<
M_\mathrm{UV} < -$16.

This evolution is entirely driven by the co-evolution of the luminosity and halo mass functions.
At the highest redshifts, the luminosity function is steep, and these
faint galaxies live in very low-mass halos.  For example, at $z\!\!=$10,
the bulk of the escaping ionizing photons are produced in halos with
$M_{UV} > -$15 (shown in purple in Figure~\ref{fig:fesc_global}).
These faint galaxies, which according to our abundance matching
results shown in Figure~\ref{fig:fig1} have log
($M_\mathrm{h}/M$\sol) $\lesssim~$9, have
globally-averaged escape fractions of 6--10\%.  At this same redshift,
brighter galaxies ($-$20 $< M_\mathrm{UV} < -$16; log
$M_\mathrm{h}/M$\sol\ $>$ 9.5) have escape fractions of only 1--3\%, shown by
the red shading in Figure~\ref{fig:fesc_global}.  While many of these
halos should have negligible escape fractions, this non-zero global
value is driven by high individual escape fractions ($f_\mathrm{esc} >$50\%) in a small
number of starbursting galaxies (see red line
in Figure~\ref{fig:fesc} and discussion in \S~\ref{sec:mcmcfesc}).  

At progressively
lower redshifts, the continued evolution of the luminosity function,
specifically the flattening of the faint-end slope, results in fewer
luminous halos with such low masses, reducing the overall escape
fraction.  This effect is also observed when exporing the ionizing
emissivity in the left-hand panel of Figure~\ref{fig:ndottau}, as the
emissivity from galaxies
decreases by nearly 1 dex from $z\!\!=$10 to $z\!\!=$4, even though the
non-ionizing $\rho_{UV}$ \emph{increases} by nearly 1 dex over this same
redshift range.

There have been a large number of observational attempts to directly
measure the ionizing photon escape fraction.  These are 
difficult endeavors for a variety of reasons.  First, 
corrections need to be made for both the galaxy SED shape and dust
attenuation to convert the observed ratio of Lyman continuum-to-1500
\AA\ flux into the
total fraction of escaping ionizing photons.  Secondly, the observed
ionizing photons have traveled through the IGM, which while
ionized still absorbs some fraction of the photons, requiring a
statistical correction with a large variance, which propagates through into
significant uncertainties on the escape fraction.  Finally, the IGM
optical depth becomes so great at $z >$ 4 to render direct measures of
the escape fraction in the epoch of reionization impossible, with the
most distant detection presently at $z\!\!=$4.0 \citep{vanzella18}.

Nonetheless, the importance of observationally constraining the escape
fraction has led to a large number of ambitious observational
programs.  A trend in recent years has been an increasing success rate
of robust direct detections of Lyman continuum radiation with $f_\mathrm{esc} \gtrsim$
10\%, both at $z \sim$ 2--4
\citep[e.g.][]{steidel01,shapley16,debarros16,bian17,vanzella18,fletcher18}, and locally
\citep{izotov16,izotov16b,izotov18}.  These galaxies were typically
selected for followup based on very high observed or inferred
ionizing environments, typically constrained via the ratio of [O\,{\sc
  iii}]/[O\,{\sc ii}] line emission, which has been proposed as a
predictor of strong ionizing photon escape \citep[e.g.][]{jaskot13,nakajima14,stasinska15}.
However, other studies have explored the characteristic global escape
fraction of galaxies by stacking larger numbers of individually
un-detected galaxies, typically finding no detection even in the
stack, setting strong upper limits on the typical escape fraction to
be as low as $<$2\%
\citep[e.g.,][]{siana10,sandberg15,rutkowski17,japelj17,grazian17,hernandez18}.

These observations only probe presently observable galaxies, and thus
are probing relatively massive halos.  However, these results
qualitatively agree with our assumed escape fraction for such halos.
These massive galaxies typically have very low escape fractions,
consistent with the non-detections in the stacks of full galaxy
samples.  However, they do have a small ($\sim$5\%) probability of
being observed with a very high escape fraction, due in the simulation
to starbursts.  This could be the origin of the small fraction of
galaxies with observed high escape fractions.  While this is a tidy
explanation, it could also be coincidental.  What is truly needed are
much more stringent escape fraction measurements for
fainter/lower-mass galaxies.  This would require a leap in our
space-based ultraviolet imaging or spectroscopic capabilities, which
is the focus of several concepts for future space missions \citep{scowen17,mccandliss17}.
However, fainter galaxies have begun to be probed.  By analyzing the
H\,{\sc i} column densities in
gamma-ray burst (GRB) host galaxies, \citet{tanvir19} find a typical
escape fraction for GRB hosts of 0.005.  As GRB hosts can be extremely faint,
this result implies that the escape fractions from even faint
galaxies in our model may be too high.

\edit1{Finally, we comment on the recent result by \citet{steidel18}, who
derived the escape fraction from bright ($\mathcal{R}$ $<$ 25.5) $z \sim$ 3 galaxies via very
deep spectroscopic followup, combined with an intensive modeling
effort.  In contrast to some previous imaging results, this
spectroscopic campaign detected escaping
Lyman continuum radiation, and derived an average escape fraction of
f$_{esc} =$ 0.09 $\pm$ 0.01 (similar results have been found via other
methods by \citealt{kakiichi18} and \citealt{fletcher18}).  This
average comes from a direct detection from 15 galaxies with f$_{esc}
=$ 0.60 $\pm$ 0.06, and a stacked detection from 109 individually
undetected galaxies, with f$_{esc} =$ 0.04 $\pm$ 0.01.  This trend is
qualitatively similar to that shown for massive galaxies in
Figure~\ref{fig:fesc}, where $\sim$10\% of the probability density
lies at high escape fractions, though in our fiducial model the
remaining probability density has f$_{esc} \ll$ 0.04.  They also find
a luminosity dependance, where the brightest 50\% of their galaxies
have escape fractions consistent with zero, and the faintest 50\% have
f$_{esc} \sim$ 0.3 (similar trends are also seen with Ly$\alpha$ EW).
They use these observations to conclude that bright galaxies are
dominating the emissivity at $z \sim$ 3 ($\sim$3$\times$ that of
quasars).  While we do not track galaxies to $z <$ 4,
extrapolating our results shows that in our fiducial model quasars should
dominate the emissivity at $z \sim$ 3.}

\edit1{While more work is required to see whether contamination from line-of-sight
interlopers is found to be minimal for this sample (available soon
from an in-progress {\it HST} program [PI Shapley]), and whether this result holds across different
fields and redshifts, we explored how our model would change if we assumed a
fixed escape fraction of 9\% for all galaxies.  Unsurprisingly, this
model completes reionization earlier at $z =$ 7.0 $\pm$ 0.8, with a total emissivity
dominated by galaxies at all epochs, which rises continuously with
decreasing redshift, reaching log $\dot{N}$ $=$ 51.25 $\pm$ 0.1 
s$^{-1}$ Mpc $^{-3}$ by $z =$
4.  This is higher than, but consistent within 1$\sigma$, of the observational
constraint at that redshift.  Due to the high emissivity from galaxies, the AGN
contribution is negligible, with the AGN emissivity matching the lower
bound allowed by our model at $z <$ 6.  
As the emissivity from galaxies should continue to rise to $z
\sim$ 3 in this model, it may begin to be in tension with the measured
value of the emissivity at that redshift of log $\dot{N}$ $=$ 50.8
s$^{-1}$ Mpc $^{-3}$,
though our model would need to be extended to lower redshifts to
explore this further.}

\subsubsection{Comparison to Other Models}\label{sec:discuss_fescb}
While we have assumed results from a particular simulation for our
escape fraction parameterization, here we explore how this differs
from other recent simulations, which find escape
fractions over nearly the full possible range, often with
a mass and/or redshift dependance
\citep[e.g.,][]{razoumov06,gnedin08,wise09,yajima11,paardekooper11,paardekooper15,kim13,wise14}.
\citet{anderson17} studied the escape fraction from several dozen
galaxies in their 25$\times$12$\times$10 Mpc simulation box, finding that their faintest
galaxies at $-$16 $< M_{UV} < -$14 had $f_\mathrm{esc} \sim$ 35\%, while
galaxies at $M_{UV} = -$18 had $f_\mathrm{esc} \sim$ 1\%.  While the
trend is qualitatively similar to what we assume, their normalization
is higher, such that utilizing the escape fractions from this
simulation would put less of an emphasis on the extreme faintest
galaxies.  However, these results are based on very few ($<$10)
galaxies fainter than $M_{UV} = -$18.   Additionally, the resolution of
this simulation is 350 pc, and unlike the
simulations of \citet{paardekooper15} the halos of interest were not
re-simulated at higher resolution, which could imply that the
important physical scales for ionizing photon escape were not
resolved (perhaps not fully resolving important physical processes such as
turbulence; \citealt{safarzadeh16}).  Nonetheless, their conclusion that the faintest galaxies in
their simulation dominate the ionizing emissivity is qualitatively consistent with
our results.

\citet{xu16} measured the ionizing escape fraction using the
Renaissance Simulation suite \citep{oshea15}, finding very high escape
fractions at $z\!\!=$8--15 of 40-60\% at log ($M_\mathrm{h}$/M\sol) $=$ 7, decreasing to
$\sim$5\% at log ($M_\mathrm{h}$/M\sol) $=$ 8--9.  Interestingly, they find
this rises to 10-20\% in their few halos log ($M_\mathrm{h}$/M\sol) $\sim$ 9.5,
presumably due to starburst activity.  These results
are in very good agreement with our assumptions, highlighting again
the dependance of extremely faint/low-mass galaxies to account for the
ionizing budget.  Alternatively, \citet{sharma16} find that if all
ionizing photons escape when the SFR surface density exceeds a
critical threshold, then it is the \emph{brighter} galaxies which
dominate the ionizing photon budget, as they exceed this critical
threshold more frequently.

\citet{ma15} explored ionizing photon escape using the FIRE
simulations, which include advanced treatment for feedback, and
utilize zoom-ins to achieve high ($<$1 pc) resolution on halos of
interest.  In this work, they find no dependance of $f_\mathrm{esc}$ on halo
mass, but they do find that the time-averaged escape fraction is
$<$5\%.  In a followup paper \citet{ma16} use these same simulations
but explore the impact of assuming binary stellar population synthesis models.
Mass transfer between binaries can result in massive stars having
significantly longer lifetimes, with the most massive ionizing-photon
producing stars extended to $>$3 Myr, long enough for the ``birth
cloud'', responsible for absorbing most of these photons, to disperse
\citep[e.g.][]{eldridge09}.  They find that this effect results in an
increase of the time-averaged escape fraction by a factor of $\sim$5--10, with
some massive simulated halos exhibiting $f_\mathrm{esc} >$ 10\%.  This effect
is amplified by the increased ionizing photon output of massive stars
in binaries, which exhibit hotter stellar photospheres due to the mass transfer.
Although these results are only available for a small number of simulated
halos, it heavily suggests that future simulations should account for
the impact of binary stars.  

\subsection{Inclusion of AGN}\label{sec:discuss_agn}
\subsubsection{Contribution of AGN versus Galaxies}\label{sec:discuss_agn}
As described in \S~\ref{sec:agnion}, we allow a contribution to the ionizing
emissivity from AGNs, with this emissivity allowed have a redshift
evolution anywhere from the steep \citet{hopkins07} evolution which
implies a minimal contribution during reionization, to the shallower
\citet{madau15} evolution which would result in AGNs dominating
reionization.  The shaded region in Figure~\ref{fig:agn} shows the
68\% confidence range of our fiducial model, compared to both of these
previously proposed trends, as well as data from the literature.  Our fiducial model
prefers an AGN ionizing emissivity in between these
previous trends at $z \sim$ 4, with a redshift evolution slope similar
to that of \citet{madau15}, albeit at a lower normalization.  

The posterior distribution of the three parameters that govern
this emissivity evolution are shown by the black lines in
Figure~\ref{fig:converge}.  The median value of these parameters are a
maximum redshift for AGN of $z_{AGN,max}\!\!=$9.2, a slope of the
emissivity with redshift of AGN$_{slope}\!\!=$$-$0.39, and a
normalization factor of AGN$_{scale}\!\!=$0.77.  However, none of these
posterior distributions has a well-constrained central value, so we
can more appropriately place one-sided 84\% lower limits of
$z_{AGN,max} >$ 6.9, AGN$_{slope} > -$0.93 and AGN$_{scale} >$ 0.47.
While these are broad lower-limits, they do constrain the emissivity
from AGNs to have a slope shallower than that from \citet{hopkins07},
kicking in at some point during the epoch of reionization.  

The AGN scale factor could be non-unity for a variety of reasons, but this is most
analogous to an ionizing photon escape fraction for AGNs.  This is
often assumed to be unity \citep[e.g.,][]{madau15}, which is
reasonable for very bright quasars, as the energetics near to the
accreting supermassive black holes likely create channels for ionizing
photon escape.  However, if the AGN emissivity evolves as shallowly as
suggested by our model, it is not these rare systems which are
dominating the emissivity, thus more pressing is the escape fractions
from fainter AGNs.  Most recently, \citet{grazian18} spectroscopically
observed 16 faint AGNs blueward of the Lyman continuum break, and
significantly detected ionizing flux from every object, implying AGN
ionizing photon escape fractions ranging from 44-100\%, with a median
of 74\%, consistent with the posterior distribution from our model
\citep[see also][]{smith18}.  \edit1{It is worth noting that the escape
fraction for ionizing photons from AGNs and from massive stars need
not be the same in a halo of fixed mass, since the central power
source of the AGN may clear a channel for those centrally-created
photons, while massive stars farther from the center still may be
subject to a high H\,{\sc i} column density.}

\begin{figure}[!t]
\epsscale{1.2}
\plotone{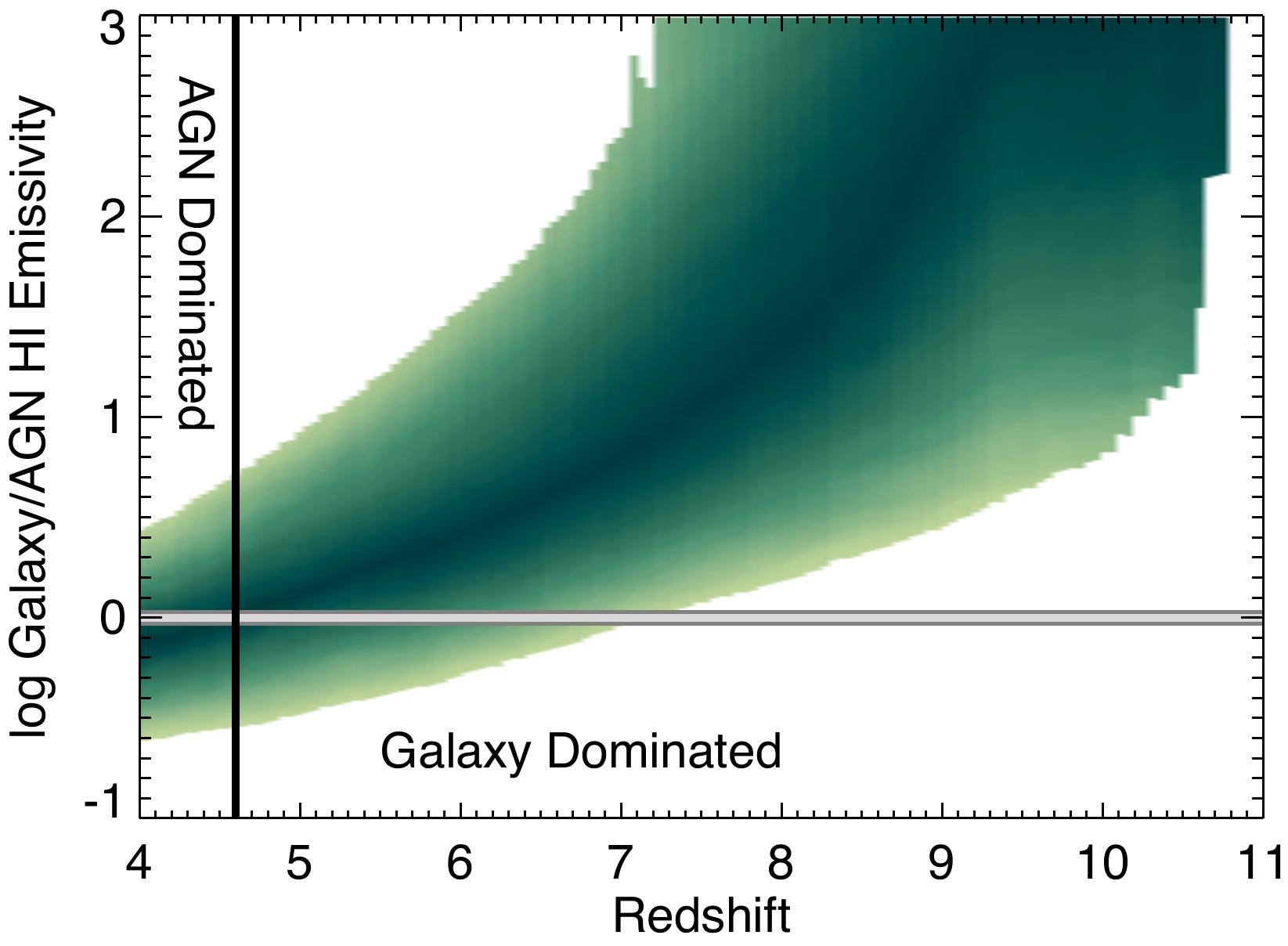}
\caption{The log of the ratio of the H\,{\sc i} ionizing emissivity from galaxies
to AGNs.  The gray bar denotes a ratio of unity, which occurs at $z
\sim$ 4.6.  At $z >$ 4.6, our fiducial model prefers a scenario
where galaxies dominate the H\,{\sc i} ionizing emissivity.}
\label{fig:ndot_ratio}
\end{figure}

In Figure~\ref{fig:ndot_ratio} we compare the emissivity from galaxies
to that from AGN in our fiducial model, with the shading denoting the
68\% confidence range, which is somewhat broad as it
includes the uncertainty in both distributions.  However, this plot
makes clear that while the emissivity from AGNs is significant,
\emph{galaxies are still the dominant driver behind reionization}.
The most conservative statement we can make is that at
68\% confidence, galaxies have a higher ionizing emissivity than AGNs
at $z >$ 7.0.  However, even at $z\!\!=$6.0, our model prefers galaxies
to dominate, with log($\dot{N}_{gal}$/$\dot{N}_{AGN}$)$=$
0.5$_{-0.4}^{+1.2}$.  As low as $z\!\!=$4.6, our model still implies
that half of the ionizing photons are coming from galaxies, with
log($\dot{N}_{gal}$/$\dot{N}_{AGN}$)$=$0.0$_{-0.3}^{+0.7}$, and it is not
until lower redshifts that the median of the posterior of our model
crosses into the AGN-dominated regime.  These results highlight that
our fiducial model of reionization is
``AGN-assisted'' rather than ``AGN-dominated.''  

Qualitatively similar models of the AGN ionizing emissivity were
considered in previous studies
\citep{2017MNRAS.468.4691D,mitra18,puchwien18}.  All of these authors
found that AGN-dominated scenarios were disfavored by contemporaneous
Ly$\alpha$ forest temperature measurements (we shall expand upon this
topic below).  This is consistent with the results of
\citet{finlator16} who found that
an AGN-dominated UV background would fail to reproduce the ratios of
observed CGM metal absorber column density distributions at $z \sim$ 6, while a small AGN
contribution could not be ruled out.
\citet{mitra18} also examined the contribution of quasars to
reionization, and similarly ruled out an AGN-dominated scenario for
hydrogren reionization.  They utilized observations of the He\,{\sc ii} Ly$\alpha$ forest to constrain their
models, finding that the contribution of AGNs to the ionizing
background must be negligible for $z >$ 6, preferring a lower
value of the AGN ionizing emissivity at $z\!\!=$4--6 than that of our model.

A significant contribution of AGNs to the $z>4$ ionizing background, while contrary to
previous expectations, can potentially resolve some interesting recent
observations.  First, large-scale opacity fluctuations have been
observed in the Ly$\alpha$ forest at $z\!\!=$5--6
\citep{fan06,becker15,bosman18}.
\citet{chardin17} proposed that this could be evidence of AGN dominating the emissivity,
and found that a $\geq$50\% contribution of AGNs at these redshifts
could explain these results, similar, though in excess of, our model
results.  However, this is not a unique explanation.  For example,
\citet{daloisio15} suggest that some of the opacity fluctuations could
owe to residual temperature fluctuations imprinted by the patchy
reionization process \citep[see however][]{becker18}. \citet{2016MNRAS.460.1328D} show that galaxies
alone could generate large fluctuations in the ionizing background, if
there are strong variations in the mean free path from location to
location \citep[see also][]{2018MNRAS.473..560D}.  Secondly,
\citet{worseck16} observed the He\,{\sc ii} Ly$\alpha$ forest
at 2.3 $< z <$ 3.5, and found evidence that He\,{\sc ii} must be
significantly ionized by $z \sim$ 3.4, earlier than previous results
which suggested $z <$ 3.  As shown in Figure~\ref{fig:q}, our model
predicts Q$_{He_{III}} \sim$ 0.60 at $z\!\!=$3.4, consistent with this
result (see discussion also in \S~\ref{sec:results_he2}).
Finally, several recent spectroscopic studies of high-redshift
galaxies have detected potential N\,{\sc v} emission, an energetic
transition which cannot be produced via starlight, and is thus
indicative of AGN activity \citep{hu17,laporte17,mainali18}.  While these detections are
tenuous, impending spectroscopy of a large sample of
reionization-era galaxies with {\it JWST} should further probe this
line of evidence.

\subsubsection{Thermal history of the IGM}\label{sec:discuss_thermal}
One way to constrain these AGN-assisted models is through the thermal
history of the IGM. In general, models with a larger contribution from
AGN at earlier times produce an earlier onset of He\,{\sc ii}
reionization.  The additional energy injected into the IGM by this
process can in principle be measured by its impact on the small-scale
statistics of the H\,{\sc i} Ly$\alpha$ forest.  Unfortunately,
modeling thermal histories for a large grid of models is currently too
computationally intensive to be incorporated into our MCMC analysis.
In addition, Ly$\alpha$ forest temperature measurements are still
subject to potentially large systematic uncertainties, which may be
reflected in the level of discord amongst existing measurements (see below).
Rather than incorporate these constraints into our pipeline, we
provide some illustrative calculations comparing our models to some of
the most recent temperature measurements. 

To compute the IGM thermal history for a given model, we adopt the
approach of \citet{2017MNRAS.468.4691D}, which is based on the
multi-zone model of \citet{2016MNRAS.460.1885U}.  We refer the reader
to those papers for technical details.  In summary, we track the
temperatures of an ensemble of gas parcels at various initial
densities. These densities evolve according to the Zel'dovich
approximation, and each gas parcel is impulsively heated at a
different time to mimic the patchy reionization process. To model the
effects of the ionizing background generated by AGN, heating from
He\,{\sc ii} reionization is separated into two regimes.  In addition to
impulsive heating, which mimics the effects of He\,{\sc iii} ionization
fronts sweeping through the IGM, gas parcels are also slowly heated
by a uniform EUV/X-ray background that is built up over time with the rise of
the AGN population.  This multi-zone model was designed as an
approximation to the heating effects that are observed in radiative
transfer simulations of the reionization process.     
Indeed, \citep{garaldi19} found good agreement between our
modeling and their radiative transfer simulations of an AGN-driven
reionization.
            
\begin{figure}[!t]
\epsscale{1.15}
\plotone{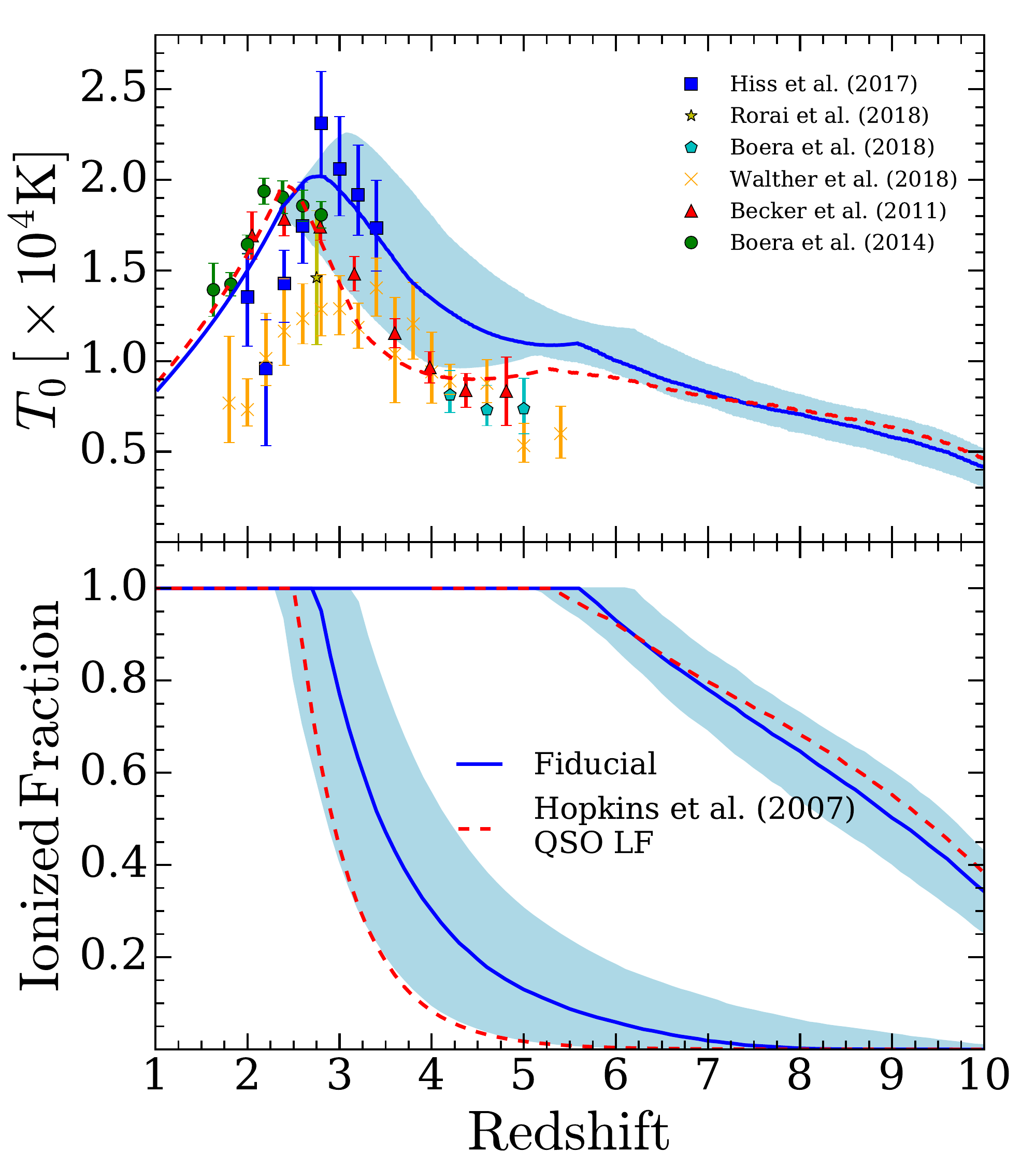}
\caption{Thermal histories of the IGM in AGN-assisted models of reionization. Top panel: gas temperature at the cosmic mean density.  The blue curve corresponds to the median emissivities from our MCMC analysis, while the shaded region corresponds to 68\% limits.  For reference, the red dashed curve corresponds to a minimal AGN model based on the QSO luminosity function of \citet[][see \S~\ref{sec:discuss_noagn} for more details]{hopkins07}.  We compare against a set of recent Ly$\alpha$ forest measurements (data points).  Bottom panel: volume-weighted mean filling factors of H {\sc ii} (right) and He {\sc iii} (left).  The thermal history contains two peaks associated with the completions of the H {\sc i} and He {\sc ii} reionization processes.  An earlier rise of ionizing emissions from AGN yields an earlier onset of He {\sc ii} reionization, which leads to earlier heating of the IGM. }
\label{fig:temp}
\end{figure}

Figure~\ref{fig:temp} shows the results of this modeling\footnote{\edit1{As noted previously, we have adopted the
clumping factor of \citet{pawlik15} for hydrogen reionization.  In our
fiducial model, we extrapolate this clumping factor to lower redshifts
and assume that $C_{\mathrm{HeIII}} = C_{\mathrm{HII}}$.   For
reference, this procedure results in $C_{\mathrm{HeIII}} =$ 5.4, 6.0,
and 6.7 at $z\!\!=$5, 4 and 3, respectively.  However, we note that
$C_{\mathrm{HeIII}}$ is uncertain, and could be considerably lower
than the values adopted here.  For example,
\citet{2012MNRAS.426.1349M} find values in the range
$C_{\mathrm{HeIII}} = 2-4$. To bracket the possibilities, we have also
explored a scenario in which $C_{\mathrm{HeIII}}=2$ for all redshifts.
In this scenario, He\,{\sc ii} reionization ends at $z\approx 3.1$.
$T_0$ follows a trajectory that is similar to our fiducial model until
$z\approx 3.1$, at which point the gas begins cooling.  Thus, in our
fiducial model, higher values of $C_{\mathrm{HeIII}}$ are necessary
for He {\sc ii} reionization to end at $z<3$.}}. The top panel shows
the evolution of the IGM temperature at the cosmic mean gas density,
$T_0$.  The blue curve and shaded region in the top panel correspond
to our median and 68 \% C.L. values of the ionizing emissivity,
respectively. The right and left blue regions in the bottom panel show
the corresponding volume-filling factors of H{\sc ii} and He{\sc
iii}.  In the top panel, we compare our calculations to
several recent Ly$\alpha$ forest temperature measurements in the
literature. We note that the curvature method employed by
\citet{2011MNRAS.410.1096B} and \citet{2014MNRAS.441.1916B} probes the
temperature at a characteristic gas density that is different from the
mean. To compare them against our $T_0$ calculations, we have
extrapolated these measurements to the mean density using the
temperature-density relation of the median model.   

The thermal history contains two peaks associated with the completion of the H{\sc i} and He{\sc ii} reionization processes. Figure~\ref{fig:temp} shows that a larger contribution of AGN to the $z>4$ ionizing background leads to an earlier onset of He{\sc ii} reionization,
which, in turn, leads to an earlier heating of the IGM.  This earlier
heating is discrepant with the lower $z\sim 5$ temperatures measured by
\citet{2011MNRAS.410.1096B}, \citet{2018arXiv180906980B}, and \citet{2018arXiv180804367W}, which imply a later and more rapid rise
of ionizing emissions from AGN compared to our models --
consistent with the conclusions of previous studies
\citep{2017MNRAS.468.4691D,mitra18,puchwien18}.  For reference, the red dashed curves show a minimal AGN model based on the QSO luminosity function of \citet[][see next sub-section for more details]{hopkins07}.  Previous studies have noted that the \citet{hopkins07} luminosity function yields temperatures that are generally consistent with the \citet{2011MNRAS.410.1096B} and \citet{2014MNRAS.441.1916B} measurements \citep{2015MNRAS.450.4081P, 2016MNRAS.460.1885U, 2017MNRAS.468.4691D}.    

The higher temperatures implied by the fiducial AGN-assisted model are in better agreement
with the recent $z<3.5$ measurements of \citet{2017arXiv171000700H}.
In principle, all Ly$\alpha$ forest temperature measurements attempt to
extract the effects of thermal broadening on the forest absorption
features. However, in practice, the discrepant sets of
measurements shown in Figure~\ref{fig:temp} utilize different
techniques for isolating those effects.  The curvature method of \citet{2011MNRAS.410.1096B} probes the shape of absorption features using the 2nd derivative of the flux, while the measurements of \citet{2017arXiv171000700H} are based on fitting Ly$\alpha$ absorption lines to extract Doppler parameters.  \edit1{On the other hand, the most recent measurements of \citet{2018arXiv180906980B} and \citet{2018arXiv180804367W} are obtained from the shape of flux power spectrum.}  The disagreement between the measurements likely owe (at least in part) to different systematics between the techniques.  Clearly, it will be important for future studies to determine the cause(s) of these discrepancies.
In addition to constraining He {\sc ii} reionization, pinning down the
thermal history at redshifts $z=2-5$ will have important implications
for the epoch of hydrogen reionization.   
\nocite{2018MNRAS.474.2871R}

\begin{figure*}[!t]
\epsscale{1.05}
\plotone{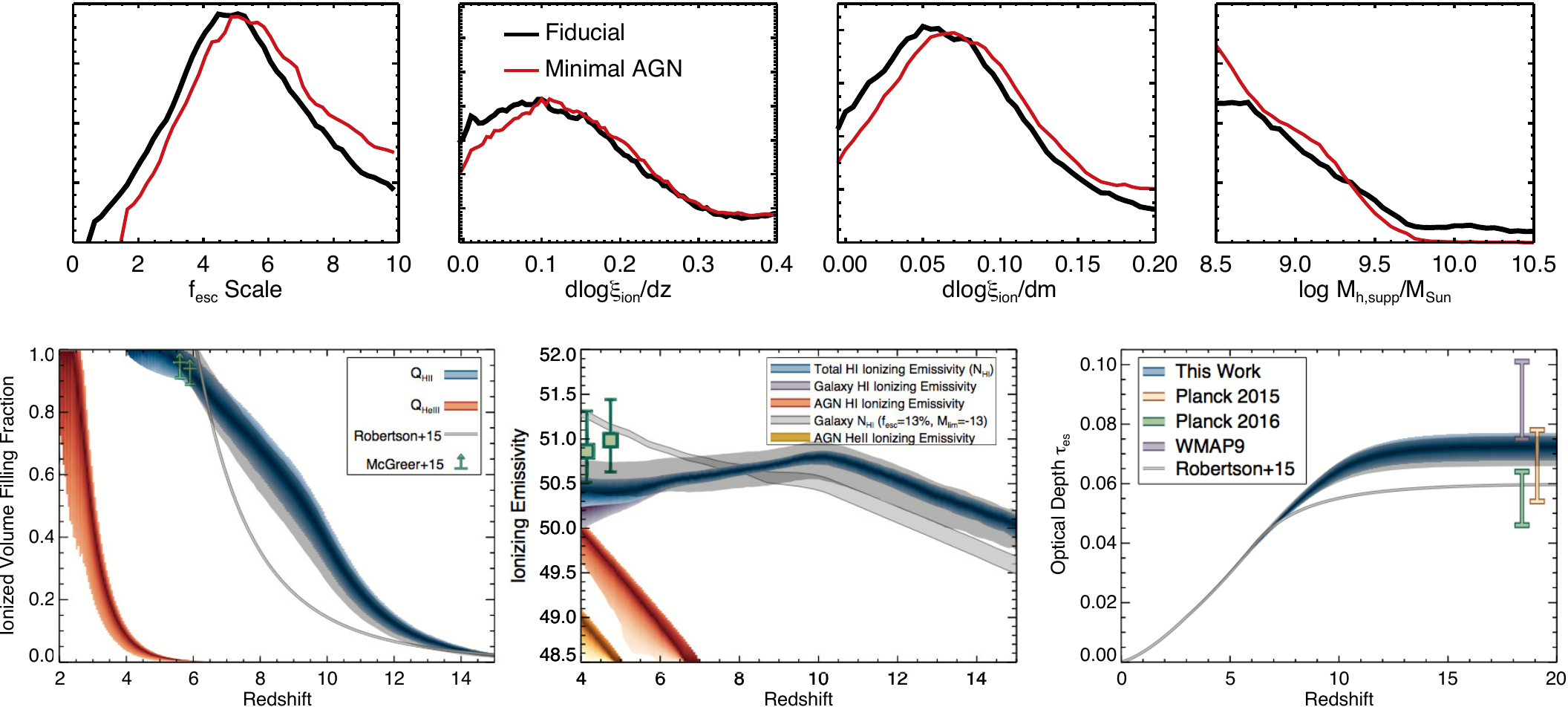}
\caption{Top) The posterior distribution for each of our non-AGN free
  parameters for our run with a minimal AGN contribution (with
  AGN$_{slope} = -$1.1), compared to that from our fiducial run.
  These parameters are moderately shifted to values which promote
  higher ionizing emissivities from galaxies.  Bottom) A comparison of
the results from this minimal AGN model to the observational
constraints, similar to Figures~\ref{fig:q} and \ref{fig:ndottau}.
The semi-transparent gray shaded regions are the results from our fiducial model.
Without a significant contribution from AGNs,
the emissivity from galaxies must be higher, which results in an
upward shift in the ionized faction at $z >$ 7, and a marginally
higher value of $\tau_{es}$.  However, at lower redshifts a
significant portion of the posterior completes reionization at $z <$
5, which is not consistent with current observations.  As observations
at $z <$ 5.5 were not used as constraints, this model is not formally
ruled out, though it does exhibit an increased tension with the
measured ionizing emissivity at $z\!\!=$4.75.}
\label{fig:noagn}
\end{figure*}

\subsubsection{A Model with a Minimal AGN Contribution}\label{sec:discuss_noagn}
With the thermal history results of the previous section in mind, it is interesting to explore how our model
changes if we fix the AGN emissivity to follow the redshift evolution
of \citet{hopkins07}.  These results are shown in
Figure~\ref{fig:noagn}.  The top panels show the distribution of the
four non-AGN related free parameters, compared to our fiducial model.
Understandably, without a significant contribution from AGNs, these
parameters are skewed (albeit only slightly) towards values which promote higher ionizing
emissivity from galaxies -- higher escape fractions, higher
$\xi_{ion}$ values, and a lower photosuppression mass.  The bottom row
compares the results of this model to our constraints, which can be
directly compared to our fiducial model in Figures~\ref{fig:q} and
\ref{fig:ndottau}, with the gray transparent shading showing
those fiducial results.  In the left-hand panel, one can see that with
a minimal AGN contribution, the reionization history is consistent
with our fiducial result, though shifted to the upper end of the
fiducial posterior.   This model requires more ionizing
photons from galaxies than the fiducial model, which results in a
shift of the free parameters to create those higher emissivities,
creating more ionizing photons at higher redshifts where
galaxies have (on average) the highest escape fractions, contributing
to the nudge to higher ionized fractions.  However, with minimal AGNs,
the slope of the reionization history exhibits a slower evolution at $z
<$ 8, resulting in $Q_{HII}=1$ at the lower redshift of $z =$
5.3$^{+0.7}_{-1.0}$.  This model has a tail extending to reionization completion down
to the low redshift of $z =$ 4.3. 
\deleted{While consistent with our
observational constraints, this type of reionization history would
further increase the tension with complementary measurements of the
IGM ionization fraction discussed in \S~\ref{sec:discuss_qcomp}.}

This very slightly earlier onset of reionization increases the electron
scattering optical depth by
$\Delta \tau =$ 0.001.  This
model increases tension with the ionizing emissivity measurements
at lower redshift, shown in the middle panel.  Though the free
parameters conspire to increase the ionizing emissivity at higher
redshifts, the flattening faint end slope towards lower redshifts results in very low escape
fractions at $z <$ 6, so galaxies cannot account for the loss of
ionizing photons from AGNs in our fiducial model at 4 $< z <$ 6,
resulting in a modest tension with the observations at $z\!\!=$4 and 4.75.  \deleted{In this minimal AGN model, the AGN emissivity does become
significant at $z <$ 4.5, so the tension with the observations at $z
=$ 4.25 is actually lower.}

\begin{figure*}[!t]
\epsscale{0.52}
\hspace{2mm}
\plotone{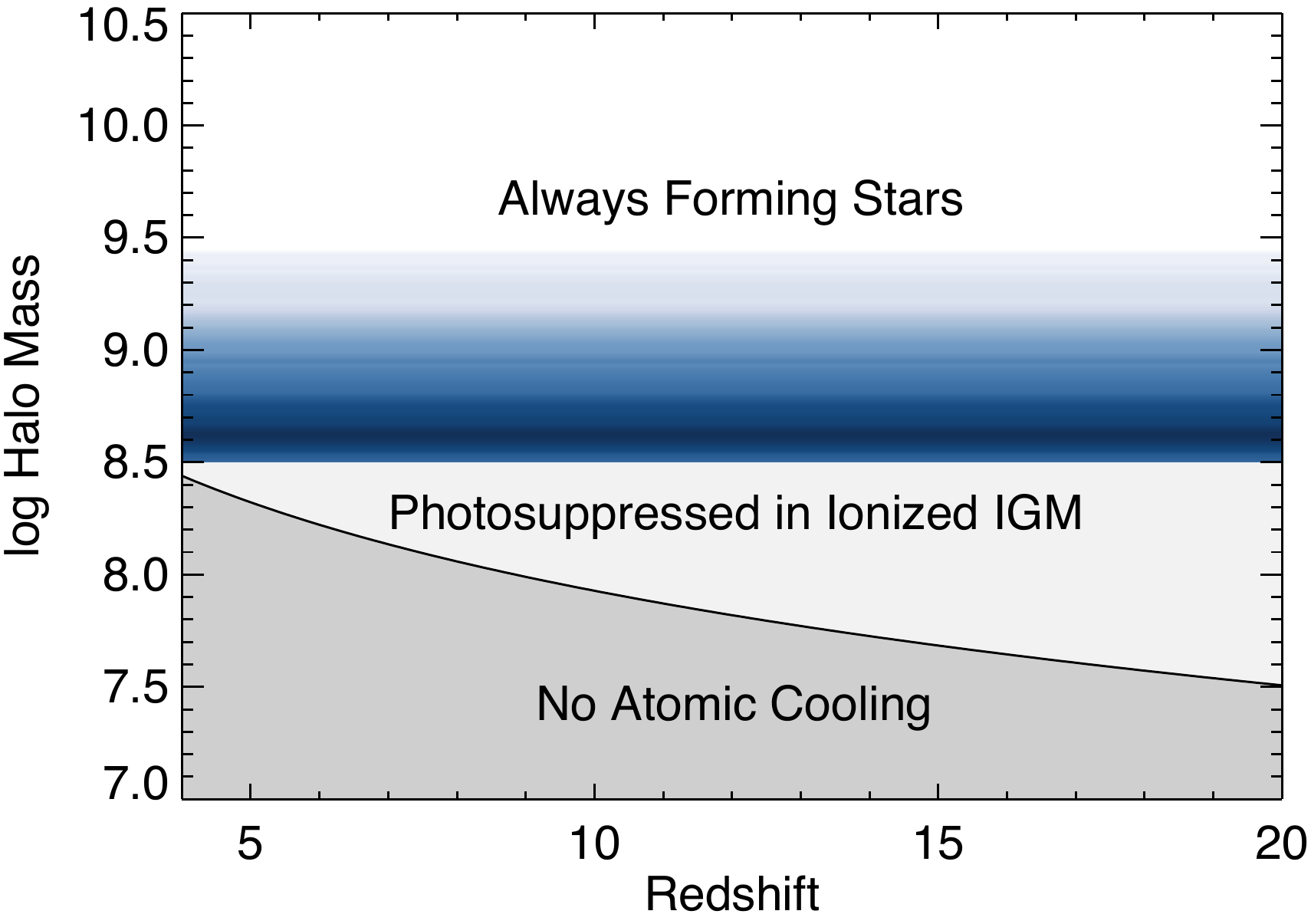}
\hspace{8mm}
\plotone{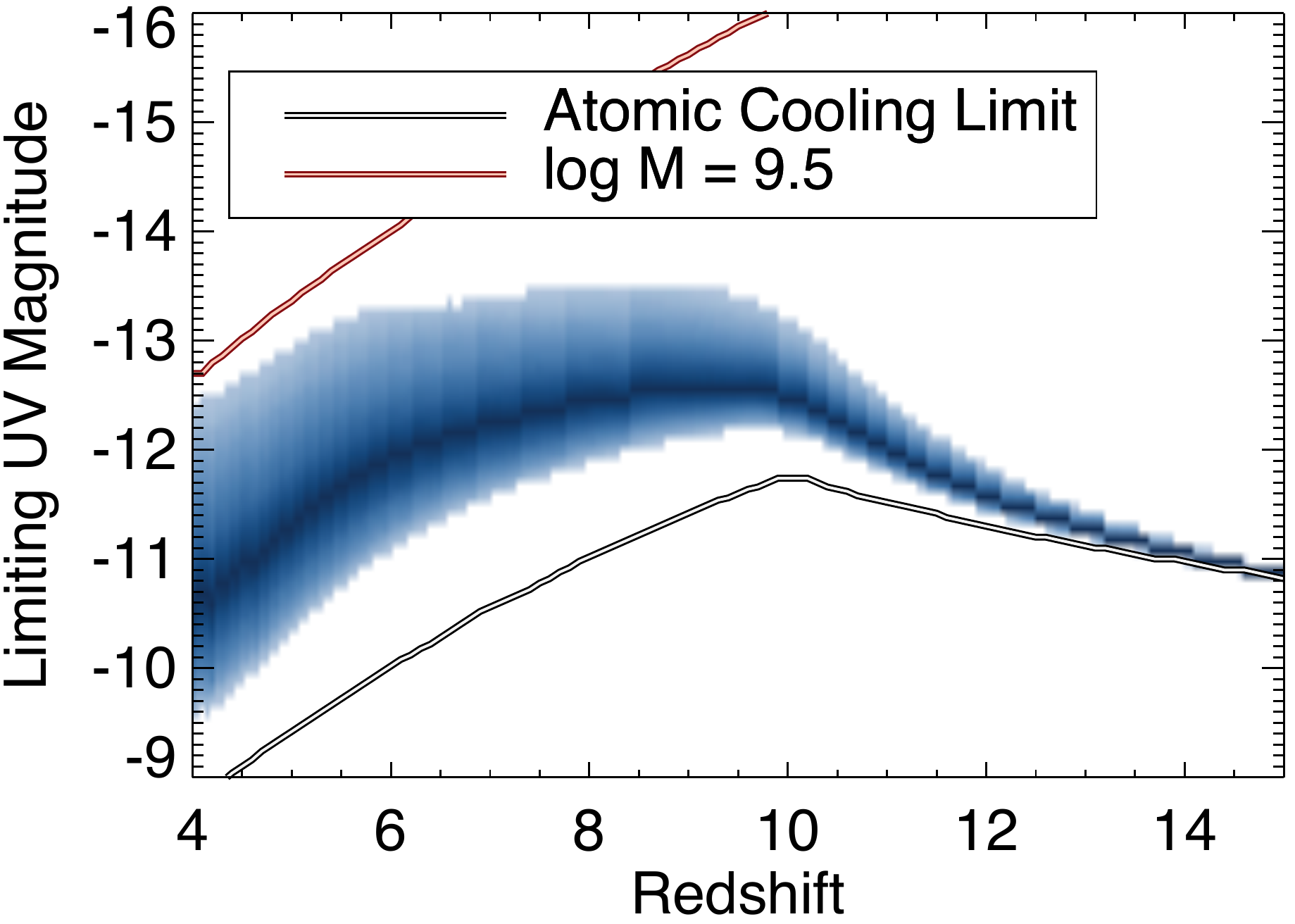}
\caption{Left) The black curve shows the halo mass corresponding to halos
  with a virial temperature of 10$^4$ K.  In a neutral IGM, halos above this curve can
  cool their gas via atomic line emission, and thus form stars efficiently.  After reionization, photosuppression will
  halt accretion onto halos with virial temperatures below the IGM
  temperature.  The shaded blue region denotes the constraints our
  model places on this photosuppression mass $M_\mathrm{h,supp}$ (darker
  denotes higher probability), which is
  close to the canonically assumed value of log ($M_\mathrm{h}$/M\sol) $=$ 9.
  Right)  The effective limiting magnitude of the UV luminosity
  function.  At very high redshifts, this corresponds to the magnitude
of halos at the atomic cooling limit (black line), while at lower redshifts this
represents the photosuppression mass (the red line denotes log
($M_\mathrm{h}$/M\sol) $=$ 9.5, the 84\% 1-sided upper limit on
this mass parameter).  In between, this shape
represents a growing fraction of halos residing in ionized regions,
thus subject to photosuppression.  At all redshifts,
this effective limiting magnitude is always fainter than the commonly
used value of $M_\mathrm{UV} = -$13.}
\label{fig:mhalo}
\end{figure*}  

Given this slight increased tension, it is instructive to compare the
goodness-of-fit from this model to our fiducial model.
\edit1{We find DIC$=$6.4 for the minimal AGN model, compared to 5.3 for our
fiducial model.}  We conclude that altering our model to
have a minimal level of AGN
contribution to reionization is not ruled out, though it does increase
the tension with observations slightly along multiple axes.  

Specifically, the portion of the posterior which has reionization
finishing at $z <$ 5 is robustly ruled out via observations
\citep[e.g.,][]{becker18}.  Since these observations were not used as
constraints on our model, this tension does not have high statistical significance (with
$\Delta$ DIC only 1.1),
predominantly due to the large observational uncertainties.  \deleted{We do
note that the ability of this model to complete reionization
``on time'' is partially be driven by our imposed prior that reionization
completes by $z\!\!=$5. We explore the effects of dropping this prior in
\S~\ref{sec:discuss_prior}.}

\subsection{Limiting Halo Mass and Magnitude}\label{sec:discuss_limmag}

In \S~\ref{sec:mcmc1b} we discussed the physical reasons behind the
need for a limiting halo mass for star formation, and thus a limiting
magnitude for the UV luminosity function.  At redshifts prior to
reionization, we allow stars to form in halos down to the
(redshift-dependent) atomic
cooling limit, while post-reionization halos are subject to Jeans
filtering at a mass below some level, which we fit as the free
parameter $M_\mathrm{h,supp}$.  The left-hand panel of Figure~\ref{fig:mhalo}
visualizes these masses, highlighting which mass regimes are allowed
to form stars at which redshifts.  Our model prefers a very low value
of the filtering mass, with a one-sided distribution peaking at the
prior minimum value of log ($M_\mathrm{h,supp}$/M\sol) $=$ 8.5, with an
1$\sigma$ (84\%) upper limit of log ($M_\mathrm{h,supp}$/M\sol) $<$ 9.5.  This
is understandable, as it is the lowest-mass halos which have the
highest escape fractions, thus maximizing star-formation in these
low-mass halos maximizes the ionizing emissivity.  This is in the
range of constraints from previous simulations, but is also consistent
with a recent analysis which shows that present-day dwarfs were
subject to the effects of reionization if their halo masses are at log
($M_\mathrm{h}$/M\sol) $<$ 8.5 \citep{tollerud18}.

The right-hand panel shows the effective UV luminosity function
limiting magnitude as a function of redshift.  The rise at very high
redshifts with negligible spread represents the evolution of the
atomic cooling limit.  As the ionized volume fraction grows, more
halos become subject to the filtering mass ($M_\mathrm{h,supp}$)
threshold.  In this
figure we approximate the typical limiting magnitude as the magnitude
corresponding to the atomic cooling limit plus the difference in that
magnitude and the magnitude corresponding to the filtering mass, with
that difference multiplied by the ionized volume filling fraction at a
given redshift, such that when $Q_{H_{II}}\!\!=$1, this is just the
magnitude corresponding to the filtering mass.  The peak in this
distribution at $z \approx$ 8--10 corresponds to transitioning from
the atomic cooling limit to this photosuppression/Jeans filtering limit.

Examining this effective limiting UV magnitude, one can see it is
essentially always fainter than $-$13, a value commonly assumed in
other studies.  Our
model's luminosity functions thus extend a redshift-dependent 1--2 magnitudes fainter than
other studies \citep[e.g.][]{finkelstein12b,robertson15,bouwens15b},
though consistent with values seen in recent simulations \citep[e.g.,][]{gnedin16,rosdahl18,yung18}.
This represents only a small increase in the non-ionizing UV
luminosity density, but as shown in Figure~\ref{fig:ncum}, a sizable
increase in the ionizing emissivity.  
The dependance of our model on these very low-mass halos
also strongly disfavors warm dark matter models with masses $\lesssim$ 2
keV \citep[e.g.,][]{menci16,dayal17}.

There are several caveats to
this result.  First, our model applies a sharp cutoff at these
values, while in reality, the luminosity function likely has a more
gentle turnover.  In fact, based on the star-formation histories in
local dwarf galaxies, it must exhibit a relatively shallow decline to
very low luminosities ($M_\mathrm{UV} = -$ 5, \citealt{weisz14}; see
also \citealt{graus18}).  This is
understandable as while the UV feedback from reionization may halt gas
accretion onto these low-mass halos, these galaxies will still form
stars for a time with the gas they have, slowly fading rather than
immediately quenching.  Second, our
model relies on abundance matching being accurate at these very small
masses, which has not yet been observationally verified.  Further
knowledge on the very faint-end of the UV luminosity function is
forthcoming, both observationally with lensing studies with {\it
  JWST}, but also theoretically, as the next generation of simulations
improves the precision of tracking both the small and large-scale
effects of reionization \citep[e.g.,][]{rosdahl18}.

\begin{figure}[!t]
\epsscale{1.2}
\vspace{4mm}
\plotone{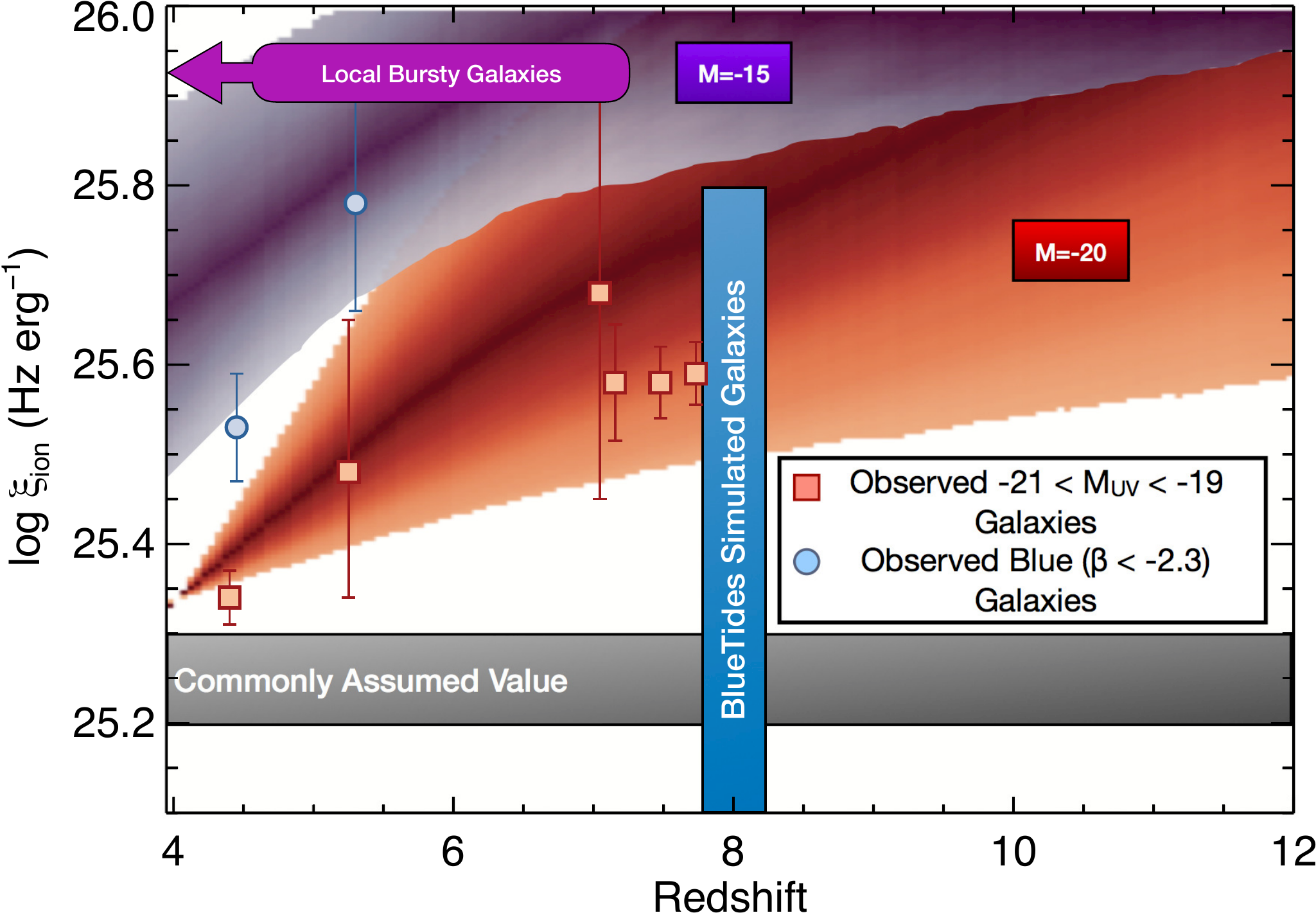}
\caption{The evolution of $\xi_{ion}$ from our fiducial model is shown
by the shaded regions, in red for $M_\mathrm{UV}=-$20, and in purple for
$M_\mathrm{UV}=-$15.  The squares show measurements of $\xi_{ion}$ for
bright galaxies ($-$21 $<$ $M_\mathrm{UV} < -$19) from \citealt{bouwens16},
\citet{stark15b} and \citet{stark17}.  The blue circles show the
inferred $\xi_{ion}$ for blue ($\beta < -$2.3) galaxies from
\citet{bouwens16}.  The blue bar shows the range of $\xi_{ion}$ values
predicted for simulated galaxies in the BlueTides simulations at $z =$
8.  The violet bubble indicates the typical $\xi_{ion}$ values for
low-redshift compact star-forming galaxies from \citet{izotov17} if
they have bursty star-formation histories, similar to that expected
for high-redshift dwarfs.  While these observations shown did not
constrain our model, they are in qualitative agreement that galaxies
at higher redshift exhibit higher ionizing photon production
efficiencies, and that values of log $\xi_{ion}$ approaching 26, as we
find for our faintest galaxies, are potentially expected for blue,
bursty galaxies.}
\label{fig:xiion}
\end{figure}  

\subsection{Evolution of $\xi_{ion}$}\label{sec:discuss_xiion}
Our model allows evolution of $\xi_{ion}$ to higher values to both
higher redshifts and fainter magnitudes (from the baseline value of
log $\xi_{ion}\!\!=$25.34 erg$^{-1}$ Hz for $M_\mathrm{UV} = -$20 galaxies at
$z\!\!=$4; see \S~\ref{sec:mcmc1c}).  Our fiducial results for
$\xi_{ion}(z,M_{UV})$ shown in
Figure~\ref{fig:xiion} prefer
evolution in $\xi_{ion}$ with both redshift and magnitude, consistent
with recent results at 4 $< z <$ 7 for modestly bright galaxies ($-21
<$ $M_\mathrm{UV} < -$19) discussed in \S~\ref{sec:mcmc1c}
\citep[e.g][]{bouwens16,stark15b,stark17}.  
Our model thus predicts a larger number of 
ionizing photons produced than previous studies which assumed log $\xi_{ion} \approx$
25.2 \citep[e.g.][]{finkelstein12b,robertson15}.

One concern with using some of these observational results to validate our
model is that the individual detected galaxies are hand-picked for
spectroscopic followup, and therefore are typically unusually bright,
and may not be representative of the general galaxy population at such
redshifts.  It becomes interesting to examine then lower-redshift analogs for
high-redshift galaxies, where less-biased spectroscopic studies can be
performed.  \citet{tang18} model the photometric and spectroscopic
measurements from galaxies at $z\!\!=$1.4--2.4 selected to have strong
[O\,{\sc iii}] emission lines, similar in strength to those observed
in many high-redshift galaxies \citep[e.g.][]{smit15}.  They find that
$\xi_{ion}$ correlates tightly and positively with [O\,{\sc iii}] EW,
such that galaxies with EW $>$ 600 \AA\ have log $\xi_{ion}$ $\sim$
25.5-25.8, similar to $z \sim$ 7 galaxies with comparable inferred
[O\,{\sc iii}] EWs \citep{stark17}.  \citet{shivaei18} examined this
quantity at $z \sim$ 2, and while they found that typical galaxies in
their sample had log $\xi_{ion}\!\!=$25.34 (assuming a SMC dust curve), they
found the bluest galaxies in their sample had $\xi_{ion}$ up to twice
as high.  This was also seen by \citet{bouwens16} at $z\!\!=$4--5, where
the bluest galaxies in their sample had log $\xi_{ion} \sim$ 25.5 at $z =$
3.8--5.0, and log $\xi_{ion} \sim$ 25.8 at $z\!\!=$5.1--5.4.
\citet{nakajima18} inferred $\xi_{ion}$ for a sample of $z
\sim$ 3 LAEs via full nebular modeling of observations of several
rest-frame UV lines.  They found that not only do LAEs have higher
values of $\xi_{ion}$ than continuum-selected galaxies, but that
fainter LAEs preferentially had higher values, with $M_\mathrm{UV} =
-$19 LAEs having log $\xi_{ion} \sim$ 25.7, compared to 25.5 for $M_\mathrm{UV} =
-$20.5 LAEs.

At very low redshifts, while \citet{schaerer16} found log $\xi_{ion}$ $\sim$
25.1-25.5 for a sample of five Lyman continuum leaking galaxies at $z
\sim$ 0.3, \citet{izotov17} found that this quantity is heavily
dependent on the assumed star-formation history.  Exploring a sample
of 14,000 low-redshift compact star-forming galaxies, they found that
the galaxies with high H$\beta$ EWs ($>$50\AA) are consistent with
having log $\xi_{ion}\!\!=$25.5--26.0, with the caveat that these high
values may only be present during a starbursting phase, and thus the
time-averaged value for a given galaxy may be lower.  However, dwarf
galaxies at high redshift likely have bursty star-formation histories
\citep[e.g.][]{jaacks12b,guo16,faucher18}, thus these values may be
representative of small galaxies at early times.  While our model does predict high values for $\xi_{ion}$
for the faint galaxies which end up dominating the ionizing emissivity (Figure~\ref{fig:ncum}), 
these observations suggest that this is consistent with their likely blue \citep[e.g.][]{finkelstein12a,bouwens14} and bursty nature.

\begin{figure*}[!t]
\epsscale{1.1}
\plotone{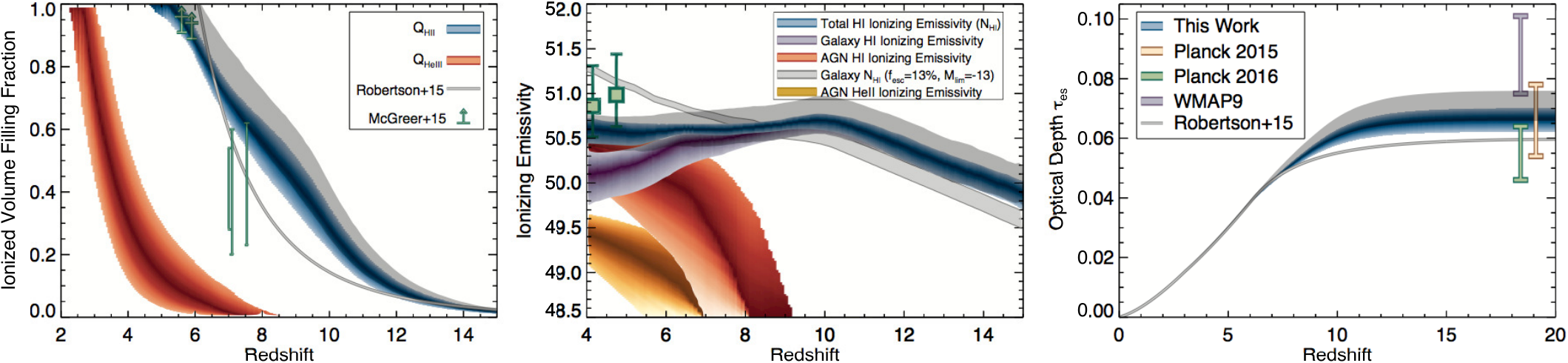}
\caption{The results of our model when the Ly$\alpha$ constraints from
\citet{mason18} as well as the constraints measured from the two $z >$
7 quasars from \citet{bolton11}, \citet{greig17b}, and
\citet{banados18} are included.  These constraints pull down the
reionization history by starting reionization slightly more
slowly, lowering the neutral fraction by $\Delta$Q$=$0.09 at $z\!\!=$7
to $Q_{H_{II}}\!\!=$0.69 ($\pm$ 0.06), and finishing slightly later, at
$z =$ 5.3$_{-0.4}^{+0.3}$.  This model does not produce $Q_{H_{II}} =$0.5 at $z\!\!=$7, thus it
will need to be revised if future observations confirm this to be the case.  Our fiducial model is shown
by the transparent gray shading for comparison,}
\label{fig:lyaqso}
\end{figure*}   

There is thus evidence at both high and low redshifts that the values
of $\xi_{ion}$ predicted by our model are not unreasonable, and
several physical effects point to them at least being
plausible.  First, galaxies at higher redshifts should have lower
stellar metallicities, and specifically lower iron opacities in their
atmospheres, than at lower redshift.  This will reduce absorption in
the outer atmospheres of stars, increasing their effective surface
temperature, and thus increasing $\xi_{ion}$.  Secondly, the
oft-ignored effects of binary stars can also increase
stellar surface temperatures and thus $\xi_{ion}$
\citep[e.g.,][]{eldridge09,wilkins16}.  Lastly, high levels of stellar
rotation ($v/v_{crit} >$ 0.4) can result in longer-lived massive stars
due to increased mixing, and thus also increase $\xi_{ion}$ \citep{choi17}.
Significant future observational efforts are required to fully explore
the redshift and luminosity dependance of this crucial quantity.

\subsection{Including Ly$\alpha$ and QSO Constraints}\label{sec:discuss_lyaqso}
The focus of this paper has been on the reionization history if
the escape fraction is significantly halo-mass dependent, which is
more extended than previous works.  This is in
slight (1--2$\sigma$) tension with the most recent Ly$\alpha$ and $z
>$ 7 QSO-based measurements, which imply $Q_{H_{II}} \sim$ 0.4--0.6 at $z
\sim$ 7.  While these observational measurements are very model dependant (see \S~\ref{sec:discuss_qcomp}), it is interesting to explore
whether our model can be made to accomodate such a significant shift
in the neutral fraction from $z\!\!=$6 to 7.

To this end, we have re-run our fiducial model, adding as constraints
the Ly$\alpha$-based measurements from \citet{mason18} at $z\!\!=$7% ,
% \citet{ouchi17} and \citet{konno17} at $z \sim$ 6
, and the QSO-based
constraints from \citet{bolton11} and \citet{greig17b} at $z\!\!=$7.1,
and \citet{banados18} at $z\!\!=$7.5, with the results shown in
Figure~\ref{fig:lyaqso}.  Compared to our fiducial model, the ionized
fraction is $\sim$10\% lower at $z\!\!=$7--10.  This is accomplished
via a tighter, and modestly lower, emissivity from galaxies at $z >$
10, with a resultant slightly lower value of $\tau_{es}$ (0.067 $\pm$
0.04), though with a similar contribution from AGNs.  While these
additional observations have large error bars, if future work confirms
that the neutral fraction is close to 50\% at $z \sim$ 7, our model
will need to be modified to allow for a slower start, and a more rapid
completion to reionization.  This scenario could imply that the basic
assumption here, that the escape fraction is dependent on halo mass,
may not hold, or that there are additional redshift-dependent effects
not considered here.

\begin{figure*}[!t]
\epsscale{0.9}
\plotone{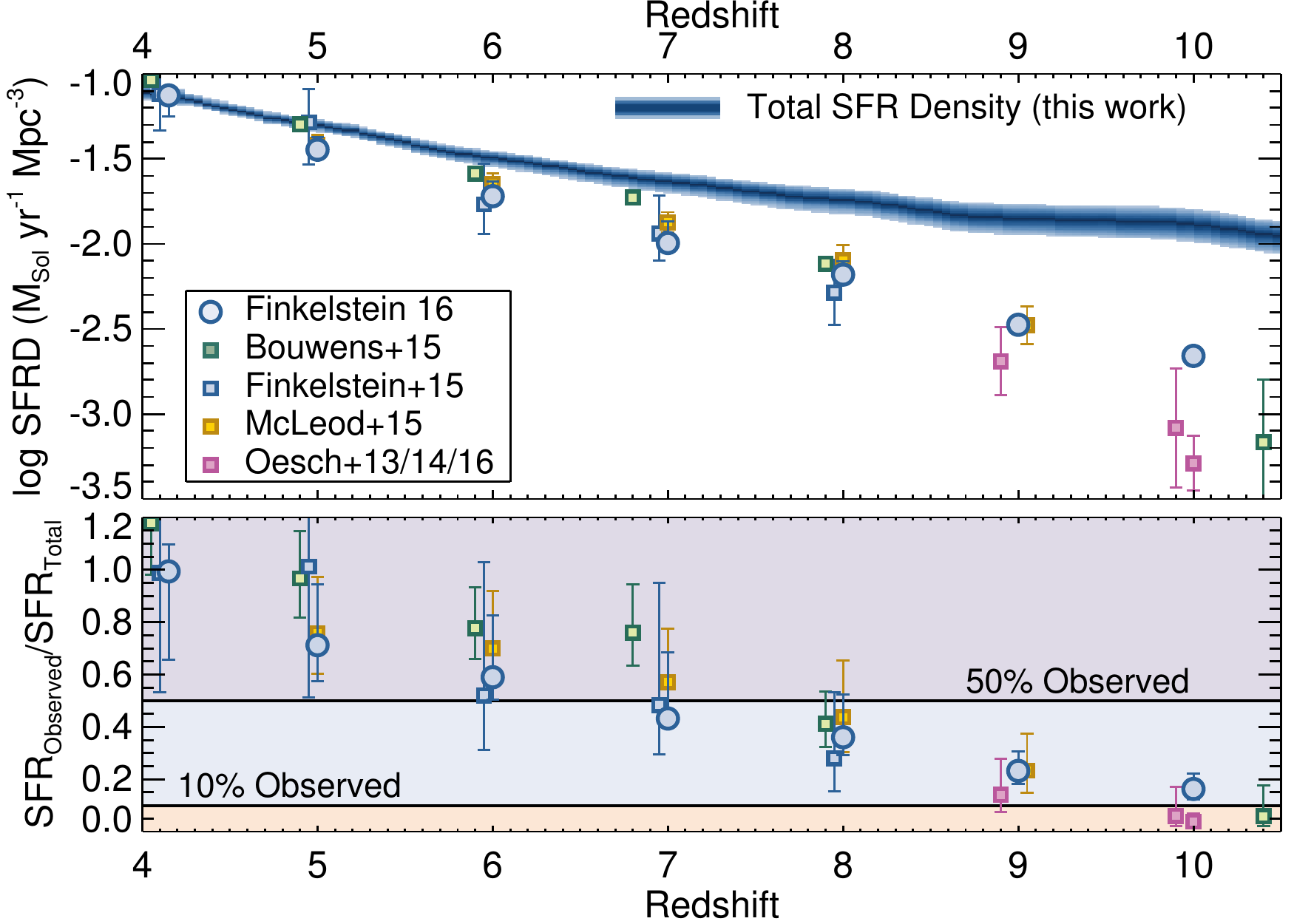}
\vspace{-2mm}
\caption{Top) The evolution of the total cosmic SFR density as a
  function of redshift from this work (in blue), compared to results
  from the literature for the \emph{observable} SFR density, defined
  as including only galaxies with M $< -$17.  All values are corrected
  for dust attenuation, and corrected to use the same conversion
  between UV luminosity and SFR.  Bottom) The ratio of the
  observable SFR density to the total SFR density estimated from this
  work.  While at $z=$4 much of the star-formation is observable, due
  to the relatively shallow faint-end slope of the UV
  luminosity function, at $z\!\!=$7 we can presently see only 50\%, and
  only $\sim$10\% of the total derived SFR density is presently
  visible at $z\!\!=$10.
}
\label{fig:sfrd}
\end{figure*}

\section{Cosmic SFR Density}
The derivation of physically motivated values for the limiting
magnitude of the UV luminosity function which evolve with redshift
affords the opportunity to take a fresh look at the cosmic SFR
density.  Our derived SFR density is shown in the top panel of Figure~\ref{fig:sfrd},
with the 68\% confidence range from this work shown as the shaded
blue region.  This was derived by integrating the UV luminosity
function at each redshift down to the limiting magnitude for that
redshift, correcting for typical dust attenuation (as described in
\S~\ref{sec:mcmc2a}).  We used the conversion factor between specific UV
luminosity
and SFR of: SFR = 1.15 $\times$ 10$^{-28}$ L$_{\nu}$ M\sol\ yr$^{-1}$ \citep{madau14}.
We compare this SFR density to the observations of \citet{oesch13},
\citet{oesch14}, \citet{finkelstein15}, \citet{bouwens15},
\citet{mcleod15}, as well as the literature compilation ``reference''
values from F16.  All of these observed values
represent the ``observable'' SFR density, defined as that obtained by
integrating the UV luminosity function to a common absolute magnitude
of M $= -$17.  As seen in this figure, while at $z\!\!=$4 the
observable values are close to our derived total value of the SFR
density, they appear to fall short at higher redshift due to the
steepening faint-end slope.

We make this
more clear in the lower panel of
Figure~\ref{fig:sfrd}, where we show the ratio between these
observable SFR density values and our derived total SFR density.  It
can be seen that present-day observations (exclusive of
gravitationally lensed sources) probe only 50\% of our estimated total
SFR density at $z\!\!=$7, and only a paltry 10\% at $z\!\!=$10.  This is
easily understood as a consequence of the evolving faint-end slope of
the luminosity function.  While at $z\!\!=$4 the relatively shallow
slope results in galaxies below the detection limit contributing very
little to the total SFR density, this is dramatically different at $z
=$ 10, where our assumed faint-end slope is $\Delta\alpha=-$0.66
steeper, resulting in the observations seeing only the very tip of the
iceberg.  This has significant consequences on the reionization
history derived from this population.  As our assumed model results in
large escape fractions coming from primarily only the smallest
galaxies, the relative paucity of these small galaxies at $z\!\!=$4
compared to $z\!\!=$10 implies that the ionizing emissivity escaping
galaxies will become nearly negligible by $z\!\!=$4, and this is exactly
what we find in Figure~\ref{fig:ndottau}, where the galaxy emissivity
at $z\!\!=$10 is nearly 1 dex higher than at $z\!\!=$4.  Clearly these
results are heavily dependent on our assumed luminosity function
evolution, which is not presently tightly constrained at $z \geq$ 8.

\begin{figure*}[!t]
\epsscale{0.9}
\plotone{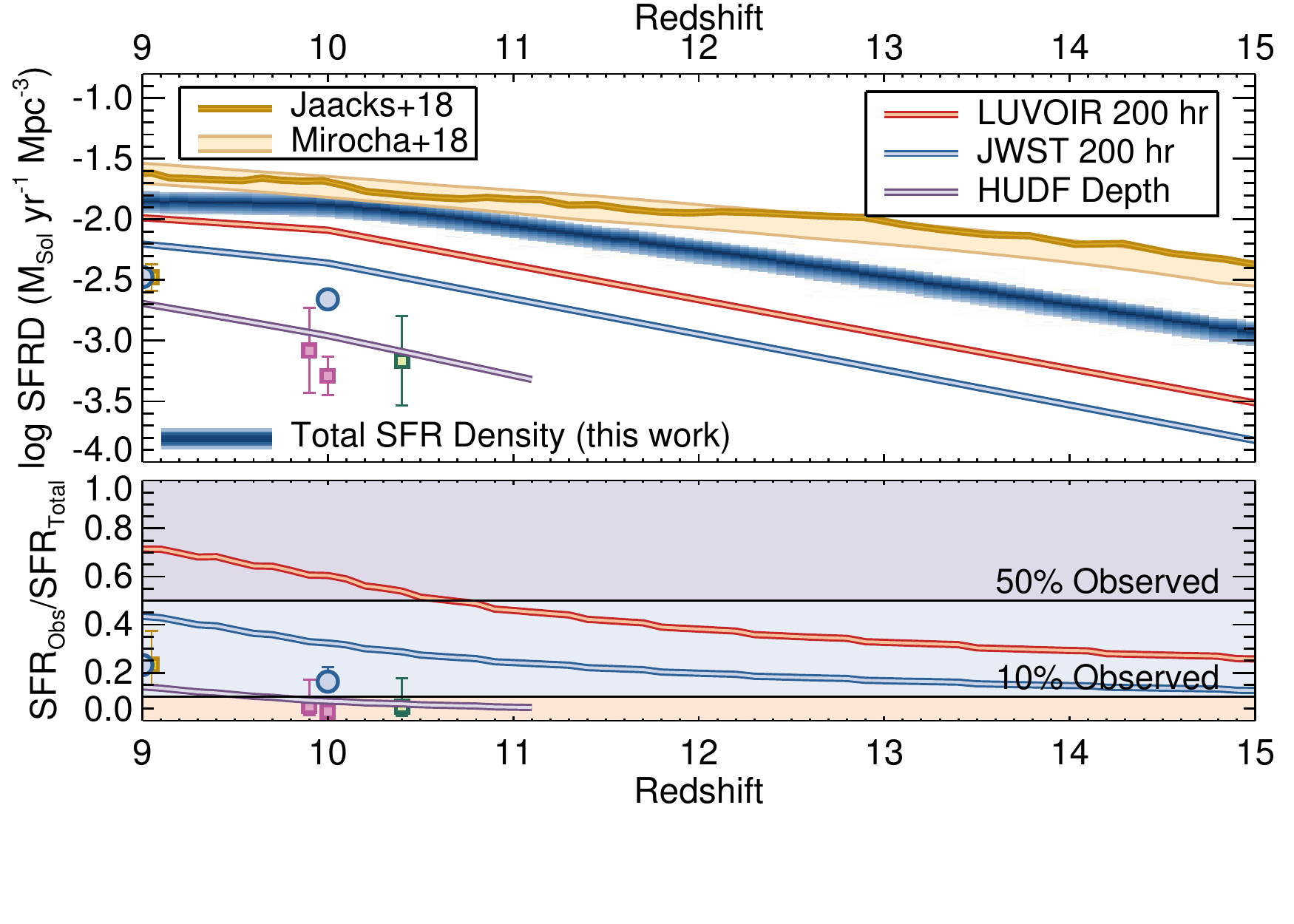}
\vspace{-14mm}
\caption{Top) The extension of our predicted SFR density to higher
  redshifts.  The total SFR density from the fiducial model from this
  work is in blue, and observations (to $M_\mathrm{UV}=-$17) are the same as
  in  Figure~\ref{fig:sfrd}.  We show the maximum limiting magnitude
  for compact sources  achieved by {\it HST} (29.5, in the HUDF), as
  well as hypothetical   200 hr integrations with {\it JWST} (32.0)
  and {\it LUVOIR} (33.5).  The bottom panel shows the ratio of these
  observable SFR densities to our predicted total value, highlighting
  that {\it JWST} will be sensitive to $\sim$10\% of the total
  SFR density out to $z \sim$ 15, and a 15.1m \emph{LUVOIR} would be
  sensitive to the majority of the star-formation activity at $z \sim$ 10.}
\label{fig:sfrdpredictions}
\end{figure*}   

\subsection{Outlook to Higher Redshifts}
In Figure~\ref{fig:sfrdpredictions} we show the extension of our
predicted SFR density to higher, mostly unexplored, redshifts.  Our
SFR density, even from our fiducial model where the faint-end slope
stops evolving at $z >$ 10, stays somewhat high, dropping only $\sim$1
dex from $z\sim$ 10 to 15.  Sophisticated model predictions for this
era are few, but we compare to two recent results which predict
similarly high SFR densities.  The first is the SPH simulation from
\citet{jaacks18c}, who ran a meso-scale simulation in a 4 Mpc box to
resolve Pop III star-formation in minihalos, and the subsequent
transition to Pop II star formation.  This model predicts SFR
densities $\sim$1 dex higher than the previous generation of simulations
\citep[e.g.,][]{johnson13,pallottini14,feng16}.  There are a variety
of reasons for these differences, but the main one stems from the
in-situ formation of Pop III stars in the \citet{jaacks18c} model,
which starts at $z =$ 26, with Pop II stars forming essentially
immediately thereafter.  This earlier start to cosmic chemical
enrichment allows the Pop II SFR density to climb more rapidly than
other models.  A similar SFR density is found by \citet{mirocha18}.
They use a semi-analytic model to explore scenarios where the observed
galaxy populations, extrapolated to fainter luminosities, can explain
the EDGES observations of 21 cm absorption at $z \sim$ 18
\citep{bowman18}.  They find that either star-formation must occur
beyond the atomic cooling limit, or that the luminosity function must
steepen at very faint luminosities -- both scenarios result in
somewhat high SFR densities at $z =$ 10--15.

The thin colored lines in these figures denote the plausible maximum
depths reached by {\it HST}, {\it JWST}, and a hypothetical 15.1m
\emph{Large Ultraviolet/Optical/Infrared} (\emph{LUVOIR}) Telescope\footnote{https://asd.gsfc.nasa.gov/luvoir/}
\citep{bolcar17}.  For {\it HST}
we assuming a limiting apparent magnitude of 29.5, the maximum depth
reached in the HUDF in the $H$-band; this curve stops at $z =$ 11.1
as that is the maximum redshift probed by {\it HST} \citep{oesch16}.
For {\it JWST}, we assume a limiting apparent magnitude of 32.  This
is achievable in $\sim$200 hr of integration per NIRCam filter,
assuming that the sizes of faint objects follow recent results of a
steepening of the size-luminosity relation \citep[e.g.,][]{shibuya16},
giving half-light radii of 50-100 pc for galaxies with $M_{UV} > -$17.
A 15m \emph{LUVOIR} would reach a 5$\sigma$ detection limit in the
$H$-band of $m_{AB}=$33.5 mag
in a total integration time of 200 hours assuming a source
radius of 0.012\arcs\ (50 pc at $z\!\!=$10; M. Postman, priv. communication).

In the bottom panel, we show the ratio of the observable SFR density
with these current and future facilities to the total SFR density from
this work.  As mentioned above, {\it Hubble} is only sensitive to
$\leq$10\% of the star-formation activity in this epoch.  The advent
of {\it JWST} will more than double this at $z \sim$ 10 to $\sim$30\%,
with {\it JWST} still being sensitive to 10\% of the total SFR density
at $z \sim$ 15.  If \emph{LUVOIR} becomes a reality, it will be able to
directly observe \emph{more than half} of the total star-formation
activity at $z \sim$ 10, and up to 25\% at $z \sim$ 15, doubling that
of {\it JWST}.  However, the baseline plan for LUVIOR is for a
passively cooled telescope, which would not be sensitive to $\lambda
>$ 2$\mu$m, leaving \emph{LUVOIR} sensitive to a similar redshift range as
{\it HST}.  Should early {\it JWST} observations indicate that high
SFR densities such as those predicted by our model are likely to be
true, then it should open the door for a discussion about whether to
extend the wavelength range of \emph{LUVOIR} to be sensitive to star
formation at much greater redshifts.

\section{Conclusions}

We have presented a semi-empirical model of reionization to explore scenarios
for completing hydrogren reionization within existent (mostly)
model-independant constraints with low average ionizing photon escape
fractions.  We have developed a MCMC algorithm to constrain the
posterior distribution of seven free parameters which, when combined
with observations of the rest-UV luminosity function, fully describe 
the ionizing source populations. The two primary successes of this
model are: 1) This model successfully completes reionization by $z
\sim$ 6, matching all utilized observational constraints with physically motivated halo-mass dependent escape
fractions.  This results in a globally averaged escape fraction $<$5\% at
all redshifts $z <$ 10 consistent with the bevy of non-detections of
ionizing photon escape in the literature.  2) Our escape fraction
parametrization naturally results in a rising emissivity with
increasing redshift throughout the epoch of reionization.
This is consistent with the boundary conditions of an observed rising
emissivity from $z\!\!=$2--5, in contrast to models with a fixed large
escape fraction, which can violate the emissivity constraints.
However, our reionization history starts
early, and implies an ionized fraction of $\sim$80\% at $z \sim$ 7, in
mild tension with inferences from Ly$\alpha$ detectability studies and the
damping wing measurements from the two known $z >$ 7 quasars.  

Our fiducial model successfully completes reionization with several
primary differences from previous work:
\begin{itemize}[leftmargin=*]

\item We tie the limiting magnitude of the UV luminosity function to
  a physically motivated limiting halo mass for star formation,
  constrained to equal the atomic cooling limit in neutral regions,
  and a free-parameter, dubbed the photosuppression mass, in ionized
  regions.  Our model prefers a photosuppression mass log ($M_\mathrm{h,supp}$/$M$\sol)
  $<$ 9.5, maximizing the amount of ionizing photons produced by
  galaxies.  This leads to limiting UV magnitudes essentially always
  fainter than the canonically assumed value of $-$13, as low as $-$11
  at some redshifts.  This results in more
  star-formation than assumed by previous reionization models, though
  consistent with recent high-resolution simulations of star-formation
  in low-mass halos \citep[e.g.,][]{paardekooper13,wise14,jaacks18c}.

\item Our model prefers evolution in the ionizing photon production
  efficiency to higher values at both higher redshifts and
  lower-luminosities.  Thus not only does our model make more
  non-ionizing UV emission, it has a larger conversion between
  non-ionizing and ionizing UV light, further increasing the intrinsic
  ionizing emissivity, compensating for the lower ionizing photon
  escape fractions.

\item Our model prefers a modest AGN contribution to the end of
  reionization, with AGNs contributing roughly one-third of the total
  ionizing photon budget at $z\!\!=$6.  AGNs do not dominate until $z
  \lesssim$ 4.6, and though this contribution is larger than many
  previous studies, it is still consistent with the limited observations of He\,{\sc ii}
reionization that are currently available, for which the
uncertainties are large.

\item Our luminosity function parametrization combined with our
  evolving limiting magnitude predicts a total SFR density which is
  very flat at $z >$ 8.  However, this is not inconsistent with the
  small number of $z >$ 9 galaxies presently known, as we find that
  only $\sim$10-20\% of the total star-formation at $z >$ 9 should be
  detectable to {\it Hubble} Ultra Deep Field depths.  Extrapolation
  of this model to the as-yet-unexplored epoch of $z >$ 10 predicts
  ample star-formation activity, with {\it JWST} and a 15.1m {\it
    LUVOIR} sensitive to $\sim$20\% and $\sim$40\% of the total
  star-formation activity at $z \sim$ 12, respectively.
\end{itemize}

\edit1{We reiterate that our model is reliant on a number of assumptions,
which will continue to be tested empirically and theoretically,
allowing the model to be improved in the future.  The most important
of these assumptions are: \emph{i}) that bright galaxies do not
significantly contribute to the ionizing emissivity, due to our
halo-mass dependent parameterization of $f_{esc}$; \emph{ii}) that
$\xi_{ion}$ varies with both redshift and luminosity, and does not
exceed a maximum value of 26.0 Hz erg$^{-1}$; \emph{iii})
star-formation in halos below the atomic cooling limit is an
insignificant contributor to the ionizing photon budget; \emph{iv}) the
faint-end slope of the galaxy UV luminosity function does not evolve
significantly at $z >$ 10; and \emph{v}) various quantities, including the
galaxy and AGN luminosity functions, evolve smoothly with redshift.}

While this model successfully completes reionization with low galaxy
escape fractions, it requires a number of physical scenarios which
have not yet been directly observationally tested.  However, this
should change in the coming years.  Deep {\it JWST} gravitational
lensing surveys should reach $\sim$2 magnitudes fainter than the
Hubble Frontier Fields, pushing robust lensing results to $M_\mathrm{UV} =
-$13, and potentially fainter, testing our predicted limiting
magnitudes.  Wide-area surveys are making progress on improving our
constraints on the faint-end slope of the AGN luminosity function, and
this will soon be directly testable with {\it JWST} spectroscopy.
Finally, this same spectroscopy will allow measurements of the
physical conditions in ionizing regions throughout the epoch of
reionization, providing empirical measures of the ionizing photon
production efficiencies.  Future models of reionization will thus have
more significant constraints on the ionizing emissivity from the
galaxy population, which coupled with future improvements on direct
measurements of the evolution of the IGM volume ionized fraction, will
lead to more robust models of reionization.

\acknowledgements
SLF acknowledges support from from the National Science Foundation
through AAG award AST 1518183.
AD acknowledges support from HST award HST-AR-15013.005-A.
JPP acknowledges support from the European Research Council under the
European Communitys Seventh Framework Programme (FP7/2007-2013) via
the ERC Advanced Grant "STARLIGHT: Formation of the First Stars"
(project number 339177).  CDV acknowledges financial support from the
Spanish Ministry of Economy and Competitiveness (MINECO) through grant
RYC-2015-1807.  We thank the anonymous referee for their
comments, which significantly improved this paper.
The authors acknowledge George Becker, Rychard Bouwens, Volker Bromm,
Richard Ellis, Andrea
Ferrara, Jason Jaacks, Harley Katz, Eiichiro Komatsu, Charlotte Mason,
Andrei Mesinger, Milos Milosavljevic, Casey Papovich, Laura
Pentericci, Alice Shapley, Kim-Vy Tran, Tomasso Treu,
Anne Verhamme and Stephen Wilkins for useful comments and conversations which improved
this paper.  We also thank Jason Jaacks, Andrei Mesinger and Joaqim Rosdahl for providing simulation
data, Dan Stark and Mengtao Tang for providing their observational
results, and Marc Postman for providing estimated \emph{LUVOIR} sensitivities.  SLF and J-PP thank Romeel Dav\'{e} and the ``Reionization: A
Multi-Wavelength Approach'' conference in Kruger Park, South Africa,
where the idea for this project was hatched.
SLF also thanks the University of Sussex for hosting him in
January 2018 while this project was being worked on, and specifically
the Rights of Man Pub in Lewes, where a significant amount of this writing was done.

 %\bibliographystyle{/Users/sf8542/Latex/files/apj}
 %\bibliography{/Users/sf8542/Latex/files/stevenf}

\appendix

In this Appendix we show the special case of a simple
reionization model, using the escape fraction results from FiBY (shown in
Figure~\ref{fig:fesc}) with no scale factor applied.
We use the same luminosity functions from F16
described in \S~\ref{sec:mcmclf}, mapping UV luminosity to halo mass
as described in \S~\ref{sec:mcmcam}.  For this simple analysis, we emulate an evolving limiting halo mass 
by making the assumption that before reionization $M_{h,supp} =$ 8, and after
$M_{h,supp} =$ 9 (similar to the reionization-epoch results
from \citealt{okamoto08}).  We then assume that reionization starts at $z =$
12, and ends at $z =$ 6, and evolve the filtering mass linearly with
redshift between those points.  Lastly, we set the limiting magnitude as
that which corresponds to the filtering mass at each redshift from our
abundance matching analysis.  

Deriving $\rho_{UV}$ by integrating the UV luminosity function down to
a limiting magnitude corresponding to the appropriate filtering
mass, the ionizing emissivity is $\dot{N}_{ion}$ = $\xi_{ion}
\rho_{UV} f_\mathrm{esc}$.  For $\xi_{ion}$ we adopt the
recent observational results from \citet{bouwens16}, assuming log($\xi_{ion}$) = 25.34 for galaxies with
$M<-$20 (similar to the value assumed by \citealt{finkelstein12b}, which
corresponds to a stellar population with a metallicity of 0.2$Z$\sol\
that is continuously forming stars), and log($\xi_{ion}$) = 25.8 for fainter galaxies, which
corresponds to the results for the bluest galaxies at $z =$ 5.1--5.4.

As our escape fraction is assumed to vary with halo mass (and thus UV
magnitude), $\rho_{UV}$ (and thus the ionizing emissivity) is calculated in magnitude bins of
$\Delta$$M_\mathrm{UV}$=0.1 down to
our adopted limiting magnitude.  In each bin we draw an escape fraction
from the distributions shown in
Figure~\ref{fig:fesc}, using the halo mass for that magnitude and
redshift from our abundance matching results shown in
Figure~\ref{fig:fig1}.
The total ionizing emissivity is then the sum of these values.  We
also track the total number of ionizing photons created ($\dot{N}_{ion,intrinsic}$ = $\xi_{ion}
\times \rho_{UV}$), such that we can follow the population-averaged escape fraction.

We use a Monte Carlo analysis to calculate the ionizing emissivity at
each redshift, sampling our assumed UV luminosity functions,
$M_\mathrm{halo}$--$M_\mathrm{UV}$ relations, and FiBY-based
$f_\mathrm{esc} (M_\mathrm{h}$).  In each of the 10$^3$
steps of the Monte Carlo, we draw a
luminosity function at random from the MCMC chains from F16 described
above, and calculate the specific non-ionizing UV luminosity density $\rho_{UV}$
by integrating the luminosity function down to a redshift-dependent
limiting magnitude.  

\begin{figure*}[!t]
\epsscale{0.38}
\plotone{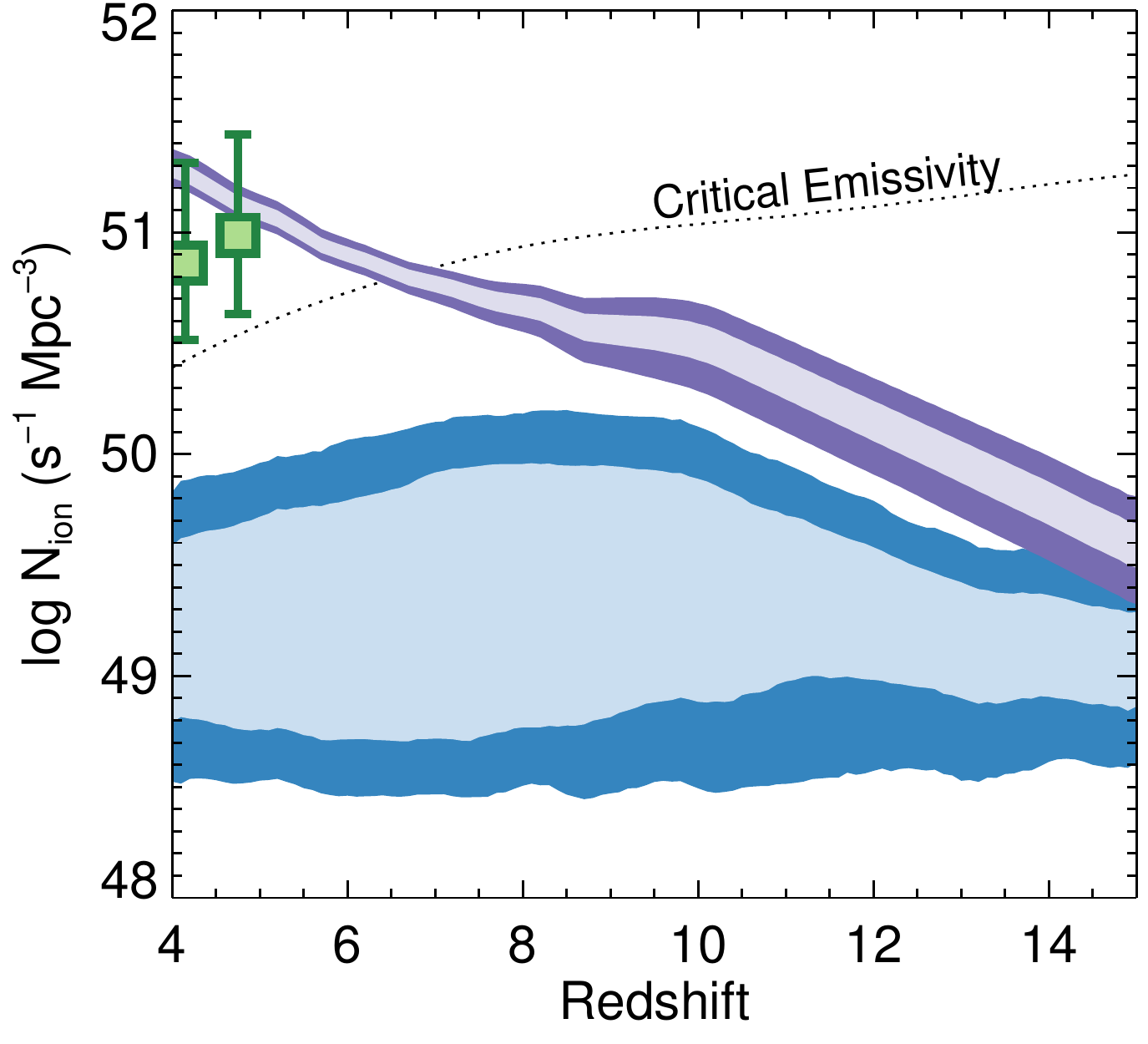}
\plotone{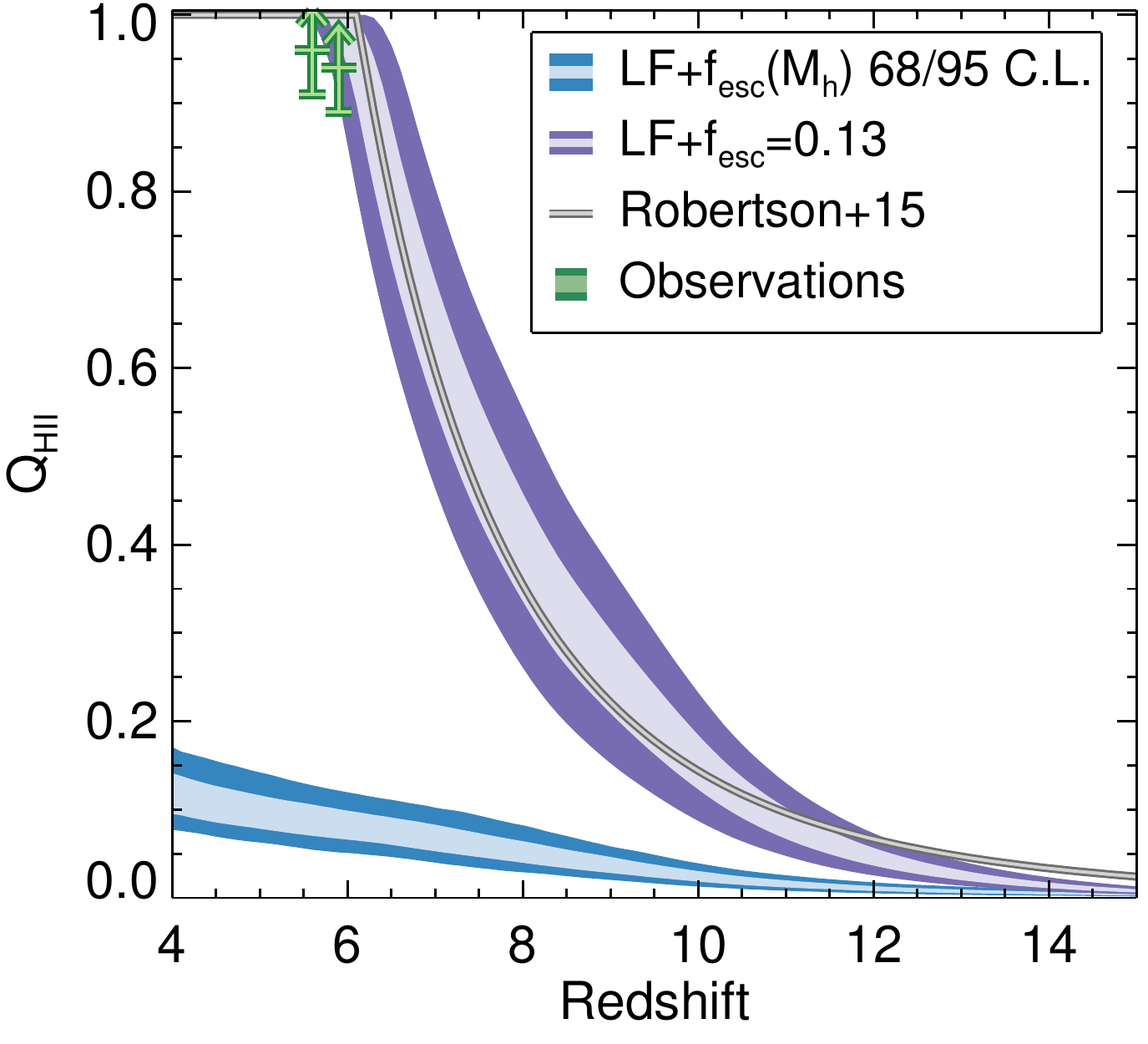}
\plotone{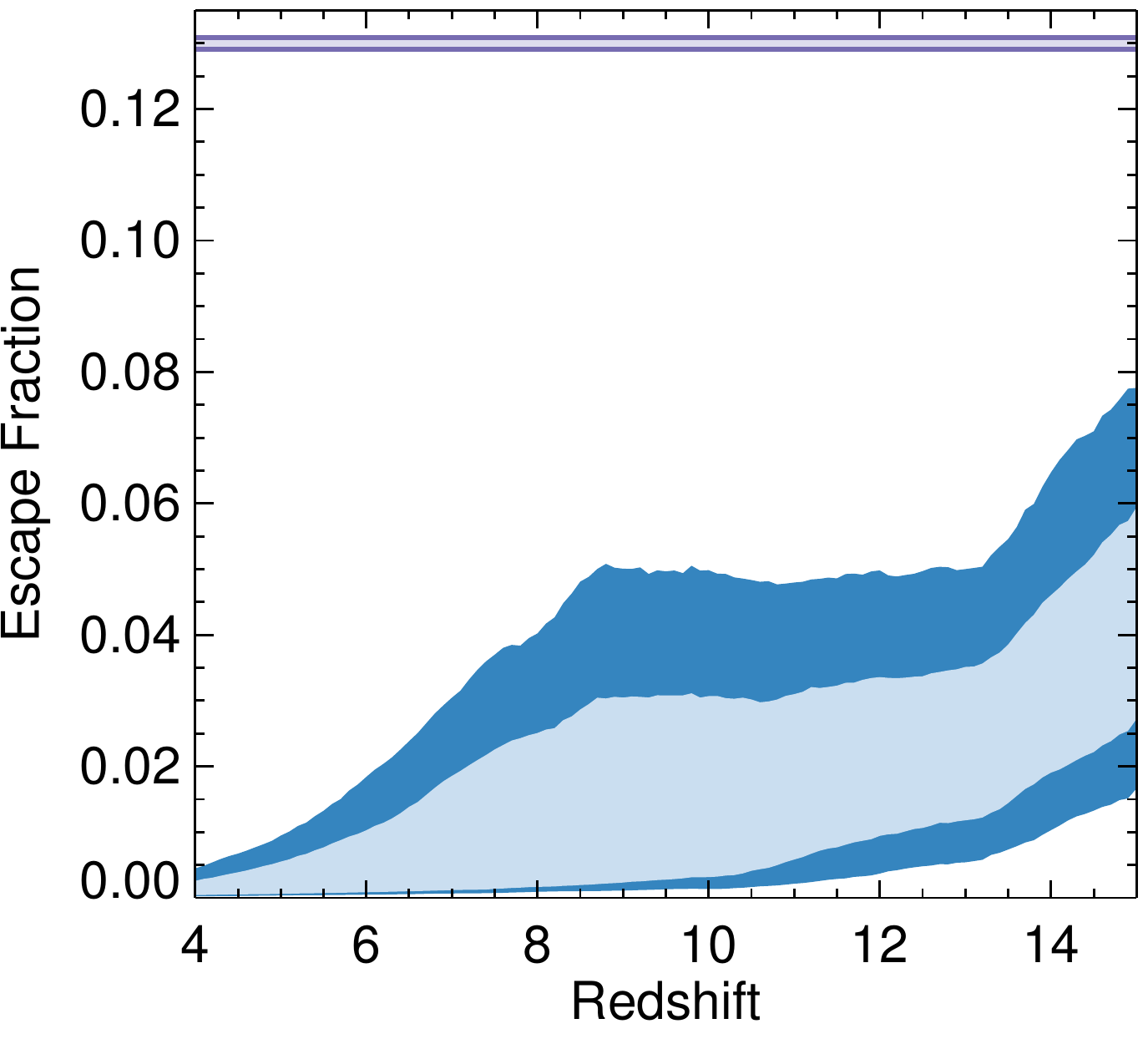}
\caption{The results for our simple model (described in this Appendix), combining the 
  \citet{paardekooper15} escape fraction results with the
  \citet{finkelstein16} luminosity functions, and somewhat standard assumptions
  on the limiting magnitude and ionizing photon production
  efficiency.  This simple model cannot complete reionization,
  motivating the more advanced modeling in the main
  paper.  Left) The co-moving ionizing emissivity, with our
  results shown as the light (68\% C.L.) and dark (95\% C.L.) blue shaded
  regions.  The purple shaded regions denotes the results if we
  instead had assumed a constant 13\% escape fraction \citep{finkelstein12b}.  The dotted black line
  denotes the required emissivity to maintain an ionized IGM from
  \citet{madau99} including a redshift-dependent clumping factor from
\citet{pawlik15}.  The green squares show the measurements of
  the ionizing emissivity from the Ly$\alpha$ forest from
  \citet{becker13}.  Center) The inferred evolution of the volume
  ionized fraction from the emissivity in the left panel, compared to model-independant constraints from
  \citet{mcgreer15}.  We also show the results from Robertson et al.\ (2015), who assumed $f_\mathrm{esc}$ =
  0.2.  Although the stochastic nature of ionizing photon
escape results in a wide range of ionizing emissivities from our
analysis, in general the they are more than an order of magnitude too
low to sustain an ionized IGM, even at $z <$ 6, and thus can be ruled
out.  Right) \emph{Population-averaged} escape fractions as a function
of redshift.  At $z <$ 10, when the bulk of reionizing photons are
expected to be produced, the average escape fraction is $\lesssim$ 2\%.}
\label{fig:appendix}
\end{figure*}

The ionizing emissivity results are shown in the left panel of
Figure~\ref{fig:appendix}, with the light and dark-blue shaded regions
denoting the 68 and 95\% confidence levels, respectively.  The large range covered
by these intervals is predominantly due to the broad distribution of
potential escape fractions, which is marginalized over by our Monte
Carlo method.  We compare our results to the observations of the
ionizing background from \citet{becker13}, which were inferred by
measurements of both the IGM temperature and opacity to Ly$\alpha$ and
ionizing photons, discussed in \S~\ref{sec:mcmc2b}.
Our computed ionizing emissivity falls well below the values observed in the IGM at $z =$ 4--4.75.  
For reference, we also show in purple the inferred ionizing emissivity
if one performs the same analysis assuming a constant 13\% escape
fraction \citep{finkelstein12b} for all galaxies at all luminosities
and redshifts, which, similar to the R15 model, matches the shown constraints.

The volume ionized fraction, calculated following \S~\ref{sec:diffeq}, is shown in the center panel of
Figure~\ref{fig:appendix} as the blue shaded regions (where purple again
denotes the special case of $f_\mathrm{esc}\!\!=$13\%).  We compare to the
model-independent observations of this quantity from
\citet{mcgreer15}, who used the dark-pixel fraction in the
Ly$\alpha$ and Ly$\beta$ forests of $z >$ 6 quasars to find lower limits
of  $Q_{H_{II}} >$ 0.96 ($\pm$0.05) at $z =$ 5.6, and $Q_{H_{II}} >$ 0.95 ($\pm$0.05) at $z =$
5.9.  While these observations imply that the IGM is predominantly
reionized by $z =$ 5.5, this simple model predicts $Q_{H_{II}}
\sim$ 0.1 at that redshift.  

However, this result should not be surprising, as indicated by the
right panel of Figure~\ref{fig:appendix}, which shows the average escape
fraction as a function of redshift from this analysis.  This was calculated by tracking the ratio
of the total number of escaping ionizing photons to
the total number of such photons created at a given redshift, and thus
is a \emph{population-averaged} escape fraction.  The
68\% confidence level on this quantity is
$<$2\% at $z <$ 8, and $<$4\% at $z <$ 14, much less than the
typically assumed values of $\geq$10\%.  As shown in purple in the left panel of
Figure~\ref{fig:appendix}, if one does an identical analysis except for assuming a flat ionizing photon escape
fraction of 13\% \citep{finkelstein12b}, one not only completes reionization by $z \sim$ 6,
but also satisfies the ionizing emissivity constraints at $z =$ 4.75.
However, it is important to note that extrapolating this simple assumption of a
constant 13\% escape fraction results to $z <$ 4 results in a galaxy ionizing emissivity
which is \emph{significantly ruled out} by the observed ionizing
emissivity at lower redshifts \citep[see also][]{becker13,stanway16}.

This simple model has a number of pitfalls, which result in a failure
to match observations.  First, it
assumed that the simulated escape fractions were properly normalized,
which may not be the case (see \S~\ref{sec:mcmcfesc}).  Second, it
assumed both a fixed pre-reionization limiting halo mass of 10$^{8}$
M\sol, while some simulations show Population II star formation occurs
in halos up to an order of magnitude lower in mass
\citep[e.g.,][]{paardekooper13,xu16}, and a fixed post-reionization
photo-suppression mass of 10$^{9}$ M\sol.  Third, we had assumed a fixed
value of the ionizing photon production efficiency, while this value
likely evolves with redshift and may also depend on host galaxy
luminosity/mass \citep[e.g.,][]{bouwens16,stark15,stark15b,stark17}.
Lastly, our initial model assumed that only galaxies were the sources
of ionizing photons, while it is possible that an AGN contribution
may be warranted, especially at the end of the reionization process.
The failure of this simple model motivates the more advanced modeling
described in this paper.

\end{document}